\documentclass[preprint2,numberedappendix,tighten]{aastex6}  
\shorttitle{Characterizing Signal Loss in the 21\,cm Reionization Power Spectrum}
\shortauthors{Cheng et al.}

\usepackage{amsmath}
\usepackage{graphicx}
\usepackage[figuresright]{rotating}
\usepackage{natbib}
\usepackage{ctable}
\usepackage{comment}
\usepackage{bm}
\usepackage{xparse}
\defcitealias{ali_et_al2015}{A15}

\NewDocumentCommand{\evalat}{sO{\big}mm}{%
  \IfBooleanTF{#1}
   {\mleft. #3 \mright|_{#4}}
   {#3#2|_{#4}}%
}

\newcommand{\x}{\mathbf{x}}

\newcommand{\C}{\mathbf{C}}
\newcommand{\Chat}{\mathbf{\widehat{C}}}
\newcommand{\F}{\mathbf{F}}

\newcommand{\Q}{\mathbf{Q}}
\newcommand{\I}{\mathbf{I}}

\newcommand{\invC}{\ensuremath{\C^{-1}}}

\DeclareMathOperator{\Tr}{tr}
\newcommand{\half}{\ensuremath{\frac{1}{2}}}
\newcommand{\PDeriv}[2]{\ensuremath{\frac{\partial #1}{\partial #2}}}

\begin{document}
\title{Characterizing Signal Loss in the 21\,cm Reionization Power Spectrum: \\A Revised Study of PAPER-64}

\author{
Carina Cheng\altaffilmark{1}$^{,\diamond}$,
Aaron R. Parsons\altaffilmark{1,2},
Matthew Kolopanis \altaffilmark{3},
Daniel C. Jacobs\altaffilmark{3}, 
Adrian Liu\altaffilmark{1,4}$^{,\dagger}$, 
Saul A. Kohn\altaffilmark{5},
James E.~Aguirre\altaffilmark{5},
Jonathan C. Pober\altaffilmark{6}, 
Zaki S. Ali\altaffilmark{1}, 
Gianni Bernardi\altaffilmark{7,8,9}, 
Richard F. Bradley\altaffilmark{10,11,12},
Chris L. Carilli\altaffilmark{13,14},
David R. DeBoer\altaffilmark{2}, 
Matthew R. Dexter\altaffilmark{2},
Joshua S. Dillon\altaffilmark{1}$^{,*}$,
Pat Klima\altaffilmark{11},
David H. E. MacMahon\altaffilmark{2},
David F. Moore\altaffilmark{5},
Chuneeta D. Nunhokee\altaffilmark{8},
William P. Walbrugh\altaffilmark{7},
Andre Walker\altaffilmark{7}
}

\altaffiltext{1}{Astronomy Dept., U. California, Berkeley, CA}
\altaffiltext{2}{Radio Astronomy Lab., U. California, Berkeley CA}
\altaffiltext{3}{School of Earth and Space Exploration, Arizona State U., Tempe AZ}
\altaffiltext{4}{Berkeley Center for Cosmological Physics, Berkeley, CA} 
\altaffiltext{5}{Dept. of Physics and Astronomy, U. Penn., Philadelphia PA} 
\altaffiltext{6}{Dept. of Physics, Brown University, Providence RI}
\altaffiltext{7}{Square Kilometer Array, S. Africa, Cape Town South Africa}
\altaffiltext{8}{Dept. of Physics and Electronics, Rhodes University, South Africa}
\altaffiltext{9}{INAF-Instituto di Radioastronomia, Bologna Italy}
\altaffiltext{10}{Dept. of Electrical and Computer Engineering, U. Virginia, Charlottesville VA}
\altaffiltext{11}{National Radio Astronomy Obs., Charlottesville VA}
\altaffiltext{12}{Dept. of Astronomy, U. Virginia, Charlottesville VA}
\altaffiltext{13}{National Radio Astronomy Obs., Socorro NM}
\altaffiltext{14}{Cavendish Lab., Cambridge UK}

	




\begin{abstract}
The Epoch of Reionization (EoR) is an uncharted era in our Universe's history during which the birth of the first stars and 
galaxies led to the ionization of neutral hydrogen in the intergalactic medium. There are many experiments investigating the 
EoR by tracing the 21\,cm line of neutral hydrogen. Because this signal is very faint and difficult to isolate, it is crucial to develop analysis techniques that maximize sensitivity and suppress contaminants in data. It is also imperative to understand the trade-offs between different analysis methods and their effects on power spectrum estimates. Specifically, with a statistical power spectrum detection in HERA's foreseeable future, it has become increasingly important to understand how certain analysis choices can lead to the loss of the EoR signal. In this paper, we focus on signal loss associated with power spectrum estimation. We describe the origin of this loss using both toy models and data taken by the 64-element configuration of the Donald C. Backer Precision Array for Probing the Epoch of Reionization (PAPER). In particular, we highlight how detailed investigations of signal loss have led to a revised, higher 21\,cm power spectrum upper limit from PAPER-64. Additionally, we summarize errors associated with power spectrum error estimation that were previously unaccounted for. We focus on a subset of PAPER-64 data in this paper; revised power spectrum limits from the PAPER experiment are presented in a forthcoming paper by Kolopanis et al. (\textit{in prep.}) and supersede results from previously published PAPER analyses.
\end{abstract}


\section{Introduction}
\label{sec:Intro}

{\let\thefootnote\relax\footnote{$^{\diamond}$\href{mailto:ccheng@berkeley.edu}{ccheng@berkeley.edu}}}
{\let\thefootnote\relax\footnote{$^{\dagger}$Hubble Fellow}}
{\let\thefootnote\relax\footnote{$^{*}$NSF AAPF Fellow}}
\setcounter{footnote}{0}

By about one billion years after the Big Bang ($z \sim 6$), the first stars and galaxies are thought to have ionized all the 
neutral hydrogen that dominated the baryonic matter content in the Universe. This transition period, during which the first 
luminous structures formed from gravitational collapse and began to emit intense radiation that ionized the cold neutral gas 
into a plasma, is known as the Epoch of Reionization (EoR). The EoR is a relatively unexplored era in our Universe's history, which spans the birth of the first stars to the full reionization of the intergalactic medium (IGM). This epoch encodes important information regarding the nature of the first galaxies and the processes of structure formation. 
Direct measurements of the EoR would unlock powerful characteristics about the IGM, revealing connections 
between the matter distribution exhibited via cosmic microwave background (CMB) studies and the highly structured 
web of galaxies we observe today (for a review, see \citet{barkana_and_loeb2001}, \citet{furlanetto_et_al2006} and \citet{loeb_furlanetto_2013}).

One promising technique to probe the EoR is to target the 21\,cm wavelength signal that is emitted and absorbed by neutral hydrogen via 
its spin-flip transition (\citealt{furlanetto_et_al2006}; \citealt{barkana_and_loeb2008}; \citealt{morales_and_wyithe2010}; \citealt{pritchard_and_loeb2010}; \citealt{pritchard_loeb2012}). This technique is powerful because it can be observed both spatially and as a function of redshift --- that is, the wavelength 
of the signal reaching our telescopes can be directly mapped to a distance from where the emission originated before 
stretching out as it traveled through expanding space. Hence, 21\,cm tomography offers a unique window into both the spatial and temporal
evolution of ionization, temperature, and density fluctuations.

In addition to the first tentative detection of our \textit{Cosmic Dawn} (pre-reionization era) made by the Experiment to Detect the Global EoR Signature (EDGES; \citealt{bowman_et_al2018}; \citealt{bowman2010}), there are several radio telescope experiments that have succeeded in using 
the 21\,cm signal from hydrogen to place constraints on the brightness of the signal. Examples of experiments investigating the 
mean brightness temperature of the 21\,cm signal relative to the CMB are the Large Aperture Experiment to Detect the Dark Ages (LEDA; \citealt{bernardi_et_al2016}), the 
Dark Ages Radio Explorer (DARE; \citealt{burns2012}), the Sonda Cosmol\'ogica de las Islas para la Detecci\'on de 
Hidr\'ogeno NeutroSciHi (SCI-HI; \citealt{voytek2014}), the Broadband Instrument for Global HydrOgen ReioNisation Signal 
(BIGHORNS; \citealt{sokolowski2015}), and the Shaped Antenna measurement of the background RAdio Spectrum (SARAS; 
\citealt{patra2015}). Radio interferometers which seek to measure statistical power spectra include the Giant Metre-wave 
Radio Telescope (GMRT; \citealt{paciga_et_al2013}), the LOw Frequency ARray (LOFAR; \citealt{van_haarlem_et_al2013}), 
the Murchison Widefield Array (MWA; \citealt{tingay_et_al2013}), the 21 Centimeter Array (21CMA; 
\citealt{peterson_et_al2004}; \citealt{wu2009}), the Square Kilometre Array (SKA; \citealt{koopmans_et_al2015}), and PAPER (\citealt{parsons_et_al2010}). The Hydrogen Epoch of 
Reionization Array (HERA), which is currently being built, is a next-generation instrument that aims to combine lessons 
learned from previous experiments and has a forecasted sensitivity capable of a high-significance power spectrum 
detection with an eventual $350$ elements using current analysis techniques (\citealt{pober_et_al2014}; \citealt{liu_parsons_2016}; \citealt{dillon_parsons2016}; \citealt{deboer_et_al2017}).

The major challenge that faces all 21\,cm experiments is isolating a small signal that is buried underneath foregrounds and 
instrumental systematics that are, when combined, four to five orders of magnitude brighter \citep[e.g.,][]{santos_et_al2005, ali_et_al2008, deOliveiraCosta_et_al2008, jelic_et_al2008, bernardi_et_al2009, bernardi_et_al2010, ghosh_et_al2011, pober_et_al2013b, bernardi_et_al2013, dillon_et_al2014, kohn_et_al2016}. A clean measurement therefore requires an intimate understanding of how data analysis choices, which are often tailored to maximize sensitivity and minimize contaminants, affect power spectrum results. More specifically, it is imperative to develop techniques that ensure 
the accurate extraction and recovery of the EoR signal, despite the analysis method chosen and how much loss (of both contaminants and the EoR signal) accompanies the method. In this paper, we specifically discuss signal loss --- the loss of the \textit{cosmological} signal --- associated with power spectrum estimation. This is an issue that is essential to investigate for a robust 21\,cm power spectrum analysis and one that has motivated a revised PAPER analysis. We first approach this topic from a broad perspective, and then perform a detailed 
case study using data from the 64-element configuration of PAPER. In this study we use a subset of PAPER-64 data to illustrate our revised analysis methods, while a related paper, Kolopanis et al. (\textit{in prep.}), builds off of the methods in this paper to present revised PAPER-64 results for multiple redshifts and baseline types.

Finally, we also highlight several additional errors made in previous PAPER analyses, including those related to bootstrapping and error estimation. This paper accompanies the erratum of \citet{ali_et_al2018} and adds to the growing foundations of lessons which have been documented, for example, in \citet{Paciga2013}, \citet{Patil2016}, and \citet{Jacobs2016}, by the GMRT, LOFAR, and MWA projects respectively. These lessons are imperative as the community as a whole moves towards higher sensitivities and potential EoR detections.

This paper is organized into three main sections. In Section \ref{sec:SiglossOverview} we use toy models to develop intuition about signal loss, its origin, and its subtleties. In Section \ref{sec:CaseStudy}, we present a case study using data from the PAPER-64 array, highlighting key changes from the signal loss methods used in the published result in \citet{ali_et_al2015}, henceforth known as \citetalias{ali_et_al2015}, which previously underestimated loss. In Section \ref{sec:OtherErrors}, we summarize additional lessons learned since \citetalias{ali_et_al2015} that have shaped our revised analysis. Finally, we conclude in Section \ref{sec:Con}.


\section{Signal Loss Toy Models}
\label{sec:SiglossOverview}

Signal loss refers to attenuation of the target cosmological signal 
in a power spectrum estimate. Certain analysis techniques can cause this loss, and if the amount of loss is not quantified accurately, it could lead to false non-detections and overly aggressive upper limits. Determining whether an analysis pipeline is lossy, and estimating the amount of loss if so, has subtle challenges but is necessary to ensure the accuracy of any result. 

One type of signal loss can occur when weighting data by itself. Broadly speaking, a dataset can be weighted to emphasize certain features and minimize others. For example, one flavor of weighting employed by previous PAPER analyses is inverse covariance weighting in frequency, which is a generalized version of inverse variance weighting that also takes into account frequency correlations (\citealt{liu_tegmark2011}; \citealt{dillon_et_al2013a}; \citealt{liu_et_al2014a}; \citealt{liu_et_al2014b}; \citealt{dillon_et_al2014}; \citealt{dillon_et_al2015}). Using such a technique enables the down-weighting of contaminant modes that obey a different covariance structure from that of cosmological modes. However, a challenge of inverse covariance 
weighting is in estimating a covariance matrix that is closest to the true covariance of the data; the discrepancy between the two, as we will see, can have large impacts on signal loss. In this paper we focus specifically on loss associated with the use of an empirically estimated covariance matrix with the ``optimal quadratic estimator'' formalism.
This loss was significantly 
underestimated in the \citetalias{ali_et_al2015} analysis and is the main reason motivating a revised power spectrum result.


\subsection{The Quadratic Estimator Method}
\label{sec:QE}


We begin with an overview of the quadratic estimator (QE) formalism used for power spectrum estimation. The goal of power spectrum analysis is to produce an unbiased estimator of the EoR power spectrum in the presence of both noise and foreground emission. Prior to power spectrum estimation, the data will often have been prepared to have minimal foregrounds by some method of subtraction, so this foreground emission may appear either directly (because it was not subtracted) or as a residual of some subtraction process not in the power spectrum domain. If an accurate estimate of the total covariance of the data is known, including both the desired signal and any contaminants, then the ``optimal quadratic estimator'' formalism provides a method of producing a minimum variance, unbiased estimator of the desired signal, as shown in 
\citet{liu_tegmark2011}, \citet{dillon_et_al2013a}, \citet{liu_et_al2014a}, \citet{liu_et_al2014b}, \citet{trott_et_al2012}, \citet{dillon_et_al2014}, \citet{dillon_et_al2015}, \citet{switzer_et_al2015}, and \citet{trott_et_al2016}. 

Suppose that the measured visibilities for a single baseline in Jy are arranged as a data vector, $\textbf{x}$. It has length $N_{t} N_{f}$,
where $N_{t}$ is the number of time integrations and $N_{f}$ is the number of frequency channels. The covariance of the data is given by 
\begin{equation}
\textbf{C} \equiv \langle\textbf{xx}^{\dagger}\rangle = \textbf{S} + \textbf{U}
\end{equation}
where the average over an ensemble of data realizations produces the true covariance, and we further assume it may be written as the sum of the desired cosmological signal $\textbf{S}$ and other terms $\textbf{U}$.  

We are interested in estimating the three-dimensional power spectrum of the EoR.  
Visibilities are measurements of the Fourier transform of the sky along two spatial dimensions (using the flat-sky approximation), and since we are interested in three-dimensional Fourier modes we only need to take one Fourier transform of our visibilities along the line-of-sight dimension.  We consider band powers $P^\alpha$ of the power spectrum of $\textbf{x}$ over some range in cosmological $\mathbf{k}$, where $\alpha$ indexes a waveband in $k_{\parallel}$ (a cosmological wavenumber $k_{\parallel}$ is the Fourier dual to frequency under the delay approximation (\citealt{parsons_et_al2012b}), which is a good approximation for the short baselines that PAPER analyzes).  The fundamental dependence of the covariance on the power spectrum band powers $P^\alpha$ is encoded as 
\begin{equation}
\textbf{S} = \sum_\alpha P^\alpha \frac{\partial\textbf{C}}{\partial P^\alpha} \equiv \sum_\alpha P^\alpha \textbf{Q}^\alpha
\end{equation}
where we define $\frac{\partial\textbf{C}}{\partial P^\alpha} \equiv \textbf{Q}^{\alpha}$. In other words, $\textbf{Q}$ describes the response of the covariance to a change in the power spectrum, relating a quadratic statistic of the data (the covariance) to a quadratic statistic in Fourier-space (the power spectrum). 

The optimal quadratic estimator prescription is then to compute
\begin{equation}
\label{eq:OQE}
\widehat{P}^{\alpha}  = \sum_\beta ({\F^{-1}})^{\alpha\beta} (\widehat{q}^{\beta} - \widehat{b}^{\beta} )
\end{equation}
where $\F$ is the Fisher matrix (which determines errors on the power spectrum estimate)
\begin{equation}
F^{\alpha \beta} \equiv \frac{1}{2} \textrm{tr} \left( \C^{-1} \textbf{Q}^{\alpha} \C^{-1} \textbf{Q}^{\beta} \right),
\end{equation}
$\widehat{q}$ is the un-normalized power spectrum estimate
\begin{equation}
\label{eq:OQEQuadratic}
\widehat{q}^\alpha =  \half \textbf{x}^\dagger \invC \textbf{Q}^{\alpha}  \invC \textbf{x},
\end{equation}
and $\widehat{b}$ is the additive bias
\begin{equation}
\label{eq:OQELinear}
\widehat{b}^{\alpha} = \half \Tr\left( \mathbf{U} \invC \textbf{Q}^{\alpha} \invC \right).
\end{equation}
The power spectrum estimator in Equation \eqref{eq:OQE} is the minimum variance (smallest error bar) estimate of the power spectrum subject to the constraint that it is also unbiased; that is, the ensemble average of the estimator is equal to its true value
\begin{equation}
\label{eq:super_unbiased}
\langle \widehat{P}^{\alpha} \rangle = P^\alpha
\end{equation}
\citep{tegmark_et_al1997a,bond_et_al1998}.

Intuitively, the estimator must be capable of ``suppressing" or ``removing" the effects of contaminants in order to obtain an unbiased estimate of the power spectrum. By construction, the subtraction of the residual foreground and noise bias accomplishes this, removing any additive bias. However, the $\mathbf{C}^{-1}$ piece of Equation \eqref{eq:OQEQuadratic} also has the effect of suppressing residual foregrounds and noise, in both the additive bias and any contributions the residuals may have to the variance. 

More specifically, the effect of the weighting in Equation \eqref{eq:OQEQuadratic} is to project out the modes of $\textbf{U}$ with a different covariance structure than $\mathbf{S}$ in the power spectrum estimate, and the effect of Equation \eqref{eq:OQELinear} is to subtract out the remaining bias. Similar effects for a realistic model of the EoR and foregrounds are shown in \citet{liu_tegmark2011}. If the covariance structure of the contaminants is sufficiently different from the desired power spectrum, then the linear bias term may be expected to be quite small, and it is only necessary to know $\C$ and $\textbf{Q}^{\alpha}$, but not $\mathbf{U}$. Since the foregrounds are expected to be strongly correlated between frequencies whereas the EoR is not, we expect different covariance structures and therefore a small linear bias. Moreover, because the linear bias is always positive and there is no multiplicative bias, the quadratic-only term will always produce an estimate which is {\it high} relative to the true value, and which can conservatively be interpreted as an upper limit. These considerations, and the difficulty of obtaining an estimate for $\mathbf{U}$, motivate the neglect of the linear bias in the rest of this analysis.

Motivated by the desire to retain the advantageous behavior of suppressing contributions of $\mathbf{U}$ to estimates of the EoR power spectrum, we note that is possible to define a modified version of the quadratic estimator 
where Equation \eqref{eq:OQEQuadratic} is replaced by
\begin{equation}
\label{eq:qhat}
\widehat{q}^{\alpha} = \frac{1}{2}\textbf{x}^{\dagger}\textbf{R}\textbf{Q}^{\alpha}\textbf{R}\textbf{x}
\end{equation}
where $\textbf{R}$ is a weighting matrix chosen by the data analyst.  For example, inverse covariance weighting (the optimal form of QE) would set $\textbf{R} \equiv \textbf{C}^{-1}$ and a uniform-weighted case would use $\textbf{R} \equiv \textbf{I}$, the identity matrix.
Again, the matrix $\textbf{Q}^{\alpha}$ encodes the dependence of the covariance on the power spectrum but in practice also also does other things, including implementing a transform of the frequency domain visibilities to $\mathbf{k}$-space, taking into account cosmological scalings, and converting the visibilities from Jansky to Kelvin.

With an appropriate normalization matrix $\textbf{M}$, the quantity
\begin{equation}
\label{eq:phat}
\widehat{\textbf{P}} = \textbf{M}\widehat{\textbf{q}}
\end{equation}
is a sensible \textit{estimate} of the \textit{true} power spectrum $\textbf{P}$. 

To ensure that $\textbf{M}$ correctly normalizes our power spectrum, one may take the expectation value of Equation \eqref{eq:phat} to obtain
\begin{eqnarray}
\langle \widehat{P}^\alpha \rangle &=& \frac{1}{2} \sum_{\beta \gamma} M^{\alpha \gamma}  \textrm{tr} \left( \textbf{R}\textbf{Q}^{\gamma}\textbf{R} \textbf{Q}^{\beta}\right) P^\beta +\frac{1}{2} \sum_{ \gamma}  \Tr\left( \mathbf{U} \textbf{R} \textbf{Q}^{\gamma} \mathbf{R} \right) \nonumber \\
&\equiv & \sum_{\beta} W^{\alpha \beta} P^\beta +\frac{1}{2} \sum_{ \gamma}  \Tr\left( \mathbf{U} \textbf{R} \textbf{Q}^{\gamma} \mathbf{R} \right), \label{eq:Wpplusbias}
\end{eqnarray}
where $W^{\alpha \beta}$ are elements of a window function matrix. Considering the first term of this expression (again, we are assuming that the linear bias term is significantly suppressed; and if this is not the case, we are simply assuming that we are setting a conservative upper limit), if $\mathbf{W}$ ends up being the identity matrix for our choices of $\mathbf{R}$ and $\mathbf{M}$, then we recover Equation \eqref{eq:super_unbiased} for the first term, and we have an estimator that has no multiplicative matrix bias. However, Equation \eqref{eq:super_unbiased} is a rather restrictive condition, and it is possible to violate it and still have a sensible (and correctly normalized) power spectrum estimate. In particular, as long as the rows of $\mathbf{W}$ sum to unity, our power spectrum will be correctly normalized. Beyond this, the data analyst has a choice for $
\textbf{M}$, and for simplicity throughout this paper we choose $\textbf{M}$ to be diagonal. In a preview of what is to come, we also stress that the derivation that leads to Equation \eqref{eq:Wpplusbias} assumes that $\mathbf{R}$ and $\mathbf{x}$ are not correlated. If this assumption is violated, a simple application of the (now incorrect) formulae in this section can result in an improperly normalized power spectrum estimator that does not conserve power, i.e., one that has signal loss.

Given the advantages of inverse covariance weighting, a question arises of how one goes about estimating $\C$.  One method is to empirically derive it from the data $\textbf{x}$ itself.
Similar types of weightings that are based on variance information in data are done in \citet{chang_et_al2010} and \citet{switzer_et_al2015}. In previous PAPER analyses, one time-averages the data to obtain:
\begin{equation}
\widehat{\textbf{C}} \equiv \langle\textbf{xx}^{\dagger}\rangle_{t} \approx \langle  \textbf{xx}^{\dagger}\rangle,
\end{equation}
assuming $\langle\textbf{x}\rangle_{t} = 0$ (a reasonable assumption since fringes average to zero over a sufficient 
amount of time), where $\langle \rangle_{t}$ denotes a finite average over time. The weighting matrix for our empirically estimated inverse covariance weighting is then $
\textbf{R} \equiv \widehat{\textbf{C}}^{-1}$, where we use a hat symbol to distinguish the empirical covariance from the true covariance $\textbf{C}$.

In the next three sections, we use toy models to investigate the effects of weighting matrices on signal loss by experimenting with different matrices $\textbf{R}$ and examining their impact on the resulting power spectrum estimates $\widehat{\textbf{P}}$. Our goal in experimenting with weighting is to suppress foregrounds and investigate EoR losses associated with it. We note that we purposely take a thorough and pedagogical approach to describing the toy model examples given in the next few sections. The specifics of how signal loss appears in PAPER's analysis is later described in Section \ref{sec:CaseStudy}.


As a brief preview, we summarize our findings in the following sections here:

\begin{itemize}

\item If the covariance matrix is estimated from the data, a strong correlation between the estimated modes and the data will in general produce an estimate of the signal power spectrum which is strongly biased {\it low} relative to the true value.  In this context, this is what we call ``signal loss''  (Section \ref{sec:toymodel}).

\item The effect of the bias is worsened when the number of independent samples used to estimate the covariance matrix is reduced (Section \ref{sec:toymodel_frf}).

\item The rate at which empirical eigenvectors converge to their true forms depends on the sample variance in the empirical estimate and the shape of the empirical eigenspectrum. In general, larger sample variances lead to more loss (Section \ref{sec:toymodel_frf}).


\item Knowing these things, there are some simple ways of altering the empirical covariance matrix to decouple it from the data and produce unbiased power spectrum estimates (Section \ref{sec:otherweight}).


\end{itemize}

\subsection{Empirical Inverse Covariance Weighting}
\label{sec:toymodel}

Using a toy model, we will now build intuition into how weighting by the inverse of the empirically estimated covariance, $\widehat{\textbf{C}}^{-1}$, can give rise to signal loss.  We construct a simple dataset that contains visibility data with $100$ time integrations and $20$ frequency channels. This model represents realistic dimensions of about an hour of PAPER data which might be used for a power spectrum analysis. For PAPER-64 (both the \citetalias{ali_et_al2015} analysis and our new analysis) we use $\sim8$ hours of data (with channel widths of $0.5$ MHz and integration times of $43$ seconds), but here we scale it down with no loss of generality. 

\begin{figure}
	\centering
	\includegraphics[trim={0cm 0cm 0cm 0cm},clip,width=\columnwidth]{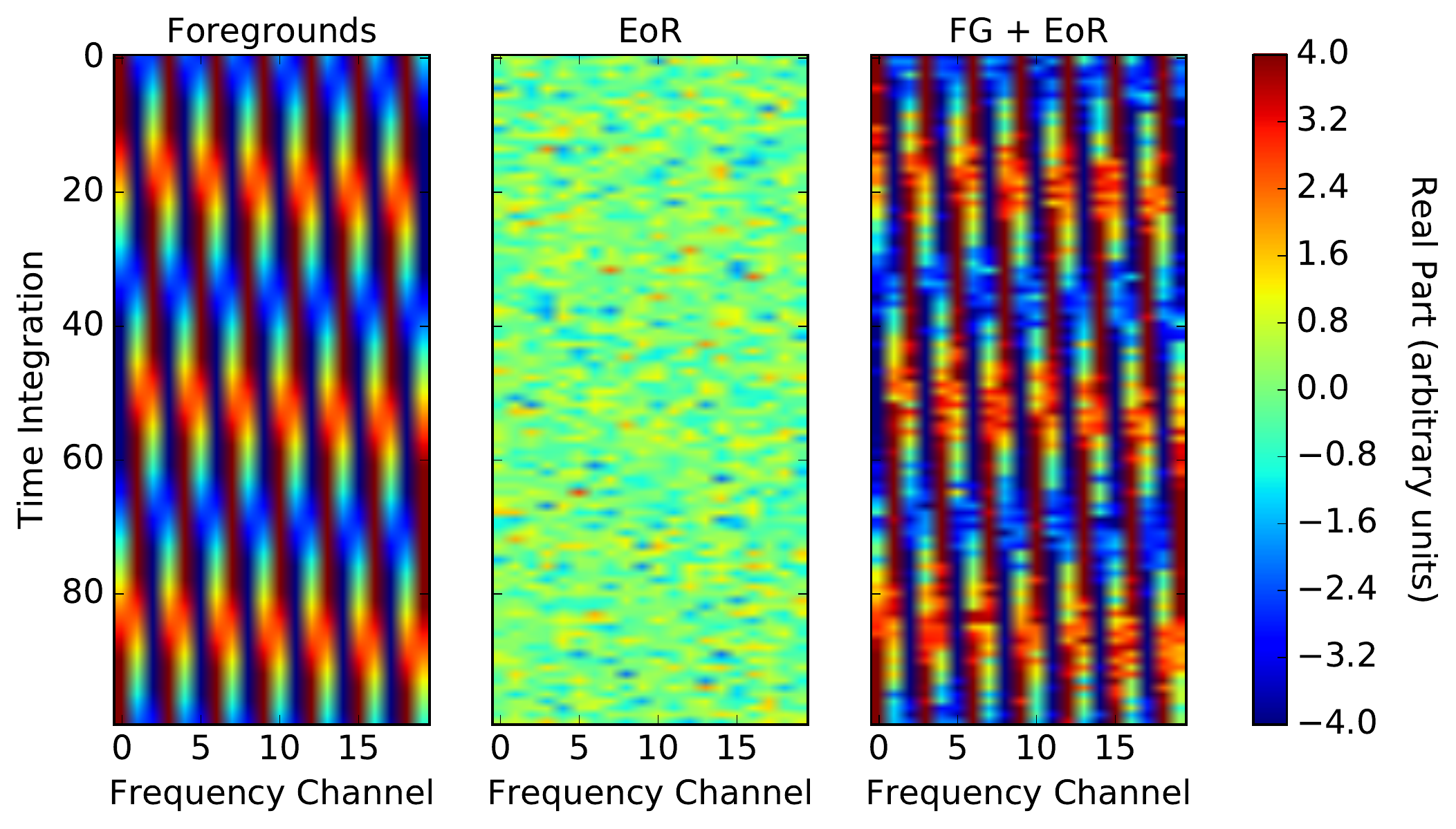}
	\caption{Our toy model dataset to which we apply different weighting schemes to in order to investigate signal loss. We model a mock foreground-only visibility with a sinusoid signal that varies smoothly in 
time and frequency. We model a mock visibility of an EoR signal as a random Gaussian signal. We add the two together to form $\textbf{x} = 
\textbf{x}_{\rm FG} + \textbf{x}_{\rm EoR}$. Real parts are shown here.}
	\label{fig:toy_sigloss1}
\end{figure}

We create mock visibilities, $\textbf{x}$, and assume a non-tracking, drift-scan observation. Hence, flat spectrum sources (away from zenith) lead to measured visibilities which oscillate in time and frequency. We therefore form a mock visibility measurement of a bright foreground signal, $\textbf{x}_{\rm FG}$, as a complex sinusoid that varies smoothly in time and frequency, a simplistic but realistic representation of a single bright source. We also create a mock visibility measurement of an EoR signal $\textbf{x}_{\rm EoR}$ as a complex, Gaussian random signal. A more realistic EoR signal would have a sloped power spectrum in $p(k)$ (instead of flat, as in the case of white noise), which could be simulated by introducing frequency correlations into the mock EoR signal. However, here we treat all $k$'s separately, so a simplistic white noise approximation can be used. Our combined data vector is then $\textbf{x} = 
\textbf{x}_{\rm FG} + \textbf{x}_{\rm EoR}$, to which we apply different weighting schemes throughout Section \ref{sec:SiglossOverview}. The three data components are shown in Figure 
\ref{fig:toy_sigloss1}. 

\begin{figure}
	\centering
	\includegraphics[trim={0cm 0cm 0cm 0cm},clip,width=\columnwidth]{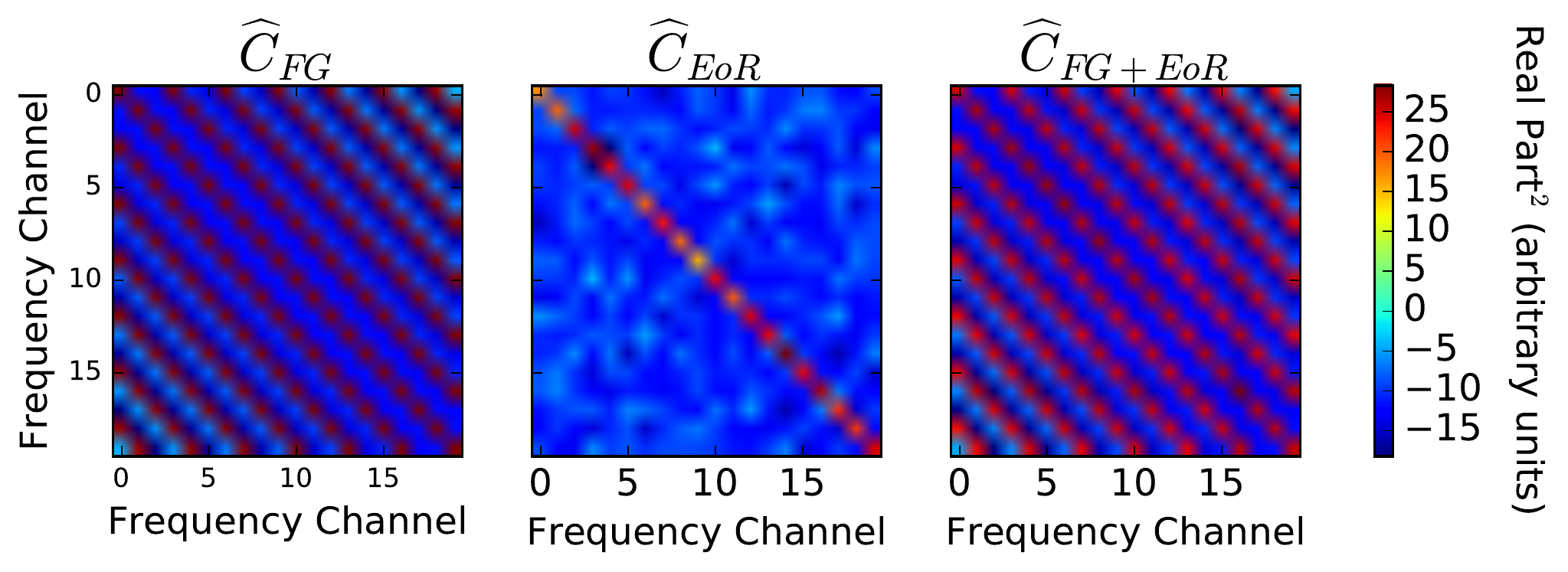}
	\includegraphics[trim={0cm 0cm 0cm 0cm},clip,width=\columnwidth]{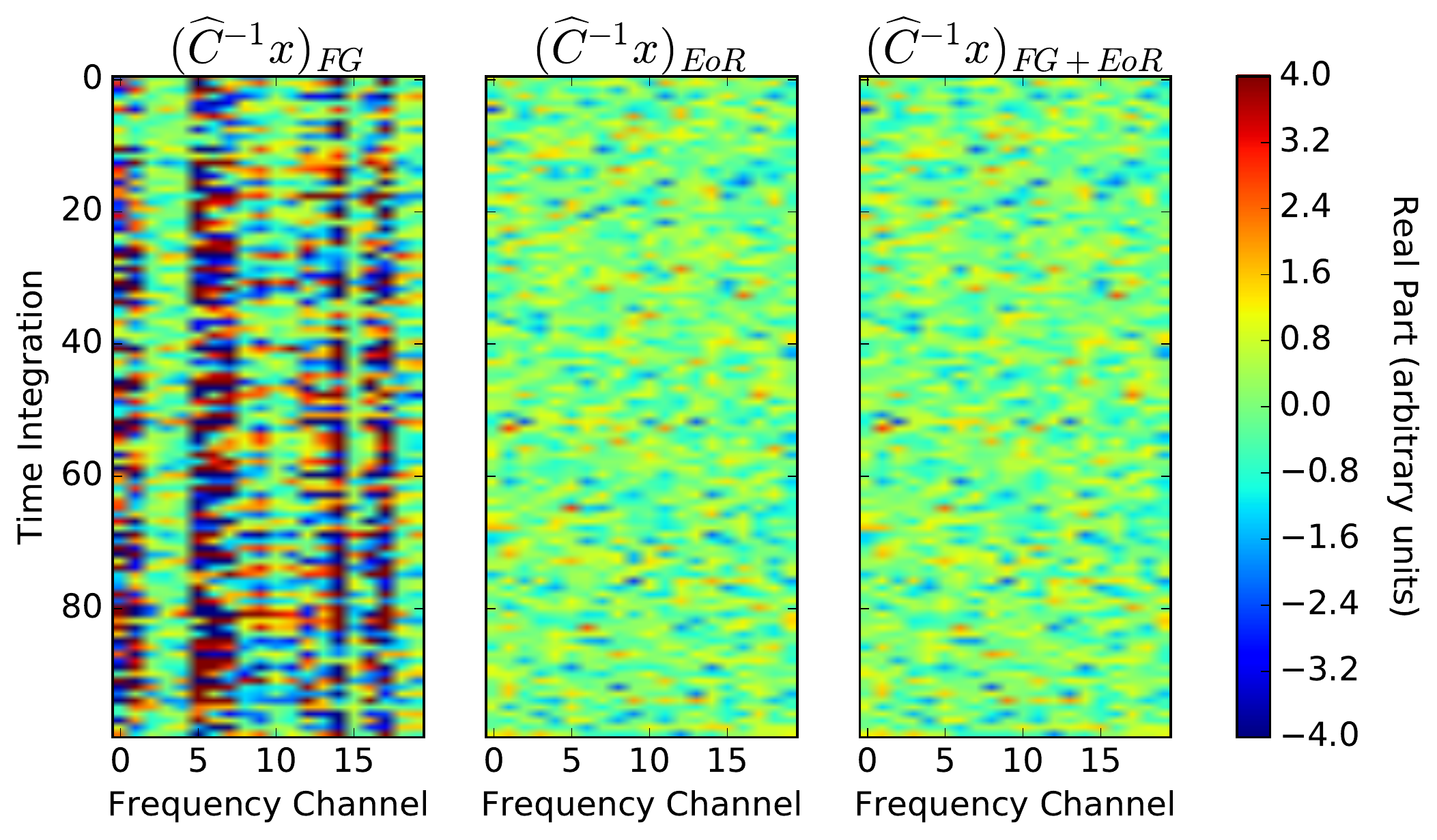}
	\caption{The estimated covariance matrices (top row) and inverse covariance-weighted data (bottom row) for FG only (left), EoR only 
(middle), and FG + EoR (right). Real parts are shown here.}
	\label{fig:toy_sigloss12}
\end{figure}

\begin{figure*}
	\centering
	\includegraphics[trim={0cm 0cm 0cm 0cm},clip,height=0.33\textwidth]{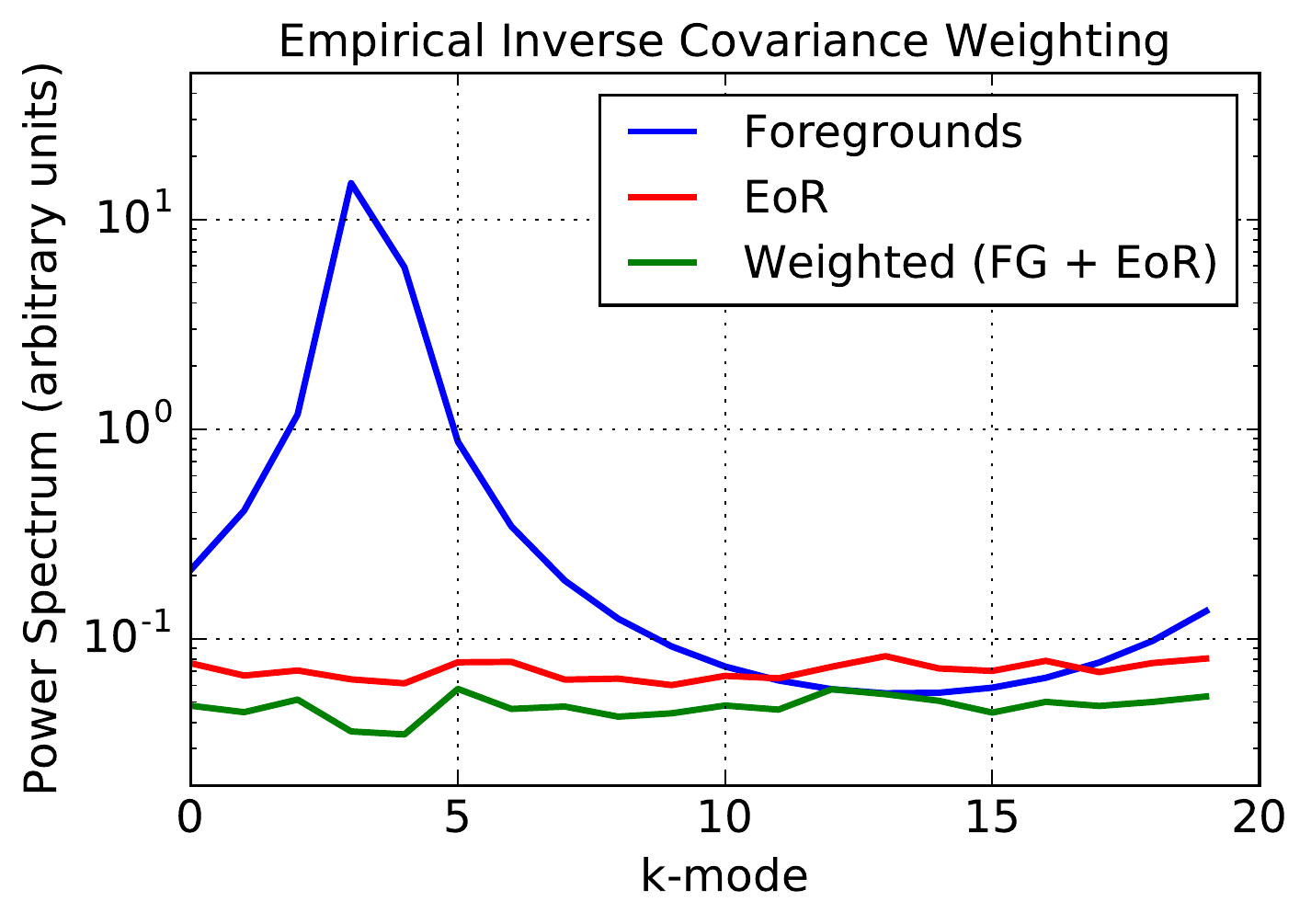}
	\includegraphics[trim={0cm 0cm 0cm 0cm},clip,height=0.33\textwidth]{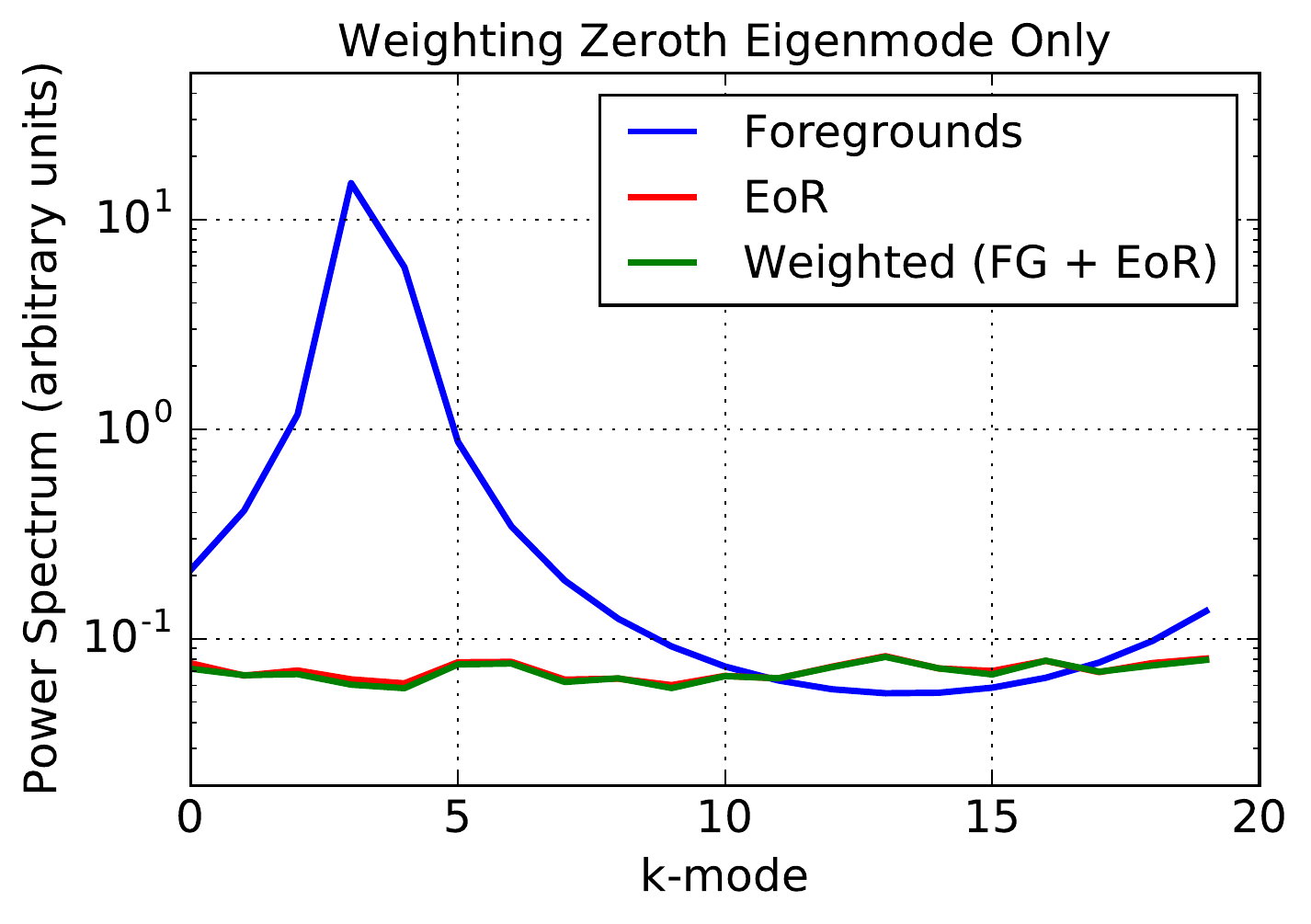}
	\caption{Resulting power spectrum estimates for the toy model simulation described in Section \ref{sec:toymodel} --- foregrounds only (blue), EoR only (red), and the weighted FG + EoR dataset (green). The power spectrum of the foregrounds peaks at a $k$-mode based on the frequency of the sinusoid used to create the mock FG signal. In the two panels, we compare using empirically estimated inverse covariance weighting where $\textbf{C}$ is derived from the data (left), and projecting out the zeroth eigenmode only (right). In the former case, signal loss arises from the coupling of the eigenmodes of $\widehat{\textbf{C}}$ to the data. 
There is negligible signal loss when all eigenmodes besides the foreground one are no longer correlated with the data.
}
	\label{fig:toy_sigloss3}
\end{figure*}

We compute the power spectrum of our toy model dataset $\textbf{x}$ using Equations \eqref{eq:qhat} and \eqref{eq:phat}, with $\textbf{R} \equiv \widehat{\textbf{C}}^{-1}$.  Figure \ref{fig:toy_sigloss12} shows the estimated covariances of our toy model datasets along with the $\widehat{\textbf{C}}^{-1}$ weighted data. The foreground sinusoid is clearly visible in $\widehat{\textbf{C}}_{\rm FG}$.  The power spectrum result is shown in green in the left plot of Figure \ref{fig:toy_sigloss3}. Also plotted in the figure are the uniform-weighted ($\textbf{R} \equiv \textbf{I}$) power spectrum of the individual components $\textbf{x}_{\rm FG}$ (blue) and $\textbf{x}_{\rm EoR}$ (red). As shown, our $\widehat{\textbf{C}}^{-1}$ weighted result successfully suppresses foregrounds,
demonstrated in Figure \ref{fig:toy_sigloss3} by the missing foreground peak in the weighted power spectrum estimate (green).  It is also evident that our result fails to recover the EoR signal --- it exhibits the correct shape, but the amplitude level is slightly low.  It is this behavior which we describe as signal loss.

As discussed in Section \ref{sec:QE}, this behavior is {\it not} expected in the case that we were to use a true $\textbf{C}^{-1}$ weighting.  Rather, we would obtain 
a nearly unbiased estimate of the power spectrum.   The key difference is that since $\widehat{\textbf{C}}$ is estimated from the data, its eigenvectors and eigenvalues are strongly coupled to the particular data realization that was used to compute it, and this coupling leads to loss.

For the case of an eigenmode which can be safely assumed to be predominantly a foreground, its presence in the true covariance matrix will result in the desired suppression via a kind of projection; whether or not it is strongly correlated with the the actual data vector is irrelevant.  However, in the case of an empirically estimated covariance matrix, the eigenmodes of $\widehat{\textbf{C}}_{\rm EoR}$ will both be incorrect and can be correlated with the data. If these incorrect eigenmodes are not correlated with the data, it will lead to non-minimum variance estimates but will not produce the suppression of the power spectrum amplitude as seen in the left plot of Figure \ref{fig:toy_sigloss3}. As described in Section \ref{sec:siglossmethod}, however, if $\widehat{\mathbf{C}}_{\rm EoR}$ $\textit{is}$ correlated with the data vector $\mathbf{x}$, there is a kind of projection of power in the {\it non}-foreground modes from the resulting power spectrum estimate, thus producing an estimate that is biased low.  In short, {\it if the covariance is computed from the data itself, it carries the risk of overfitting information in the data and introducing a multiplicative bias (per $k$) to estimates of the signal.}

The danger of an empirically estimated covariance matrix comes mostly from not being able to describe the EoR-dominated eigenmodes of $\textbf{C}$ accurately, for which the EoR signal is brighter than foregrounds. In such a case, the coupling between these modes to the data realization leads to the overfitting and subtraction of the EoR signal. More specifically, the coupling between the estimated covariance and the data is anti-correlated in nature (which is explained in more detail in Section \ref{sec:siglossmethod}), which leads to loss. Mis-estimating $\textbf{C}$ for EoR-dominated eigenmodes is therefore more harmful than for FG-dominated modes, and since the lowest-valued eigenmodes of an eigenspectrum are typically EoR-dominated, using this part of the spectrum for weighting is most dangerous.

Armed with this information,
we can tweak the covariance in a simple way to suppress foregrounds and yield minimal signal loss. Recall that our toy model foreground 
can be perfectly described by a single eigenmode. Using the full dataset's (foreground plus EoR signal) empirical covariance, we can 
project out the zeroth eigenmode and 
then take the remaining covariance to be the identity matrix.  
This decouples the covariance from the data for the EoR modes.  The resulting power spectrum estimate for this case is shown in the right plot of Figure \ref{fig:toy_sigloss3}. 
In this case we recover the EoR signal, demonstrating that if we can disentangle the foreground-dominated modes and EoR-dominated modes, we can suppress
foregrounds with negligible signal loss. 

Altering $\widehat{\textbf{C}}$ as such is one specific example of a regularization method for this toy model, in which we are changing $\widehat{\textbf{C}}$ in a way that reduces its coupling to the data realization. There are several other simple ways to regularize $\widehat{\textbf{C}}$, and we will discuss some in Section 
\ref{sec:otherweight}.

\subsection{Fringe-Rate Filtering}
\label{sec:toymodel_frf}

We have shown how signal loss can arise due to the coupling of EoR-dominated eigenmodes to the data. We will next show how this effect is exacerbated by reducing the total number of independent samples in a dataset. 

A fringe-rate filter is an analysis technique designed to maximize sensitivity by integrating in time (\citealt{parsons_et_al2016}). Rather than a traditional box-car average in time, a time domain filter can be designed to up-weight temporal modes consistent with the sidereal motion on the sky, while down-weighting modes that are noise-like. 

\begin{figure}
	\centering
	\includegraphics[width=\columnwidth]{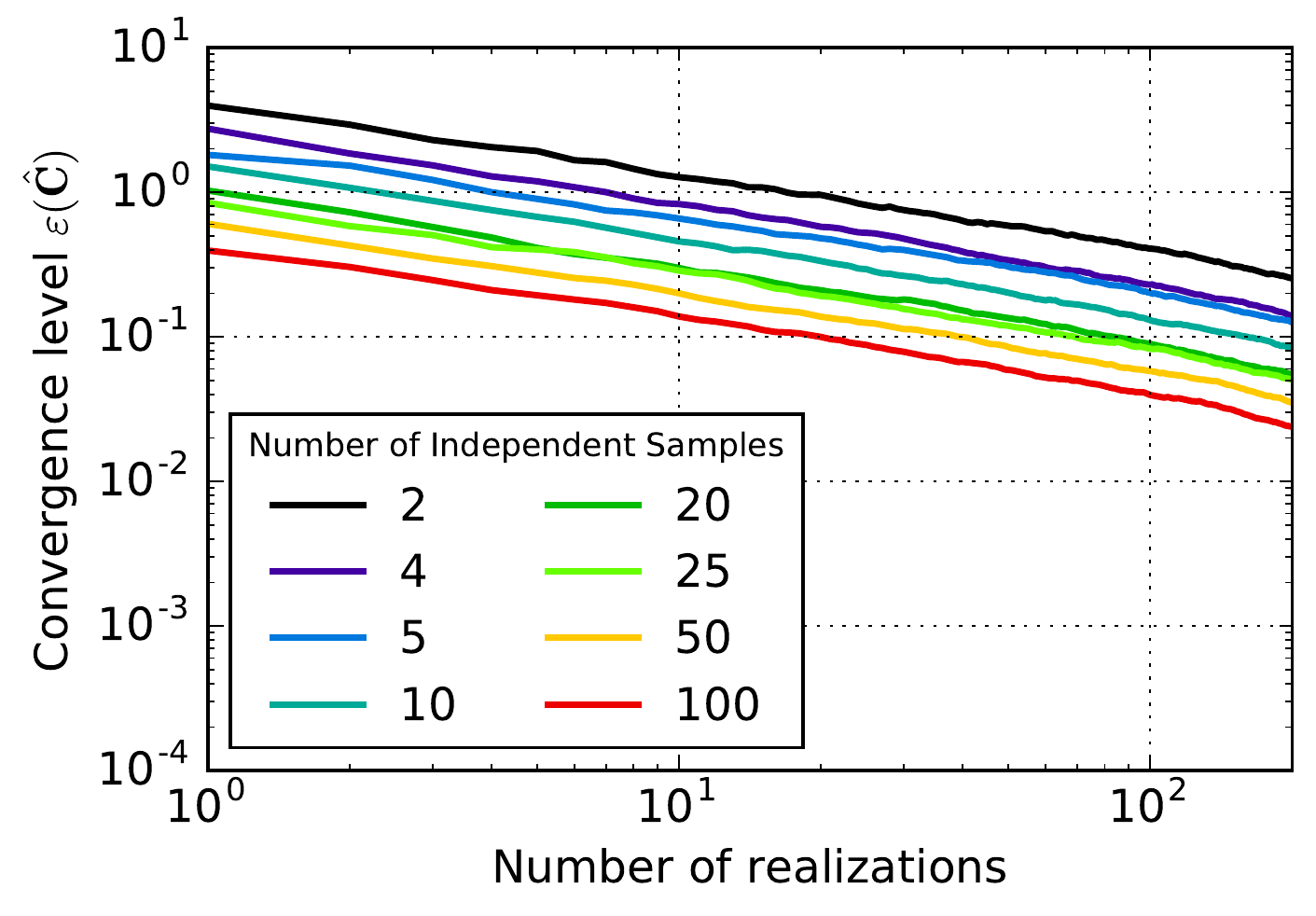}
	\caption{The convergence level, as defined by Equation \eqref{eq:converge}, of empirically estimated covariances of mock EoR signals with different numbers of independent samples. In red, the mock EoR signal is comprised entirely of independent samples (100 of them). Subsequent colors show time-averaged signals. As the number of realizations increases, we see that the empirical covariances approach the true covariances. With more independent samples, the quicker an empirical covariance converges (i.e., the quicker it decouples from the data), and the less signal loss we would expect to result.}
	\label{fig:toy_sigloss16}
\end{figure}

Because fringe-rate filtering is analogous to averaging in time, it comes at the cost of reducing the total number of independent samples in the data. With fewer independent modes, it becomes more difficult for the empirical covariance to estimate the true covariance matrix of the fringe-rate filtered data. We can quantify this effect by evaluating a convergence metric $\varepsilon(\Chat)$ for the empirical covariance, which we define as


\begin{equation}
\label{eq:converge}
\varepsilon (\Chat) \equiv \sqrt{\frac{\sum_{ij} (\widehat{C}_{ij} - {C}_{ij})^2}{\sum_{ij} {C}_{ij}^2}},
\end{equation}
where $\C$ is the true covariance matrix. To compute this metric, we draw different numbers of realizations (different draws of Gaussian noise) of our toy model EoR measurement, $\textbf{x}_{\rm EoR}$, and take their ensemble average. We then compare this to the ``true" covariance, which in our simulation is set to be the empirical covariance after a large number ($500$) of realizations. As shown in Figure \ref{fig:toy_sigloss16}, we perform this computation for a range of total independent ensemble realizations (horizontal axis) and number of independent samples in the data following time-averaging, or ``fringe-rate filtering" (different colors). With more independent time samples (i.e., more realizations) in the data, one converges to the true fringe-rate filtered covariance more quickly. 

The situation here with using a finite number of time samples to estimate our covariance is analogous to a problem faced in galaxy surveys, where the non-linear covariance 
of the matter power spectrum is estimated using a large --- but finite --- number of expensive simulations. There, the limited 
number of independent simulations results in inaccuracies in estimated covariance matrices 
\citep{dodelson_schneider2013,taylor_joachimi_etal2014}, which in turn result in biases in the final parameter constraints 
\citep{hartlap_et_al2007}. In our case, the empirically estimated covariances are used for estimating the power spectrum, and 
as we discussed in the previous section (and will argue more thoroughly in Section \ref{sec:siglossmethod}), couplings between these covariances and the data can lead to power spectrum estimates that are biased 
\emph{low}---which is precisely signal loss. In future work, it will be fruitful to investigate whether advanced techniques from the 
galaxy survey literature for estimating accurate covariance matrices can be successfully adapted for $21\,\textrm{cm}$ 
cosmology. These techniques include the imposition of sparsity priors \citep{padmanabhan_et_al2016}, the fitting of 
theoretically motivated parametric forms \citep{pearson_samushia2016}, covariance tapering \citep{paz_sanchez2015}, 
marginalization over the true covariance \citep{sellentin_heavens2016}, and shrinkage methods 
\citep{pope_szapudi2008,joachimi_2017}.

\begin{figure}
	\centering
	\includegraphics[width=\columnwidth]{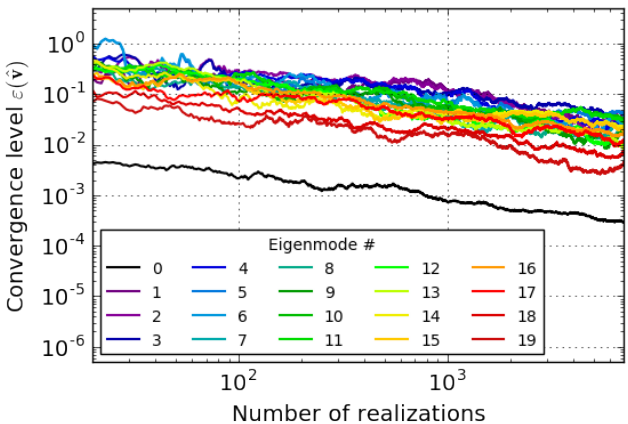}
	\caption{The convergence level, as defined by Equation \eqref{eq:converge_eig}, of empirically estimated eigenvectors for different numbers of mock data realizations. The colors span from the 0th eigenmode (has the highest eigenvalue) to the 19th eigenmode (has the lowest eigenvalue), where they are ordered by eigenvalue in descending order. This figure shows that the zeroth eigenmode converges the quickest, implying that eigenvectors with eigenvalues that are substantially different than the rest (the FG-dominated mode has a much higher eigenvalue than the EoR modes) are able to converge to the true eigenvectors the quickest. On the other hand, eigenmodes $1$-$19$ have similar eigenvalues and are slower to converge because of degeneracies between them.}
	\label{fig:toy_sigloss17}
\end{figure}

The overall convergence of the covariance is important, but also noteworthy is the fact that different eigenvectors converge to their true forms at different rates. This is illustrated by Figure \ref{fig:toy_sigloss17}, which shows the convergence of eigenvectors in an empirical estimate of a covariance matrix. For this particular toy model, we construct a covariance whose true form combines the same mock foreground from the previous toy models with an EoR component that is modeled as a diagonal matrix with eigenvalues spanning one order of magnitude (more specifically, we construct the EoR covariance as a diagonal matrix in the Fourier domain, where the signal is expected to be uncorrelated; its Fourier transform is then the true covariance of the EoR in the frequency domain, or $\textbf{C}_{\rm EoR}$). For different numbers of realizations, we draw random EoR signals that are consistent with $\textbf{C}_{\rm EoR}$, add them to the mock foreground data, and compute the combined empirical covariance by averaging over the realizations. The eigenvectors of this empirical covariance are then compared to the true eigenvectors $\widehat{\textbf{v}}$, where we use as a convergence metric $\varepsilon(\widehat{\textbf{v}})$, defined as:
\begin{equation}
\label{eq:converge_eig}
\varepsilon (\widehat{\textbf{v}}) \equiv \sqrt{\sum_{i}^{N_{f}}|\textbf{v}-\widehat{\textbf{v}}|_{i}^2},
\end{equation}
where $N_{f}$ is the number of frequencies ($20$) in the mock data. The eigenmode convergence curves in Figure \ref{fig:toy_sigloss17} are ranked ordered by eigenvalue, such that ``Eigenmode \#0" illustrates the convergence of the eigenvector with the largest eigenvalue, ``Eigenmode \#1" for the second largest eigenvalue, and so on. We see that the zeroth eigenmode --- the mode describing the foreground signal --- is quickest to converge.

Our numerical test reveals that the convergence rates of empirical eigenvectors is related to the sample variance in our empirical estimate. In general, computing an empirical covariance from a finite ensemble average means that the empirical eigenmodes have sample variances. Consider first a limiting case where all eigenvalues are equal. In such a scenario, any linear combination of eigenvectors is also an eigenvector, and thus there is no sensible way to define the convergence of eigenvectors. In our current test, aside from the zeroth mode, the eigenvalues have similar values but are not precisely equal. Hence, there is a well-defined set of eigenvectors to converge to. However, due to the sample variance of our empirical covariance estimate, there may be accidental degeneracies between modes, where some modes are mixing and swapping with others. Therefore, the steeper an eigenspectrum, the easier it is for the eigenmodes to decouple from each other and approach their true forms. A particularly drastic example of this can be seen in the behavior of mode $0$ (the foreground mode), whose eigenvalue differs enough from the others that it is able to converge reasonably quickly despite substantial sample variance in our empirical covariance estimate. To break degeneracies in the remaining modes, however, requires many more realizations.

While the connection between the rate of convergence of an empirical eigenvector with the sample variance of an eigenspectrum is interesting, it is also important to note that regardless of convergence rate, any mode that is coupled to the data is susceptible to signal loss. The true eigenvectors are not correlated with the data realizations; thus, if our empirical eigenvectors are converged fully, there will not be any signal loss. However, an unconverged eigenvector estimate will retain some memory of the data realizations used in its generation, leading to signal loss.

In the toy models throughout Section \ref{sec:SiglossOverview}, we exploit the fact that the strongest eigenmode (highest eigenvalue mode) is dominated by foregrounds in order to purposely incur signal loss for that mode. Even for the case of real PAPER data (Section \ref{sec:CaseStudy}), we make the assumption that the strongest eigenmodes are likely the most contaminated by foregrounds. However, in general, foregrounds need not be restricted to the strongest eigenmodes, and as we have seen, it is really the degeneracies between modes that determines how quickly they converge, and hence how much signal loss can result.



\begin{figure}
	\centering
	\includegraphics[width=\columnwidth]{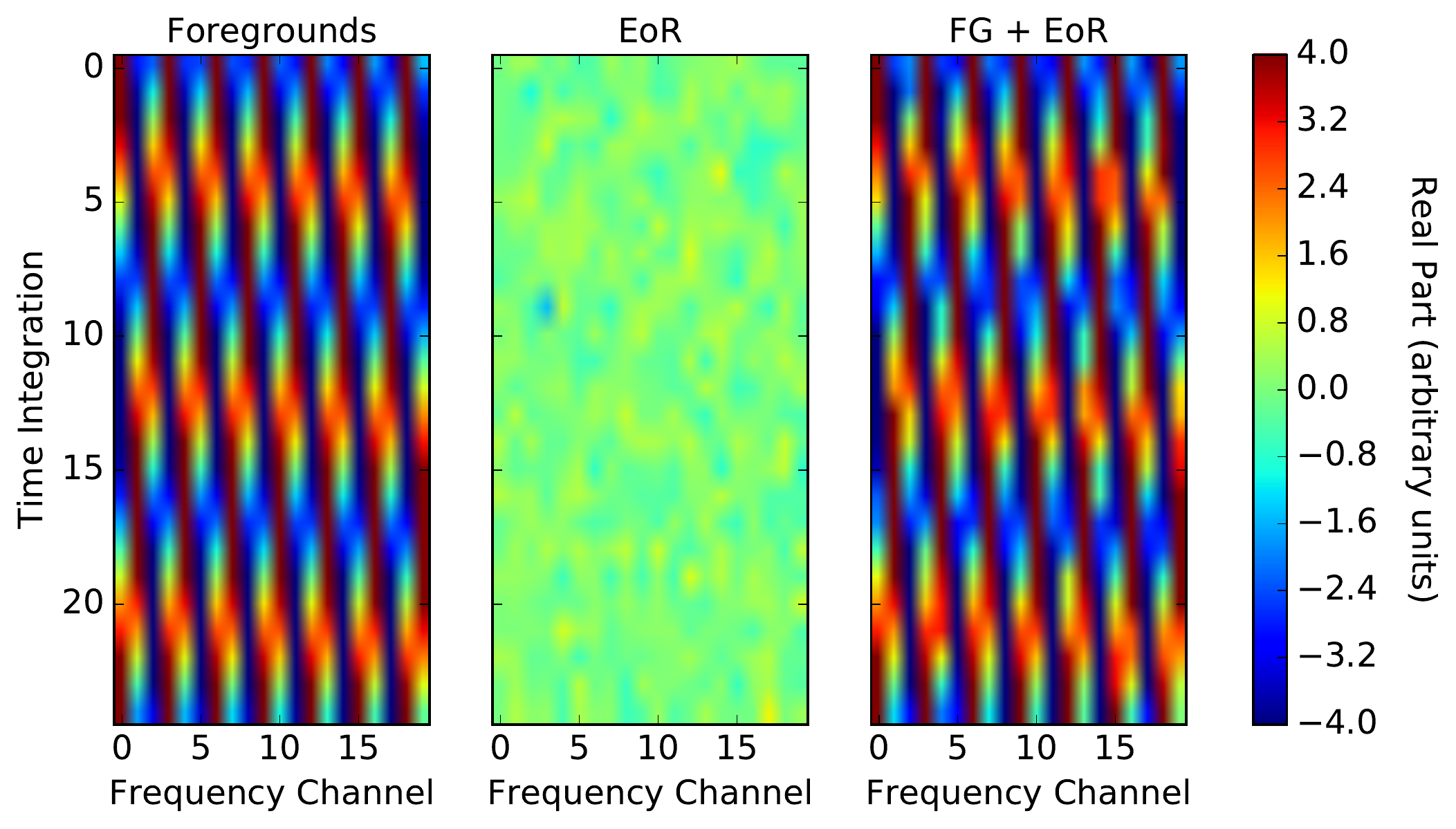}
	\caption{Our ``fringe-rate filtered" (time-averaged) toy model dataset. We average every four samples together, 
yielding $25$ independent samples in time. Real parts are shown here.}
	\label{fig:toy_sigloss5}
\end{figure}

\begin{figure}
	\centering
	\includegraphics[trim={0cm 0cm 0cm 0cm},clip,width=\columnwidth]{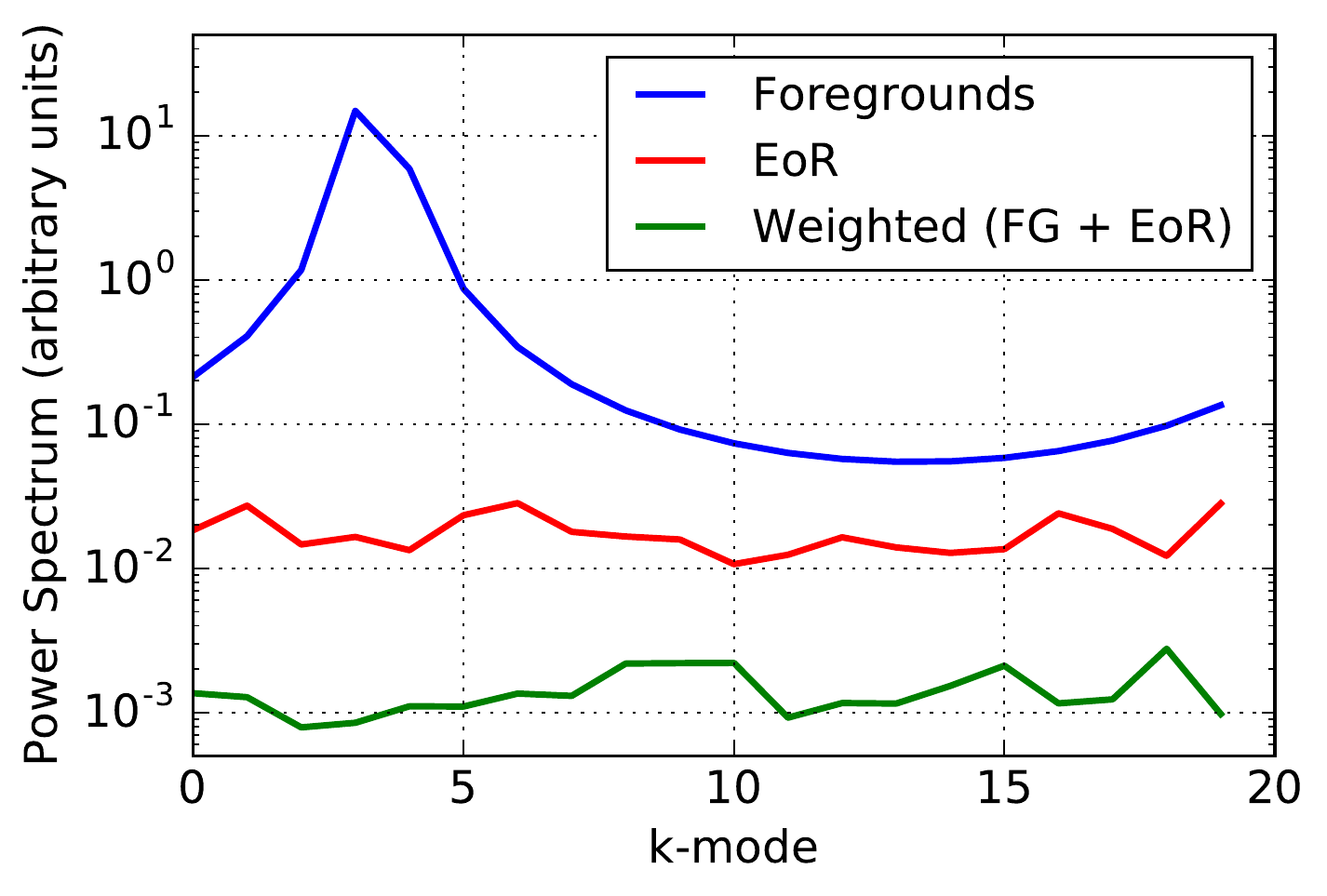}
	\caption{Resulting power spectrum estimate for the ``fringe-rate filtered" (time-averaged) toy model simulation --- foregrounds only (blue), 
EoR only (red), and the weighted FG + EoR dataset (green). We use empirically estimated inverse covariance weighting where $\textbf{C}$ is 
computed from the data. There is a larger amount of signal loss than for the non-averaged data, a consequence of weighting by eigenmodes that are more strongly coupled to the data due to there being fewer independent modes in the data.}
	\label{fig:toy_sigloss7}
\end{figure}

With Figures \ref{fig:toy_sigloss16} and \ref{fig:toy_sigloss17} establishing the connection between convergence rates (of empirical covariances and eigenvectors) and number of realizations, we now turn back to our original toy model used in Section \ref{sec:toymodel}, which is comprised of a mock foreground and mock EoR signal. We mimic a fringe-rate filter by averaging every four time integrations of our toy model dataset together, yielding $25$ independent samples in time (Figure \ref{fig:toy_sigloss5}). We choose these numbers so that the total number of independent samples is similar to the number of frequency channels --- hence our matrices will be full rank. We use this ``fringe-rate filtered" mock data for the remainder of Section \ref{sec:SiglossOverview}.

The power spectrum results for this model are shown in Figure \ref{fig:toy_sigloss7}, and as 
expected there is a much larger amount of signal loss for this time-averaged dataset since we do a worse job estimating the true covariance. In addition, as a result of having fewer independent samples, we obtain an estimate with more scatter. This is evident by noticing that the 
green curve in Figure \ref{fig:toy_sigloss7} fails to trace the shape of the uniform-weighted EoR power spectrum (red). 

Using our toy model, we have seen that a sensitivity-driven analysis technique like fringe-rate filtering has trade-offs of signal 
loss and noisier estimates when using data-estimated covariance matrices. Longer integrations increase sensitivity but reduce 
the number of independent samples, resulting in 
eigenmodes correlated with the data
that can overfit signal greatly. We 
note that a fringe-rate filter does have a range of benefits, many described in \citet{parsons_et_al2016}, so it can still be 
advantageous to use one despite the trade-offs.

\subsection{Other Weighting Options}
\label{sec:otherweight}

\begin{figure*}
	\centering
	\includegraphics[trim={0cm 0cm 0cm 0cm},clip,height=0.3\textwidth]{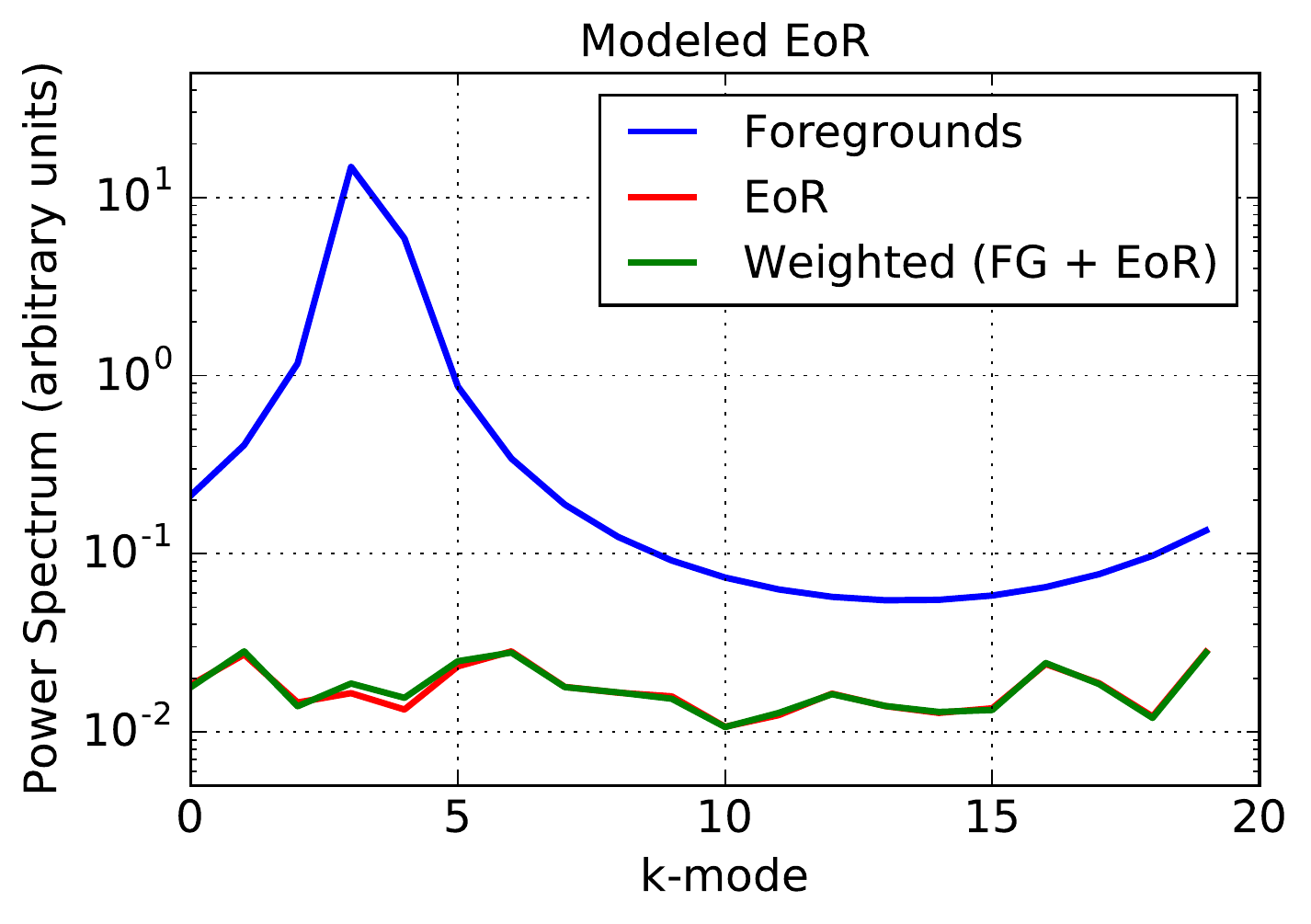}
	\includegraphics[trim={0cm 0cm 0cm 0cm},clip,height=0.3\textwidth]{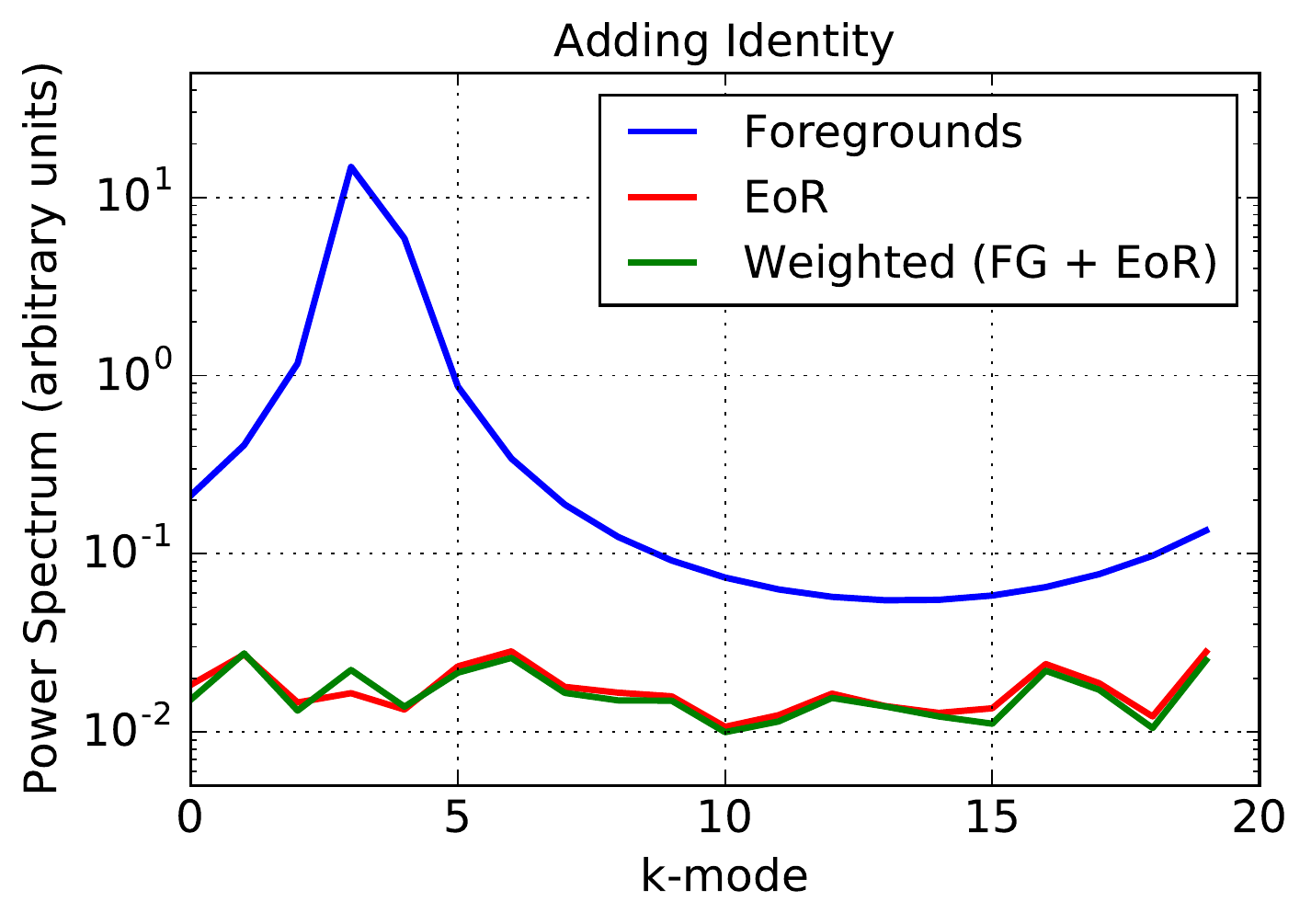}
	\includegraphics[trim={0cm 0cm 0cm 0cm},clip,height=0.3\textwidth]{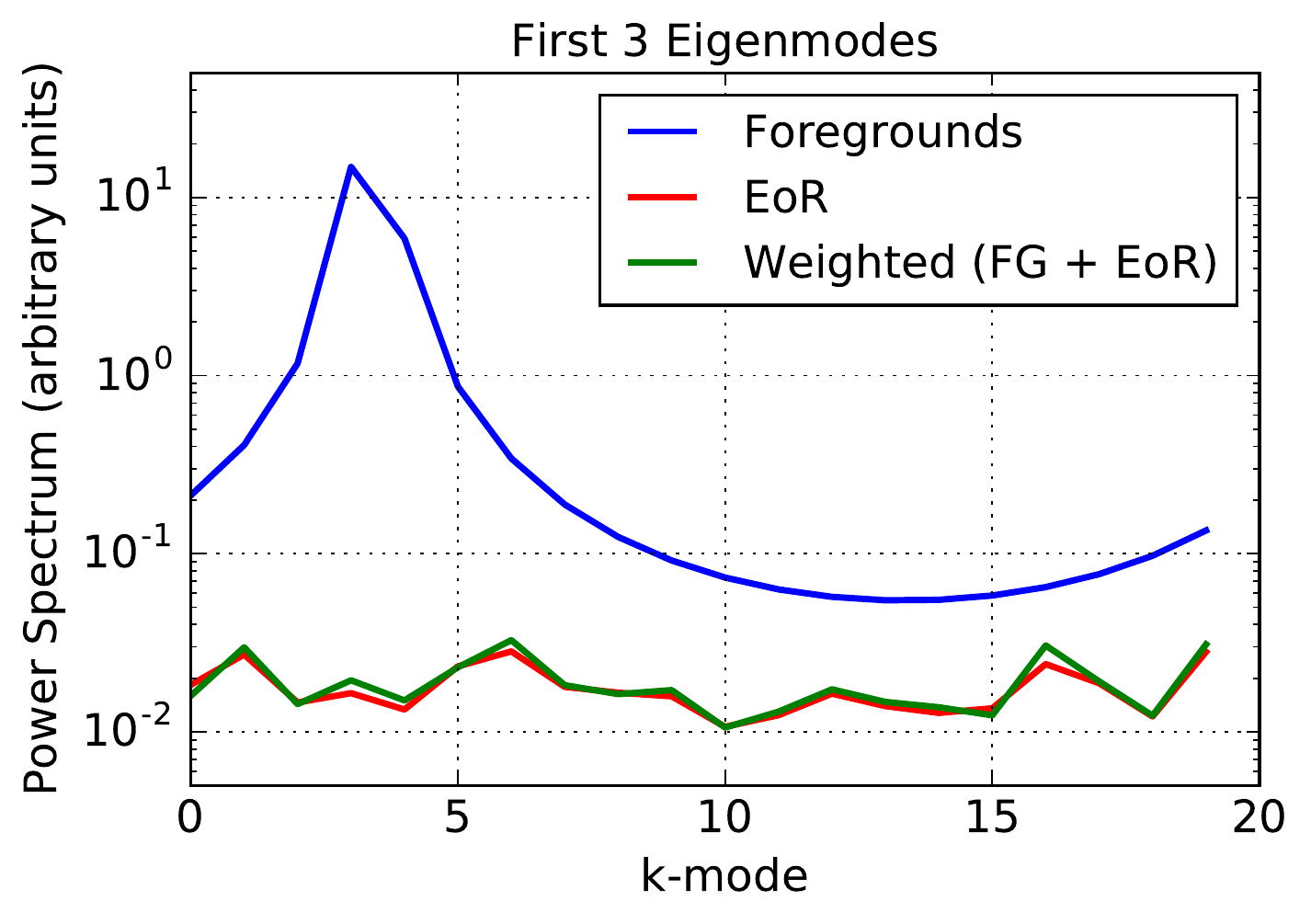}
	\includegraphics[trim={0cm 0cm 0cm 0cm},clip,height=0.3\textwidth]{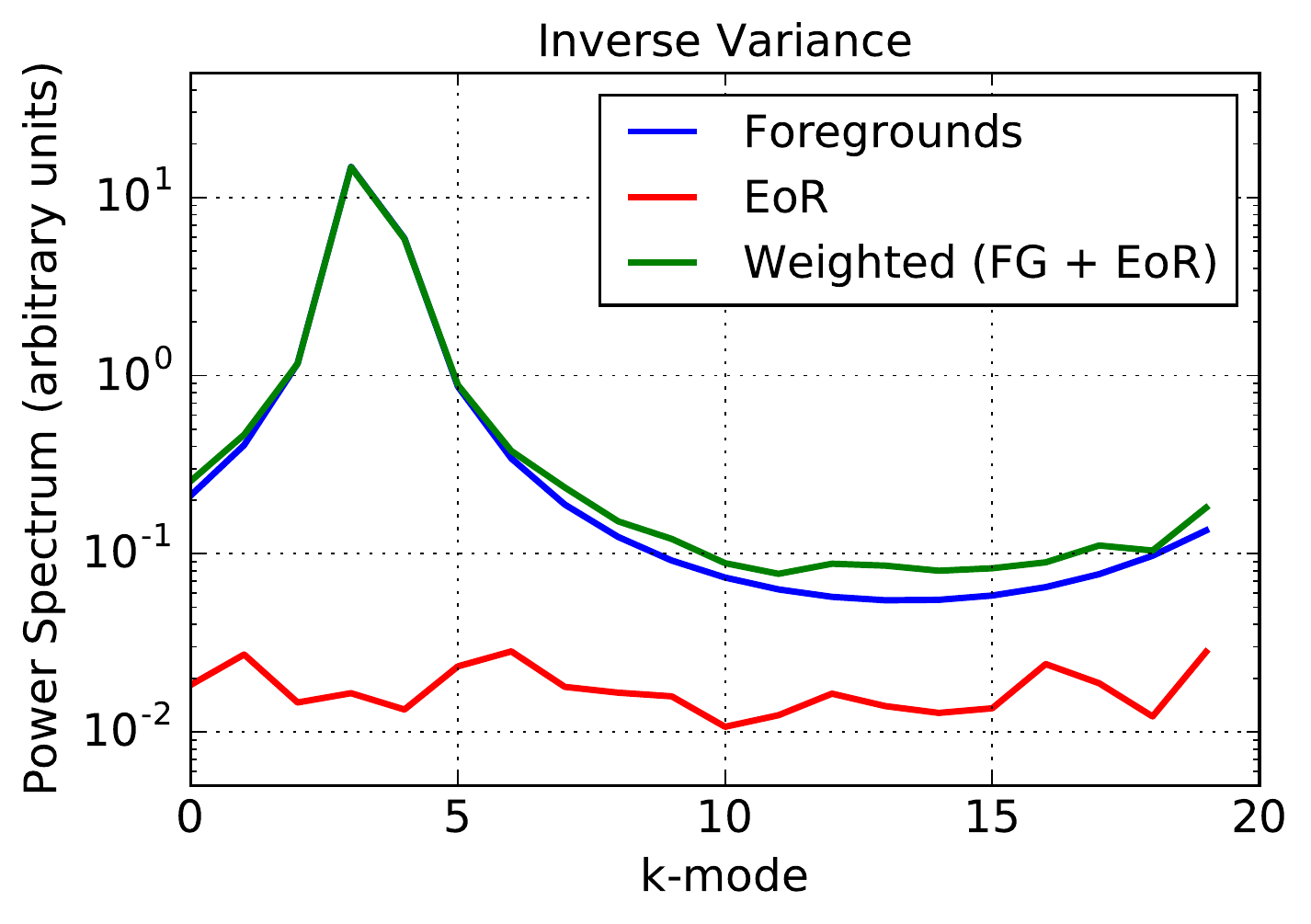}
	\caption{Resulting power spectra estimates for our ``fringe-rate filtered" (time-averaged) toy model simulation --- foregrounds only (blue), EoR only (red), and the weighted FG + EoR dataset (green). We show four alternate weighting options that each minimize signal loss, including modeling the covariance matrix of EoR (upper left), regularizing $\widehat{\textbf{C}}$ by adding an identity matrix to it (upper right), using only the first three eigenmodes of $\widehat{\textbf{C}}$ (lower left), and keeping only the diagonal elements of $\widehat{\textbf{C}}$ (lower right). The first case (upper left) is not feasible in practice since we do not know $\textbf{C}_{\rm FG}$ and $\textbf{C}_{\rm EoR}$ like we do in the toy model.}
	\label{fig:toy_sigloss8}
\end{figure*}

In Section \ref{sec:toymodel} we showed one example of how altering $\widehat{\textbf{C}}$ can 
make the difference between nearly zero and some signal loss. We will now use our toy model to describe several other ways to tailor $\widehat{\textbf{C}}$ 
in order to minimize signal loss. We choose four independent regularization methods to highlight in this section, which have 
been chosen due to their simplicity in implementation and straightforward interpretations. We illustrate the resulting power 
spectra for the different cases in Figure \ref{fig:toy_sigloss8}.
These examples are not meant to be taken as suggested analysis methods but rather as illustrative cases. 

As a first test, we model the covariance matrix of EoR as a proof of concept that if perfect models are known, signal loss can be 
avoided. We know that our simulated EoR signal should have a covariance matrix that mimics the identity matrix, with its 
variance encoded along the diagonal. We model $\textbf{C}_{\rm EoR}$ as such (i.e., the identity), instead of computing it based on $\textbf{x}
_{\rm EoR}$ itself. Next, we add $\textbf{C}_{\rm EoR} + \widehat{\textbf{C}}_{\rm FG}$ (where $\widehat{\textbf{C}}_{\rm FG} = \langle\textbf{x}_{\rm FG}
\textbf{x}_{\rm FG}^{\dagger}\rangle_{t}$) to obtain a final $\widehat{\textbf{C}}_{\rm reg}$ (regularized empirical covariance matrix) to use in weighting. In Figure \ref{fig:toy_sigloss8} (upper 
left), we see that there is negligible signal loss. This is because by modeling $\textbf{C}_{\rm EoR}$, we avoid overfitting EoR fluctuations in the data that our model doesn't know about (but, an empirically derived $\widehat{\textbf{C}}_{\rm EoR}$ would know about the fluctuations). 
In practice such a weighting option is not feasible, as it is difficult to model $\textbf{C}_{\rm EoR}$, and $\widehat{\textbf{C}}_{\rm FG}$ is unknown because we do not know how to separate out the foregrounds from the EoR in our data.

%
The second panel (top right) in Figure \ref{fig:toy_sigloss8} uses a regularization method of setting $\widehat{\textbf{C}}_{\rm reg} \equiv 
\widehat{\textbf{C}} + \gamma\textbf{I}$, where $\gamma = 5$ (an arbitrary strength 
of $\textbf{I}$ for the purpose of this toy model). By adding the identity matrix, element-wise, we are weighting the diagonal 
elements of the estimated covariance matrix more heavily than those off-diagonal. Since the identity component does not know anything about the data realization, it alters the covariance to be less coupled to the data and there is no loss. 

The third panel (bottom left) in Figure \ref{fig:toy_sigloss8} minimizes signal loss by only using the first three eigenmodes of the estimated covariance. Recalling that our toy model foregrounds can be described entirely by the zeroth eigenmode, this 
method intentionally projects out the highest-valued modes only by replacing all but the three highest weights in the 
eigenspectrum with $1$'s (equal weights). Again, avoiding the overfitting of EoR-dominated modes which are coupled to the data results in negligible signal loss. While this case is illuminating for the toy model, in practice it is not obvious which eigenmodes are foreground or EoR dominated (and they could be mixed as well), so determining which subset of modes to down-weight is not trivial. We experiment with this idea using PAPER data in Section \ref{sec:Weight}.

The last regularization scheme we are highlighting here is setting $\widehat{\textbf{C}}_{\rm reg} \equiv \widehat{\textbf{C}} \circ \textbf{I}$ (element-wise multiplication), or inverse variance weighting (i.e., keeping only the diagonal elements of $\widehat{\textbf{C}}$). In the bottom right 
panel of Figure \ref{fig:toy_sigloss8}, we see that this method does not down-weight the foregrounds at all --- this regularization altered $\widehat{\textbf{C}}$ in a way where it is no longer coupled to \textit{any} of the empirically estimated eigenmodes, including the FG-dominated one. To understand this, we recall that our foregrounds are spread out in frequency and therefore have non-negligible frequency-frequency correlations. Multiplying by 
the identity matrix, element-wise, results in a diagonal matrix, meaning we do not have any correlation information. Because of this, we do a poor job 
suppressing the foreground. But because we decoupled the whole eigenspectrum from the data, we also avoid signal loss. Although this method did not successfully recover the EoR signal for this particular simulation, it is important that we show that there 
are many options for estimating a covariance matrix, and some may down-weight certain eigenmodes more effectively than others based on the spectral nature 
of the components in a dataset. 

In summary, we have shown how signal loss is caused by weighting a dataset by itself, and in particular how estimated covariances can overfit EoR modes when they are coupled to data and not converged to their true forms. We have also seen that there are trade-offs between a chosen weighting method, its foreground-removal effectiveness, the number of independent samples in a dataset, and the amount of resulting signal loss.


\section{Signal Loss in PAPER-64}
\label{sec:CaseStudy}

We now turn to a detailed signal loss investigation using a subset of the PAPER-64 dataset from \citetalias{ali_et_al2015}. In the previous section we showed how signal loss arises when weighting data with empirically estimated covariances; in this section we highlight how the amount of this loss was underestimated in the previous analysis. Additionally, we illustrate how we have revised our analysis pipeline in light of our growing understandings.

\begin{figure}
	\centering
	\includegraphics[trim={0cm 0cm 0cm 0cm},width=\columnwidth]{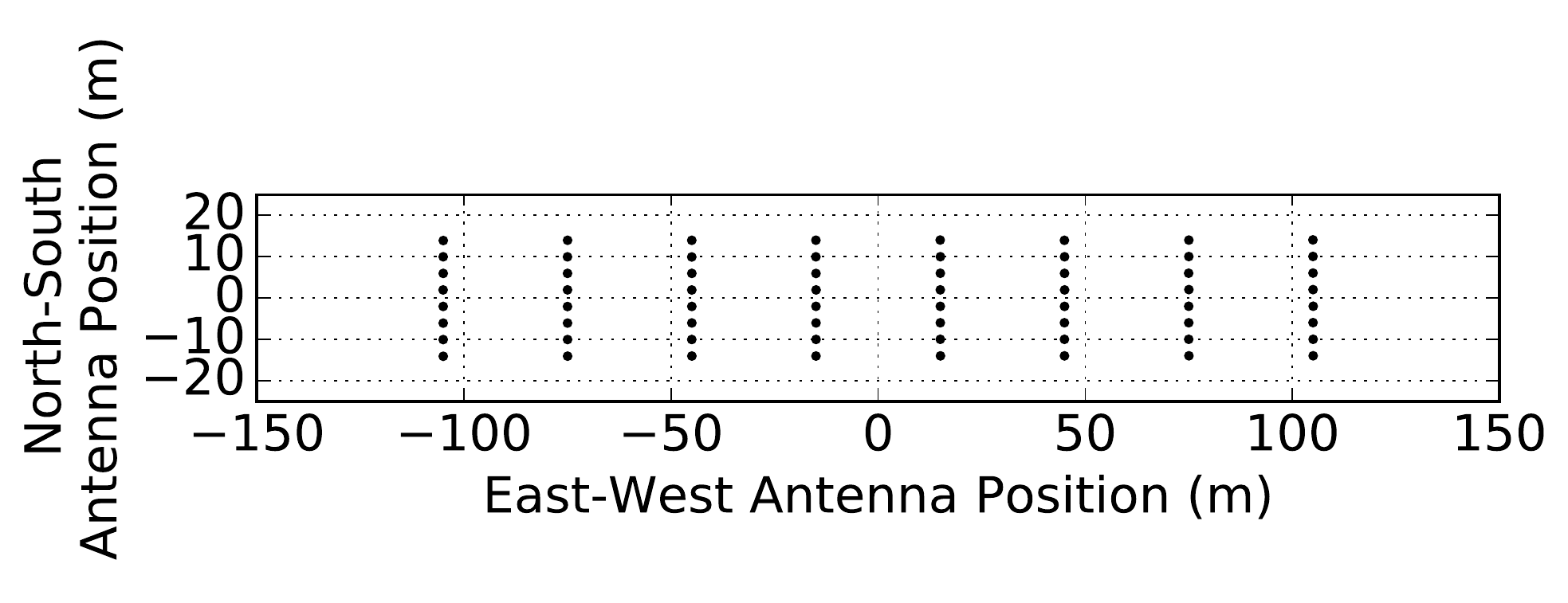}
	\caption{The PAPER-64 antenna layout. We use only $10$ of the $30$ m East/West baselines for the analysis in this 
paper (i.e., a subset of the shortest horizontal spacings).}
	\label{fig:ant_layout}
\end{figure}

As a brief review, PAPER is a dedicated 21\,cm experiment located in the Karoo Desert in South Africa. The PAPER-64 
configuration consists of 64 dual-polarization drift-scan elements that are arranged in a grid layout. For our case study, we 
focus solely on Stokes I estimated data \citep{moore_et_al2013} from PAPER's $30$ m East/West baselines (Figure 
\ref{fig:ant_layout}). All data is compressed, calibrated (using self-calibration and redundant calibration), delay-filtered (to remove foregrounds inside the wedge), LST-binned, and fringe-rate filtered. For detailed information about the backend system of PAPER-64, its observations, and data reduction pipeline, we 
refer the reader to \citet{parsons_et_al2010} and \citetalias{ali_et_al2015}. We note that all data processing steps are identical to those in \citetalias{ali_et_al2015} until after the LST-binning step in Figure 3 of \citetalias{ali_et_al2015}.

The previously best published 21\,cm upper limit result from \citetalias{ali_et_al2015} placed a $2\sigma$ upper limit 
on $\Delta^{2}(k)$, defined as

\begin{equation}
\Delta^{\textbf{2}}(k) = \frac{k^{3}}{2\pi^{2}}\,\hat{P}(k),
\end{equation}

\noindent of $(22.4$ mK$)^{2}$ in the range $0.15 < k < 0.5$\,$h$ Mpc$^{-1}$ at $z = 8.4$. The need to revise this limit stems mostly from previously underestimated signal loss, which we 
address in this section.

For the analysis in this paper, we use $8.1$ hours of LST, namely an RA range of $0.5$-$8.6$ hours (\citetalias{ali_et_al2015} uses a slightly longer RA 
range of $0$-$8.6$ hours; we found that some early LSTs were more severely foreground contaminated). We also use only $10$ baselines, a subset of the $51$ total East/West baselines used in \citetalias{ali_et_al2015}, in order to illustrate our revised methods. All power spectrum results are produced for a center frequency of 151\,MHz using a width of 10\,MHz ($20$ channels), identical to the analysis in \citetalias{ali_et_al2015}. In the case study in this paper, we only use one baseline type instead of the three as in 
\citetalias{ali_et_al2015}, but Kolopanis et al. (\textit{in prep.}) uses the full dataset presented in \citetalias{ali_et_al2015} to revise the result and place limits on the EoR at multiple redshifts (using a straightforward and not lossy approach to avoid many of the issues that will be made clear later on).

\begin{figure}
	\centering
	\includegraphics[width=\columnwidth]{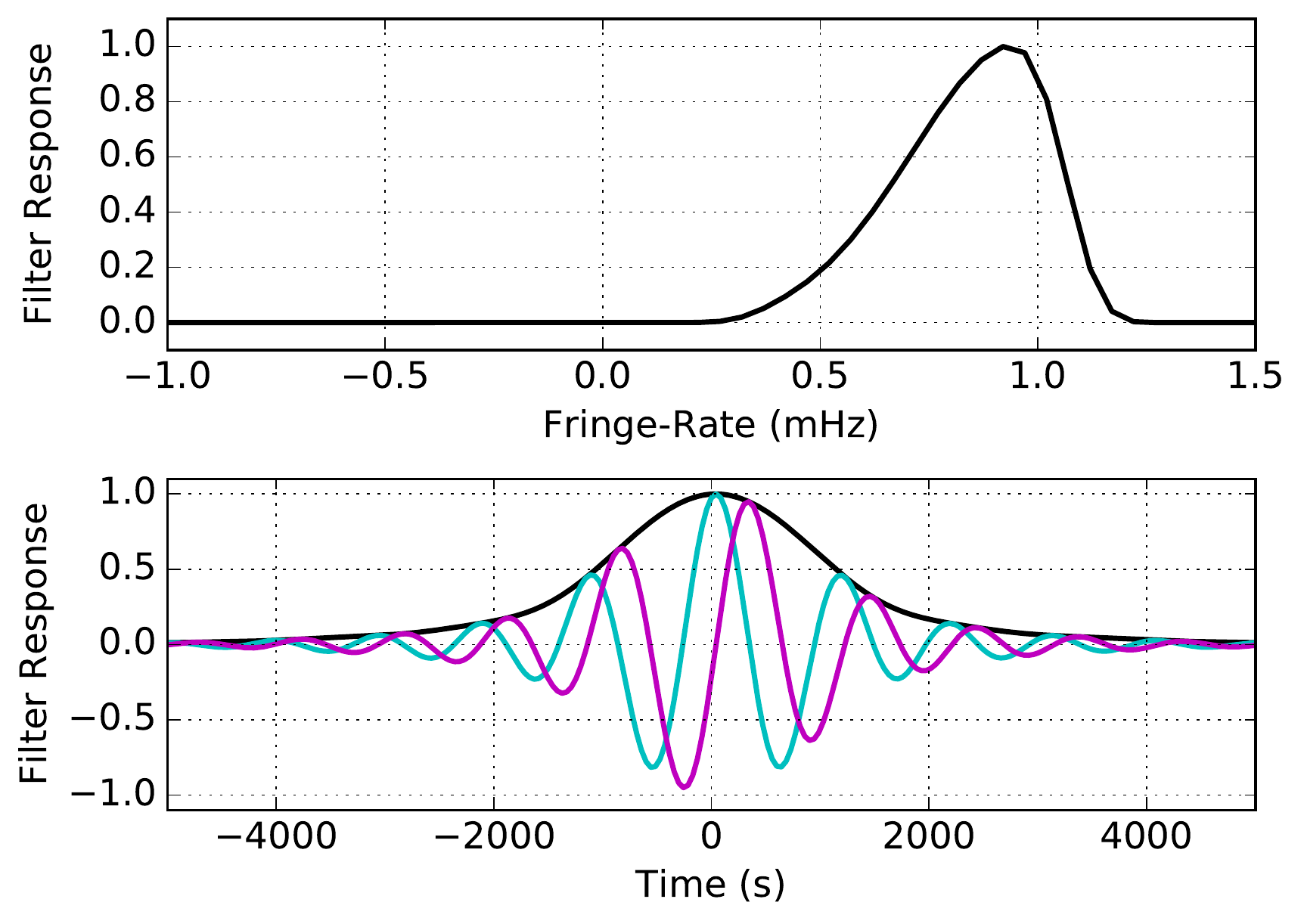}
	\caption{Top: the normalized optimal power-spectrum sensitivity weighting in fringe-rate space for our fiducial baseline and 
Stokes I polarization beam. Bottom: the time domain convolution kernel corresponding to the top panel. Real and imaginary 
components are illustrated in cyan and magenta, respectively, with the absolute amplitude in black. The fringe-rate filter acts as 
an integration in time, increasing sensitivity but reducing the number of independent samples in the dataset.}
	\label{fig:frp}
\end{figure}

The most significant changes from \citetalias{ali_et_al2015} occur in our revised power spectrum analysis, which is explained in the rest of this paper, but we also note that the applied fringe-rate filter is also slightly different. In \citetalias{ali_et_al2015}, the 
applied filter was not equivalent to the optimal fringe-rate filter (which is designed to maximize power spectrum sensitivity). Instead, the optimal filter was degraded slightly by widening it in fringe-rate space. This was chosen in order to increase the number of independent 
modes and reduce signal loss associated with the quadratic estimator, though as we will explain in the next section, this signal loss was still underestimated. With the development of a new, 
robust method for assessing signal loss, we choose to use the optimal filter in order to maximize sensitivity. This filter is 
computed for a fiducial 30\,m baseline at 150\,MHz, the center frequency in our band. The filter in both the fringe-rate 
domain and time domain is shown in Figure \ref{fig:frp}.

Finally, we emphasize that the discussion that follows is solely focused on signal loss associated with empirical covariance weighting. As mentioned in Section \ref{sec:SiglossOverview}, there are a number of steps in our analysis pipeline which could lead to loss, including gain calibration, delay filtering, and fringe-rate filtering, which have been investigated at various levels of detail in \citet{parsons_et_al2014} and \citetalias{ali_et_al2015} but are clearly the subject of future work. Here we only focus on the most significant source of loss we have identified and note that Kolopanis et al. (\textit{in prep.}) and other future work will consider additional sources of signal loss and exercise increased caution in reporting results.

We present our PAPER-64 signal loss investigation in three parts. We first give an overview of our signal injection framework which is used to estimate loss. In this framework (and as in \citetalias{ali_et_al2015}), we inject simulated cosmological signals into our data and test the recovery of those signals (an approach also taken by \citet{masui_et_al2013}). As we will see, correlations between the injected signals and the data are significant complicating factors which were previously not taken into account. Next, we describe our methodology in practice and detail how we map our simulations into a posterior for the EoR signal. Finally, we build off of the previous section by experimenting with different regularization schemes on PAPER data in order to minimize loss. Throughout each section, we also highlight major differences from the signal loss computation used in \citetalias{ali_et_al2015}.


\subsection{Signal Loss Methodology} 
\label{sec:siglossmethod}
In short, our method for estimating signal loss consists of adding an EoR-like signal into visibility data and then measuring how much of this injected signal would be detectable given any attenuation of this signal by the (lossy) data analysis pipeline.  To capture the full statistical likelihood of signal loss, one requires a quick way to generate many realizations of simulated 21\,cm signal visibilities. Here we use the same method as in \citetalias{ali_et_al2015}, where mock Gaussian noise visibilities (mock EoR signals) 
are filtered in time using an optimal fringe-rate filter to retain only ``sky-like" modes. Since the optimal filter has a shape that matches the rate of the sidereal motion of the sky, this transforms the Gaussian noise into a measurement that PAPER could make. This signal is then added to the visibility data.\footnote{One 
specific change from \citetalias{ali_et_al2015} is that we add this simulated signal - which has been fringe-rate filtered once already in order to transform it into a ``sky-like" signal - into the analysis pipeline before a fringe-rate filter is 
applied to the data (i.e., prior to the analysis step of fringe-rate filtering). Previously, the addition was done after the fringe-rate filter analysis step.  This change results in an increased 
estimate of signal loss, 
likely due to the use of the fringe-rate filter as a simulator. However, this pipeline difference, while significant, is not the dominant reason why signal loss was underestimated in \citetalias{ali_et_al2015} (the dominant reason is explained in the main text in Section \ref{sec:siglossmethod}).}

Mathematically, suppose that $\textbf{e}$ is the mock injected EoR signal (at some amplitude level). We do not know the true EoR signal contained within our visibility data, $\textbf{x}$, so $\textbf{e}$ takes on the role of the true EoR signal (for which we measure its loss). Furthermore, one can make the assumption that the true EoR signal is small within our measured data, so the data vector $\textbf{x}$ itself is representative of mostly contaminants. Using this assumption, the sum of $\textbf{x}$ and $\textbf{e}$, defined as $\textbf{r}$:

\begin{equation}
\label{eq:rxe}
\textbf{r} = \textbf{x} + \textbf{e},
\end{equation}
can be thought of as the sum of contaminants plus EoR. The quantity $\textbf{r}$ then becomes the dataset for which we are measuring how much loss of $\textbf{e}$ there is due to our power spectrum pipeline.

We are interested in quantifying how much variance in $\textbf{e}$ is lost after weighting $\textbf{r}$ and estimating the power 
spectrum according to QE formalism. We investigate this by comparing two quantities we call the input power spectrum and 
output power spectrum: $\widehat{P}_{\rm in}$ and $\widehat{P}_{\rm out}$, estimated using QE as

\begin{equation}
\label{eq:Pin}
\widehat{P}_{\rm in}^{\alpha} \equiv \text{M}^{\alpha}_{\rm in}\textbf{e}^{\dagger}\textbf{I}\textbf{Q}^{\alpha}\textbf{I}\textbf{e}
\end{equation}

\noindent and

\begin{eqnarray}
\label{eq:sigloss}
\widehat{P}_{\rm out}^{\alpha} &\equiv& \widehat{\textbf{P}}_{r}^{\alpha} \nonumber\\
&=& \text{M}^{\alpha}_{r}\textbf{r}^{\dagger}\textbf{R}_{r}\textbf{Q}^{\alpha}\textbf{R}_{r}\textbf{r},
\end{eqnarray}
where, for illustrative purposes and notational simplicity, we have written these equations with scalar normalizations M, even though for our numerical results we choose a diagonal matrix normalization using $\mathbf{M}$ as in Equation \eqref{eq:phat}.

The quantity $\widehat{P}_{\rm in}$, defined by Equation \eqref{eq:Pin}, is a uniformly weighted estimator of the power spectrum of $\mathbf{e}$. It can be considered the power spectrum of this particular realization of the EoR; alternatively, it can be viewed as the true power spectrum of the injected signal up to cosmic variance fluctuations. The role of $\widehat{P}_{\rm in}$ in our analysis is to serve as a reference for the power spectrum that would be measured if there were no signal loss or other systematics. The input power spectrum is then to be compared to $\widehat{P}_{\rm out}$, which approximates the (lossy) power spectrum estimate that is output by our analysis pipeline prior to any signal loss adjustments. 



\begin{figure*}
	\centering
	\includegraphics[width=1\textwidth]{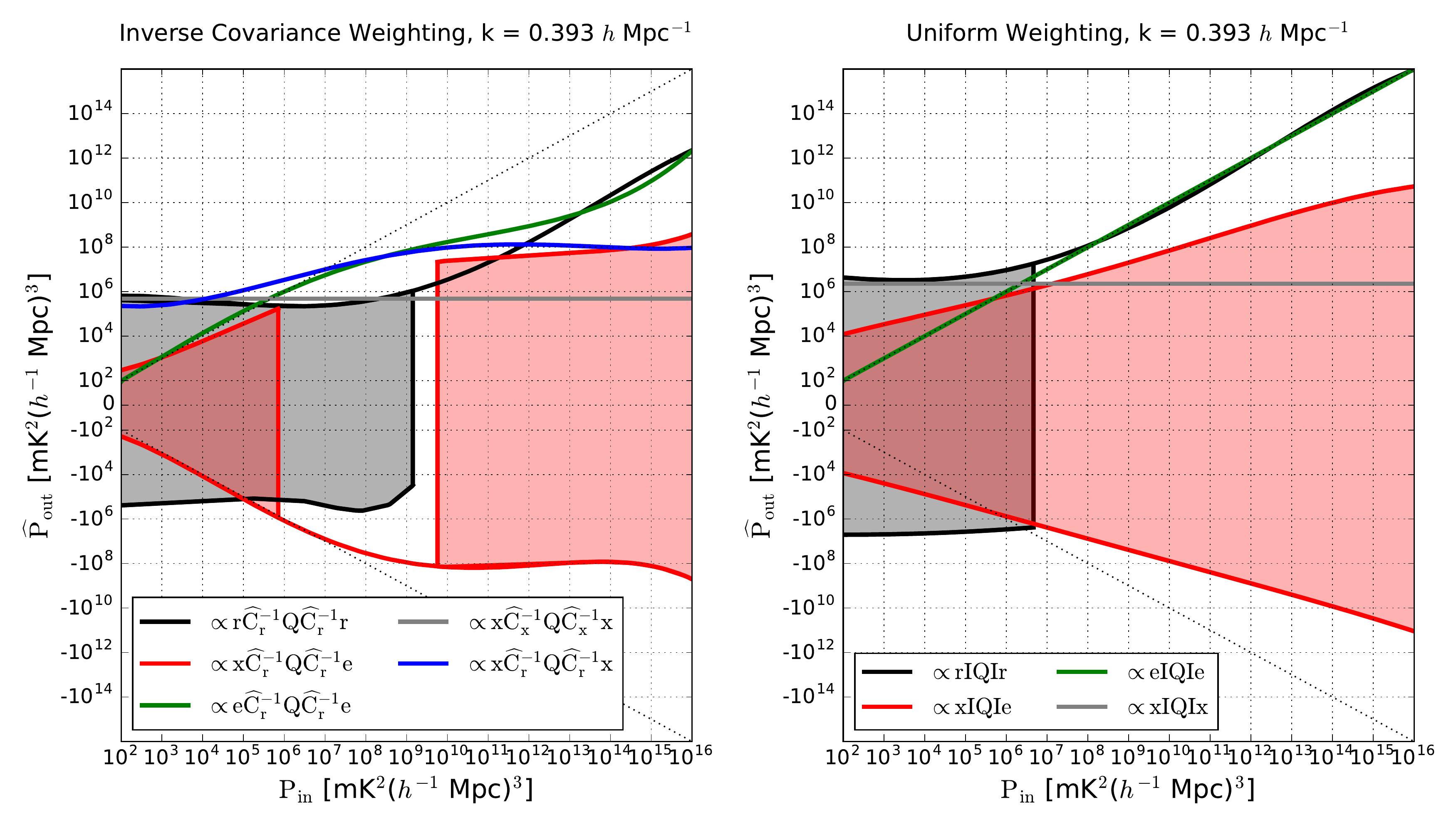}
	\caption{Illustration of the power spectrum amplitude of five different power spectrum terms, each a function of visibility data ($\textbf{x}$), simulated injected EoR signal ($\textbf{e}$), or both ($\textbf{r}$). This figure shows how these quantities behave as the power level of the injected EoR signal increases (along the x-axis).  The details of the simulation used to generate the figure is explained in Section \ref{sec:Practice}; here we sample a larger $P_{\rm in}$ range and fit smooth polynomials to our data points to make an illustrative example. We emphasize that the output power spectrum in black ($\widehat{P}_{\rm out}=\widehat{\textbf{P}}_r$) approximates the (lossy) power spectrum estimate that is output by our analysis pipeline prior to any signal loss adjustments. Roughly speaking, it can be compared to the input signal level ($P_{\rm in}$) to estimate the amount of signal loss. Left: Empirical inverse covariance weighting is used in power spectrum estimation, as done in \citetalias{ali_et_al2015}. The dotted diagonal black line indicates perfect 1:1 input-to-output mapping (no signal loss). The gray horizontal line is the power spectrum value of data alone, $\widehat{\textbf{P}}_{x}$ (it does not depend on injected power). The green signal-signal component is the term used in \citetalias{ali_et_al2015} to estimate signal loss. It is significantly higher than $\widehat{\textbf{P}}_{r}$ (black) when the cross-terms (red) are large and negative (black $=$ green $+$ red $+$ blue). In the regime where cross-correlations between signal and data are not dominant (small and large $P_{\rm in}$), the cross-terms have a noise-like term with width $\sqrt{\widehat{\textbf{P}}_e}\sqrt{\widehat{\textbf{P}}_x}$. However, at power levels comparable to the data (the middle region), the cross-terms can produce large, negative estimates due to couplings between $\textbf{x}$ and $\textbf{e}$ which affect $\widehat{\textbf{C}}_{r}$. This causes the difference between the green curve (which exhibits negligible loss at the data-only power spectrum value) and the black curve (which exhibits $\sim4$ orders of magnitude of loss). Right: The same power spectrum terms illustrated for the uniform weighted case.}
\label{fig:sigloss_terms}
\end{figure*}

Under this injection framework, we can begin to see explicitly why there can be large signal loss. Expanding out Equation \eqref{eq:sigloss}, $\widehat{P}_{\rm out}$ becomes:

\begin{eqnarray}
\label{eq:crossterm_full}
\widehat{P}_{\rm out}^{\alpha} &=& \text{M}^{\alpha}_{r}(\textbf{x}+\textbf{e})^{\dagger}\textbf{R}_{r}\textbf{Q}^{\alpha}\textbf{R}_{r}(\textbf{x}+\textbf{e}) \nonumber \\
&=& \text{M}^{\alpha}_{a}\textbf{x}^{\dagger}\textbf{R}_{r}\textbf{Q}^{\alpha}\textbf{R}_{r}\textbf{x} + \text{M}^{\alpha}_{b}\textbf{e}^{\dagger}\textbf{R}_{r}\textbf{Q}
^{\alpha}\textbf{R}_{r}\textbf{e} \nonumber \\
&+& \text{M}^{\alpha}_{c}\textbf{x}^{\dagger}\textbf{R}_{r}\textbf{Q}^{\alpha}\textbf{R}_{r}\textbf{e} + \text{M}^{\alpha}_{d}\textbf{e}^{\dagger}\textbf{R}_{r}\textbf{Q}^{\alpha}\textbf{R}_{r}\textbf{x}. 
\end{eqnarray}
Assuming \textbf{R}$_{r}$ is symmetric, the two cross-terms (terms with one copy of $\textbf{e}$ and one copy of $\textbf{x}$) can be summed together as:

\begin{eqnarray}
\label{eq:crossterm}
\widehat{P}_{\rm out}^{\alpha} &= &  \text{M}^{\alpha}_{a}\textbf{x}^{\dagger}\textbf{R}_{r}\textbf{Q}^{\alpha}\textbf{R}_{r}\textbf{x} + \text{M}^{\alpha}_{b}\textbf{e}^{\dagger}\textbf{R}_{r}\textbf{Q}
^{\alpha}\textbf{R}_{r}\textbf{e} \nonumber \\
&+& 2 \text{M}^{\alpha}_{c}\textbf{x}^{\dagger}\textbf{R}_{r}\textbf{Q}^{\alpha}\textbf{R}_{r}\textbf{e}. 
\end{eqnarray}
One of the key takeaways of this section is that the \citetalias{ali_et_al2015} analysis estimated signal loss by comparing \textit{only} the signal-only term (second term in Equation \eqref{eq:crossterm}) with $\widehat{P}_{\rm in}$, whereas in fact the cross-term (third term in Equation \eqref{eq:crossterm}) can substantially lower $\widehat{P}_{\rm out}$. In order to investigate the effect of each of these terms on signal loss, all three components are plotted in Figure \ref{fig:sigloss_terms} for two cases: empirically estimated inverse covariance weighting ($\textbf{R}_{r} \equiv \widehat{\textbf{C}}_{r}^{-1}$) and uniform weighting ($\textbf{R}_{r} \equiv \textbf{I}$). We will now go into further detail and examine the behavior of this equation in three different regimes of the injected signal - very weak (left ends of the $P_{\rm in}$ axes in Figure \ref{fig:sigloss_terms}), very strong (right ends), and in between (middle portions).

{\bf Small injection:}
In this regime, the cross-terms (red) behave as noise averaged over a finite number of samples. Output values are Gaussian distributed around zero, spanning a range of values set by the injection level. This is because $\widehat{\textbf{R}}_{r}$ is dominated by the data $\textbf{x}$, avoiding correlations with $\textbf{e}$ that can lead to solely negative power (explained further below). In fact, for the uniformly weighted case, the cross-term  $\text{M}^{\alpha}_{x}\textbf{x}^{\dagger}\textbf{I}\textbf{Q}^{\alpha}\textbf{I}\textbf{e}$ is well modeled as a symmetric distribution with zero mean and width $\sqrt{\widehat{\textbf{P}}_e}\sqrt{\widehat{\textbf{P}}_x}$. We also note that in this regime, $\widehat{\textbf{P}}_{r}$ (black) approaches the data-only power spectrum value (gray) as expected. 

{\bf Large injection:}
When the injected signal is much larger than the measured power spectrum, the data-only components can 
be neglected as they are many orders of magnitude smaller. We include a description of this regime for completeness in our discussion, but note that the upper limits that we compute are typically not determined by simulations in this regime (i.e., in using an empirical weighting scheme we've assumed the data to be dominated by foregrounds rather than the cosmological signal).  However, it is useful as a check of our system in a relatively simple case. As we can see from Figure \ref{fig:sigloss_terms}, the cross-terms (red) are small in comparison to the signal-only term (green). Here only does the signal-only term used in \citetalias{ali_et_al2015} dominate the total power output. We again see that, in the empirical inverse covariance weighted case, the cross-terms behave as noise (positive and negative fluctuations around zero mean). This is for the same reason as at small injections --- here $\widehat{\textbf{C}}_{r}$ is dominated by the signal $\textbf{e}$. The cross-correlation can again be modeled as a symmetric distribution of zero mean and width $\sqrt{\widehat{\textbf{P}}_e}\sqrt{\widehat{\textbf{P}}_x}$.

{\bf In between:}
When the injected signal is of a similar amplitude to the data by itself, the situation becomes less straightforward. We see that 
the weighted injected power spectrum component mirrors the input power indicating little loss (i.e., the green curve follows the dotted black line), eventually 
departing from unity when the injected amplitude is well above the level of the data power spectrum. However, 
in this regime the cross-term (red) has nearly the same amplitude, but with a negative sign. As explained below, this negativity is the result of cross-correlating inverse covariance weighted terms.  This negative component drives down the $\widehat{P}_{\rm out}$ estimator (black). Again, we emphasize that in \citetalias{ali_et_al2015}, signal loss was computed by only looking at the second term in Equation \eqref{eq:crossterm} (green), which incorrectly implies no loss at the data-only power spectrum level. Ignoring the effect of the negative power from the cross-terms is the main reason for underestimating power spectrum limits in \citetalias{ali_et_al2015}.

The source of the strong negative cross-term is not immediately obvious, however it is an explainable effect. 
When $\textbf{R}_{r}$
is taken to be $\widehat{\textbf{C}}_{r}^{-1}$, the third term of Equation \eqref{eq:crossterm} is a cross-correlation between $\widehat{\textbf{C}}_{r}^{-1}\textbf{x}$ and
$\widehat{\textbf{C}}_{r}^{-1}\textbf{e}$. As shown in \citet{switzer_et_al2015}, this cross-correlation term is non-zero, and in fact negative in expectation. 
This negative cross-term power arises from a coupling between the inverse of 
$\widehat{\textbf{C}}_{r}$ and $\mathbf{x}$. 
Intuitively, we can see this by expanding the empirical covariance of $\textbf{r}=\textbf{x}+\textbf{e}$:

\begin{eqnarray}
\widehat{\textbf{C}}_{r} &=& \langle \textbf{rr}^{\dagger} \rangle_{t} \nonumber \\ 
&=& \langle \textbf{xx}^{\dagger} \rangle_{t} + \langle \textbf{xe}^{\dagger} \rangle_{t} + \langle \textbf{ex}^{\dagger} \rangle_{t} + \langle 
\textbf{ee}^{\dagger} \rangle_{t},
\end{eqnarray}

\noindent where we can neglect the first term because $\textbf{x}$ is small (i.e., the large negative cross-term power in the left panel of Figure \ref{fig:sigloss_terms} occurs when the injected amplitude surpasses the level of the data-only power spectrum).  Without loss of generality, we will assume
an eigenbasis of $\textbf{e}$, so that $\langle 
\textbf{ee}^{\dagger} \rangle_{t}$ is diagonal. The middle 
two terms, however, can have power in their off-diagonal terms due to the fact that, when averaging over a finite
ensemble, $\langle\textbf{xe}^\dagger\rangle_t$ is not zero.  As shown in Appendix C of \citet{parsons_et_al2014}
, to leading order the inversion of a diagonal-dominant matrix like $\widehat{\textbf{C}}_{r}$ (from $\langle 
\textbf{ee}^{\dagger} \rangle_{t}$) with smaller
off-diagonal terms results in a new diagonal-dominant matrix with negative off-diagonal terms. These off-diagonal
terms depend on both $\textbf{x}$ and $\textbf{e}$. Then, when $\widehat{\textbf{C}}^{-1}_{r}$ is multiplied into $\textbf{x}$,
the result is a vector that is similar to $\textbf{x}$ but
contains a residual correlation to $\textbf{e}$ from the off-diagonal components of $\widehat{\textbf{C}}^{-1}_{r}$. The
correlation is negative because the product $\widehat{\textbf{C}}_r^{-1}\textbf{x}$ effectively squares the $\textbf{x}$-dependence
of the off-diagonal terms in $\widehat{\textbf{C}}^{-1}_{r}$ while retaining the negative sign that arose from the inversion
of a diagonal-dominant matrix.



{\bf In general:} Another way to phrase the shortcoming of the empirical inverse covariance estimator is that it is not properly normalized. Signal loss due to couplings between the data and its weightings arise because our unnormalized quadratic estimator from Equation \eqref{eq:qhat} ceases to be a quadratic quantity, and instead contains higher order powers of the data. However, the normalization matrix $\mathbf{M}$ is derived assuming that the unnormalized estimator is quadratic in the data. The power spectrum estimate will therefore be incorrectly normalized, which manifests as signal loss. We leave a full analytic solution for $\mathbf{M}$ for future work, since our simulations already capture the full phenomenology of signal loss and have the added benefit of being more easily generalizable in the face of non-Gaussian systematics.

\subsection{Signal Loss in Practice}
\label{sec:Practice}

We now shift our attention towards computing upper limits on the EoR signal for the fringe-rate filtered PAPER-64 dataset in a way that accounts for signal loss. While our methodology 
outlined below is independent of weighting scheme, here we demonstrate the computation using empirically estimated inverse covariance weighting 
($\textbf{R} \equiv \widehat{\textbf{C}}^{-1}$), the weighting scheme used in \citetalias{ali_et_al2015} which leads to substantial loss. 

One issue to address is how one incorporates the randomness of $\widehat{P}_{\rm out}$ into our signal loss corrections. A different realization of the mock EoR signal is injected with each bootstrap run, causing the output to vary in three ways ---  there is noise variation from the bootstraps, there is cosmic variation from generating multiple realizations of the mock EoR signal, and there is a variation caused by whether the injected signal looks more or less ``like'' the data (i.e., how much coupling there is, which affects how much loss results). 

For each injection level, the true $P_{\rm in}$ is simply the average of our bootstrapped estimates $\widehat{P}_{\rm in}$, since $\widehat{P}_{\rm in, \alpha}$ is by construction an unbiased estimator. Phrased in the context of Bayes' rule, we wish to find the posterior probability distribution $p(P_{\rm in} | 
\widehat{P}_{\rm out})$, which is the probability of $P_{\rm in}$ given the uncorrected/measured power spectrum estimate $\widehat{P}_{\rm out}$.  Bayes' rule relates the posterior, which we don't know, to the likelihood, which we can forward model. In other words,

\begin{equation}
\label{eq:Bayes}
p(P_{\rm in} | \widehat{P}_{\rm out}) \propto {\mathcal{L} (  \widehat{P}_{\rm out} | P_{\rm in})}\,p(P_{\rm in}) ,
\end{equation}

\noindent where $\mathcal{L} $ is the likelihood function defined 
as the distribution of data plus signal injection ($\widehat{P}_{\rm out}$) given the injection $P_{\rm in}$.  We construct this distribution  
by fixing $P_{\rm in}$ and simulating our analysis pipeline for many realizations of the injected EoR signal 
consistent with this power spectrum. The resulting distribution is normalized such that the sum over $\widehat{P}_{\rm out}$ is unity, and the 
whole process is then repeated for a different value of $P_{\rm in}$. 


The implementation details of the injection process require some more detailed explanation. In our code, we add a new realization of EoR to each independent bootstrap of data (see Section \ref{sec:Boot} for a description of PAPER's bootstrapping routine) with the goal of simultaneously capturing cosmic variance, noise variance, and signal loss. To limit computing time we perform $20$ realizations of each $P_{\rm in}$ level. We also run $50$ total EoR injection levels, yielding $P_{\rm in}$ values that range from $\sim$$10^{5}$\,mK$^{2}$ ($h^{-1}$ Mpc)$^{3}$ to $\sim$10$^{11}$\,mK$^{2}$ ($h^{-1}$ Mpc)$^{3}$, resulting in a total of $1000$ data points on our $P_{\rm in}$ vs. $\widehat{P}_{\rm out}$ grid. 

Going forward, we treat every $k$-value separately in order to determine an upper limit on the EoR signal per $k$. We bin our simulation outputs along the $P_{\rm in}$ axis (one bin per injection level) and, since they are well-approximated by a Gaussian distribution in our numerical results, we smooth the distribution of $\widehat{P}_{\rm out}$ values by fitting Gaussians for each bin based on its mean and variance (and normalize them). Stitching all of them together results in a 2-dimensional transfer function --- the likelihood function in Bayes' rule, namely $\mathcal{L} (  \widehat{P}_{\rm out} | P_{\rm in})$. We then have a choice for our prior, $p(P_{\rm in})$, and we choose to invoke a Jeffreys prior (\citealt{jaynes1968}) because it is a true uninformative prior. 

\begin{figure*}
	\centering
	\includegraphics[width=1\textwidth]{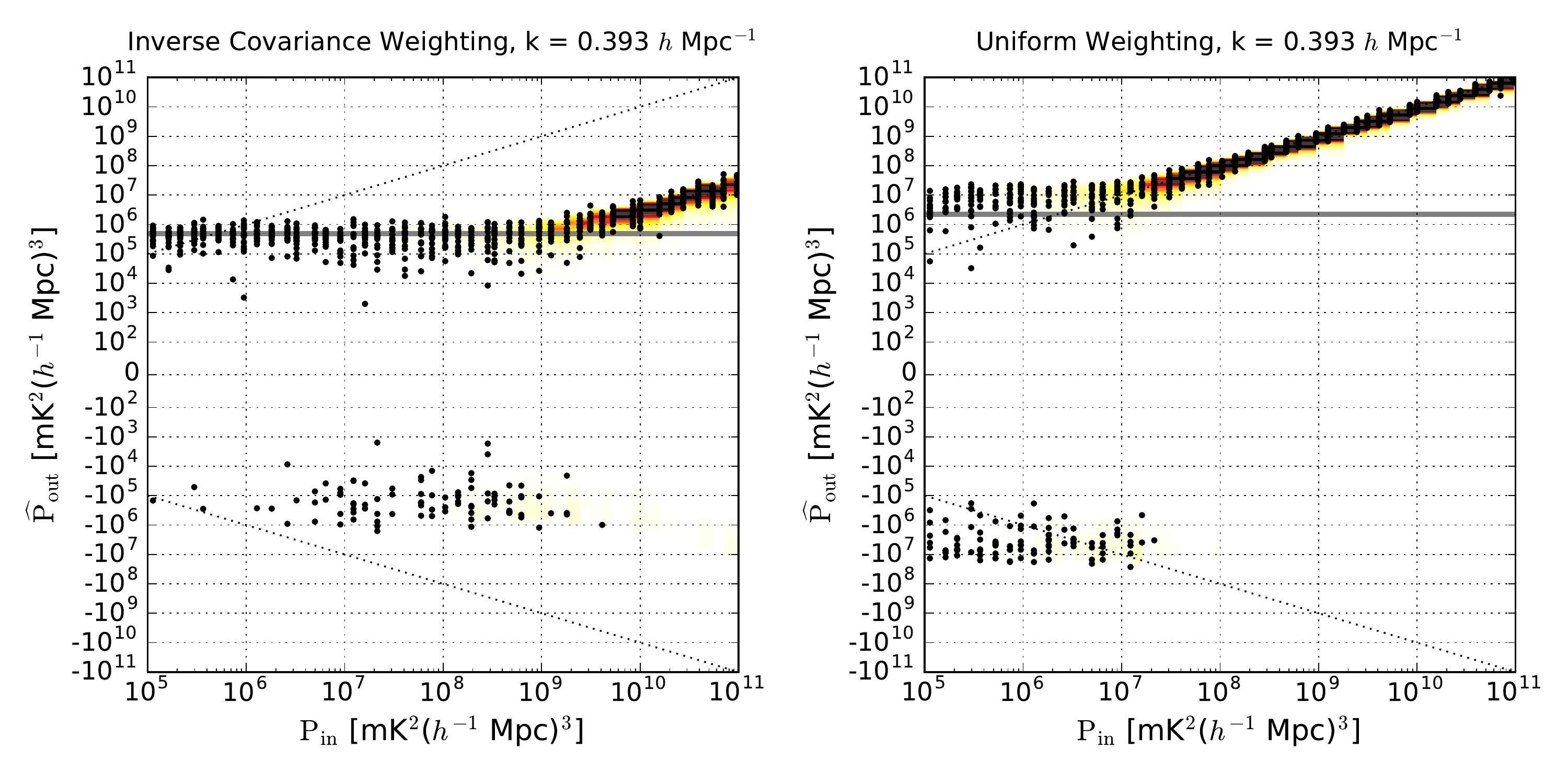}
	\caption{Signal loss transfer functions showing the relationship of $P_{\rm in}$ and $\widehat{P}_{\rm out}$, as defined by Equations \eqref{eq:Pin} and \eqref{eq:sigloss}. Power spectra values (black points) are generated for $20$ realizations of $\textbf{e}$ per signal injection level. Since our $\widehat{P}_{\rm out}$ values are well-approximated by a Gaussian distribution, we fit Gaussians to each injection level based on the mean and variance of the simulation outputs. This entire likelihood function is then multiplied by a Jeffreys prior for $p(P_{\rm in}$), with the final result shown as the colored 
heat-maps on top of the points. Two cases are displayed: empirically estimated inverse covariance weighted PAPER-64 data (left) and uniform-weighted data (right). The dotted black 
diagonal lines mark a perfect unity mapping, and the solid gray horizontal line denotes the power spectrum value of the data $\widehat{\textbf{P}}_{x}$, from which a posterior distribution for the signal is extracted. From these plots, it is clear that the weighted case results in $\sim4$ orders of magnitude of signal loss at the data-only power spectrum value, whereas the uniform-weighted case does 
not exhibit loss. The general shape of these transfer functions are also shown by the black curves in Figure \ref{fig:sigloss_terms} for comparison.}
	\label{fig:sigloss_transfercurve}
\end{figure*}

Finally, our transfer functions are shown in Figure \ref{fig:sigloss_transfercurve} for both the weighted (left) and unweighted (right) cases. Our bootstrapped power spectrum outputs are shown as black points and the colored heat-map overlaid on top is the likelihood function modified by our prior. Although we only show figures for one $k$-value, we note that 
the shape of the transfer curve is similar for all $k$'s. We then invoke Bayes' interpretation and re-interpret it as the posterior $p(P_{\rm in}|\widehat{P}_{\rm out})$ where we recall that $\widehat{P}_{\rm out}$ represents a (lossy) power spectrum. To do this we make a horizontal cut across at the data value $\widehat{\textbf{P}}_{x}$ (setting $\widehat{P}_{\rm out} = \widehat{\textbf{P}}_{x}$), shown by the gray solid line, to yield a posterior distribution for the signal. We normalize this final distribution and compute the $95\%$ confidence interval (an upper limit on EoR).


By-eye inspection of the transfer function in Figure \ref{fig:sigloss_transfercurve} gives a sense of what the signal loss result should be. The power spectrum value of our data, $
\widehat{\textbf{P}}_{x}$ is marked by the solid gray horizontal lines. From the left plot (empirically estimated inverse covariance weighting), one can eyeball that a data value of $10^{5} \,$mK$^{2}$ ($h^{-1}$ Mpc)$^{3}$, for example, would map approximately to an upper limit of $\sim10^{9} \,$mK$^{2}$ ($h^{-1}$ Mpc)$^{3}$, implying a signal loss factor of $\sim10^{4}$.

%

\begin{figure*}
	\centering
	\includegraphics[width=0.4\textwidth]{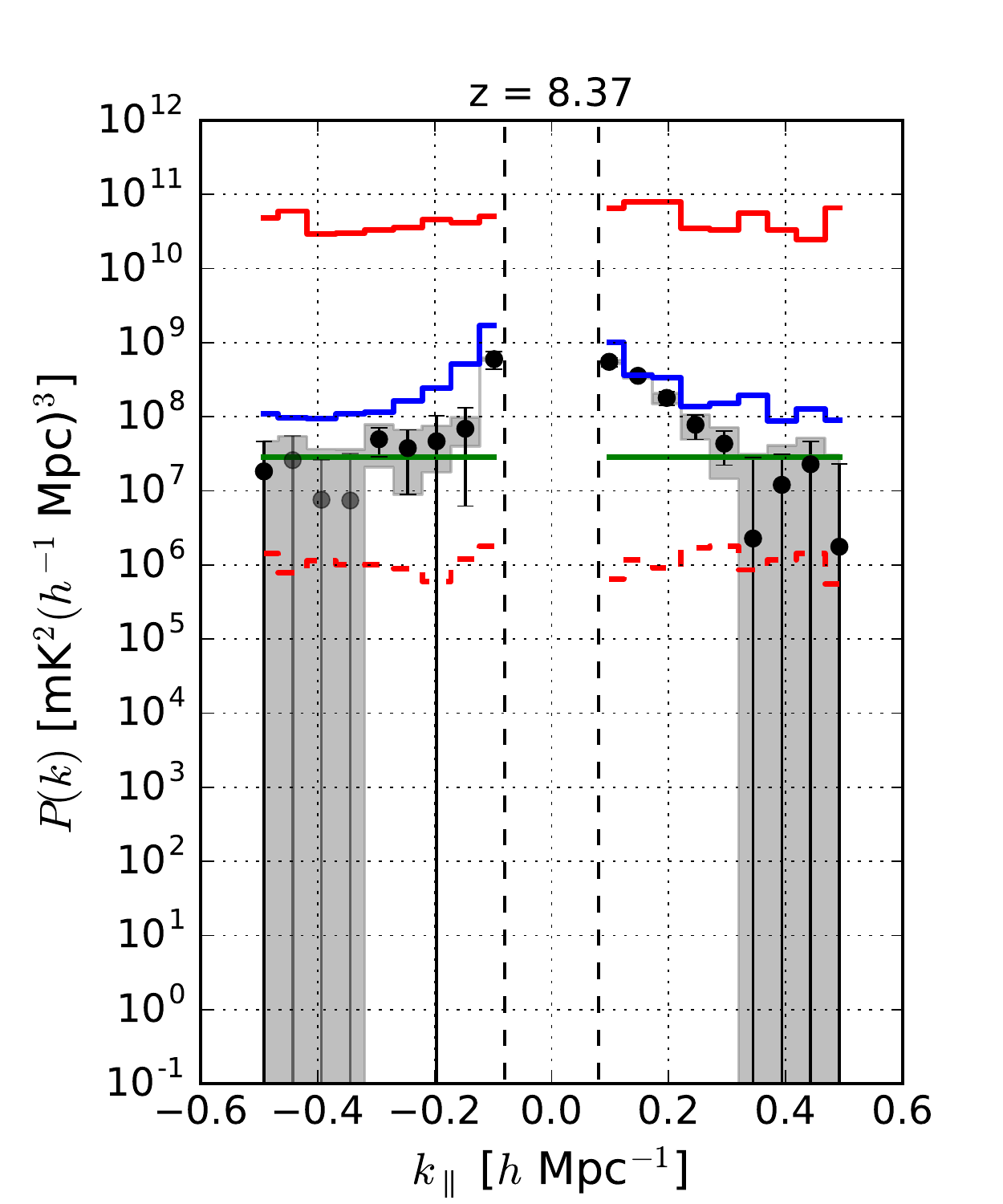}
	\includegraphics[width=0.4\textwidth]{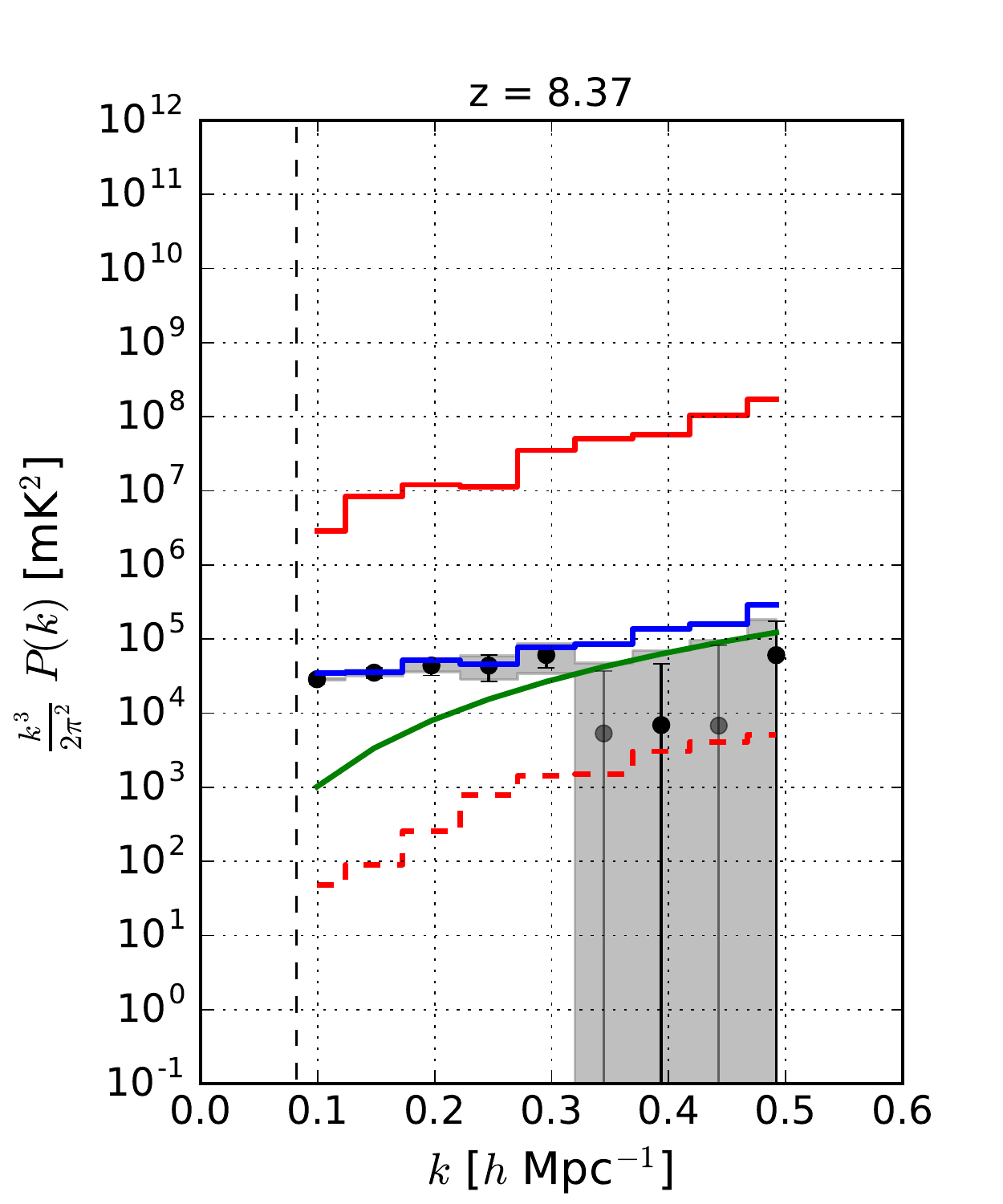}
	\caption{A power spectrum of a subset of PAPER-64 data illustrating the use of empirical inverse covariance weighting. The solid red curve is the $2\sigma$ upper limit on the EoR signal estimated from our signal injection framework using empirical inverse covariance weighting. Shown for comparison is the lossy limit prior to signal loss estimation (dashed red). The theoretical $2\sigma$ thermal noise level prediction based on observational parameters is in green, whose calculation is detailed in Section \ref{sec:Error}. Additionally, the power spectrum result for the uniform weighted case is shown in three different ways: power spectrum values (black and gray points as positive and negative values, respectively, with $2\sigma$ error bars from bootstrapping), the $2\sigma$ upper limit on the EoR signal using our full signal injection framework (solid blue), and the measured power spectrum values with $2\sigma$ thermal noise errors (gray shaded regions). The vertical dashed black lines signify the horizon limit for this analysis using $30$\,m baselines. In this example, we see that the lossy power spectrum limit is $\sim 4$ orders of magnitude too low when using empirical inverse covariance weighting.}
\label{fig:ps2_data}
\end{figure*}

The loss-corrected power spectrum limit for empirically estimated inverse covariance weighted PAPER-64 data is shown in Figure \ref{fig:ps2_data} (solid red), which we can compare to the original lossy result (dashed red). 
Post-signal loss estimation, the power spectrum limits are higher than both the theoretical noise level (green) and uniform-weighted power spectrum (which is shown three ways: black and gray points are positive and negative power spectrum values, respectively, with $2\sigma$ error bars from bootstrapping, the solid blue is the upper limit on the EoR signal using the full signal injection framework, and the shaded gray is the power spectrum values with thermal noise errors). We elaborate on this point in the next section, as well as investigate alternate 
weighting schemes to inverse covariance weighting, with the goal of finding one that balances the aggressiveness of down-weighting contaminants and minimizing the loss of the EoR signal. 

\subsection{Minimizing Signal Loss}
\label{sec:Weight}

With a signal loss formalism established, we now have the capability of experimenting 
with different weighting options for $\textbf{R}$. Our goal here is to choose a weighting method that successfully down-weights 
foregrounds and systematics in our data without generating large amounts of signal loss as we have seen with the inverse covariance estimator. We have found that the balance 
between the two is a delicate one and requires a careful understanding and altering of empirical covariances. 

\begin{figure*}
	\centering
	\includegraphics[width=1\textwidth]{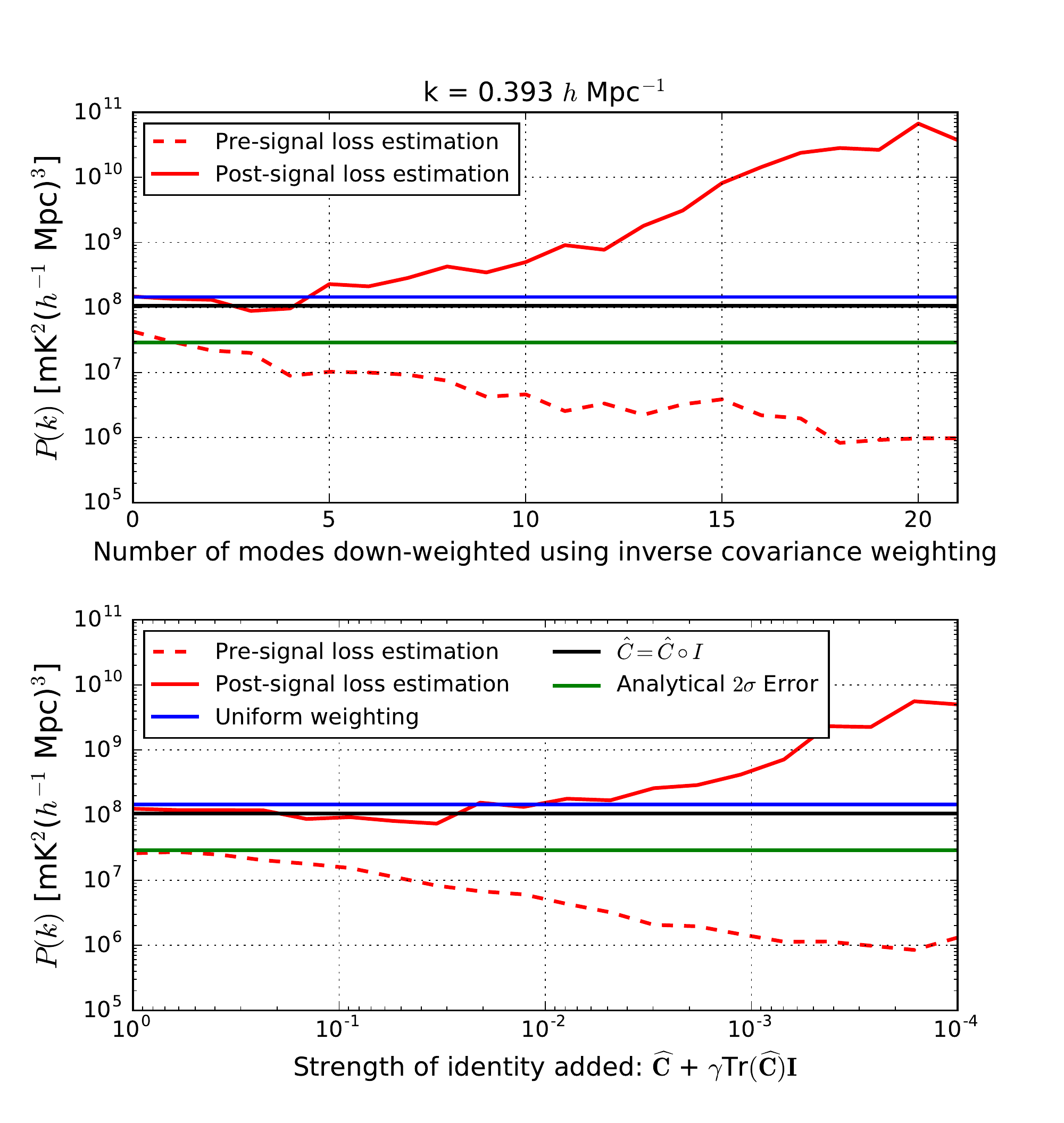}
	\caption{Power spectra $2\sigma$ upper limits for $k=0.393$\,$h$ Mpc$^{-1}$ for fringe-rate filtered PAPER-64 data. Top: Values 
are shown before (dashed red) and after (solid red) signal loss estimation via our signal injection framework as a function of number of eigenmodes of $\widehat{\textbf{C}}$ that 
are down-weighted. This regularization knob is tuned from $0$ modes on the left (i.e., unweighted) to $21$ modes on the right (i.e., the full inverse 
covariance estimator). $\sim4$ orders of magnitude of signal loss results when using empirically estimated inverse covariance weighting. Bottom: Power spectrum upper limits before (dashed red) and after (solid red) signal loss estimation as a function of identity added to the empirical covariance. This regularization knob is tuned from $\gamma = 10^{-4}$ on the right (i.e., very little regularization) to $\gamma = 1$ on the left (see main text for the definition of $\gamma$). Also 
plotted in both panels for comparison are $2\sigma$ power spectrum upper limits for the uniform-weighted case (blue) and inverse variance 
weighted case (black); both are after signal loss estimation. Finally, a theoretical prediction for noise ($2\sigma$ error) is plotted 
as green. In the PAPER-64 analysis in this paper, we choose to use a regularization scheme of $\widehat{\textbf{C}}_{\rm eff} \equiv 0.09 \, $Tr($\widehat{\textbf{C}})\textbf{I} + \widehat{\textbf{C}}$ ($\gamma = 0.09$) as a simple example of regularization that minimizes loss, and note that the power spectrum limits using this type of regularization are roughly constant across a large range of values of $\gamma$.}
	\label{fig:sigloss_modeloop}
\end{figure*}

We saw in Section \ref{sec:otherweight} how limiting the number of down-weighted eigenmodes (i.e., flattening out part of the 
eigenspectrum and effectively decoupling the lowest-valued eigenmodes, which are typically EoR-dominated, from the data) can help minimize signal loss. We experiment with this idea on PAPER-64 data, dialing the number of modes 
that are down-weighted from zero (which is equivalent to identity-weighting, or the uniform-weighted case) to $21$ (which is the full inverse 
covariance estimator). The power spectrum results for one $k$-value, both before and after signal loss 
estimation, are shown in the top panel in Figure \ref{fig:sigloss_modeloop}. We see that the amount of signal loss increases as weighting 
becomes more aggressive (dashed red). In other words, more EoR-dominated fluctuations are being overfit and 
subtracted as more modes are down-weighted. We also find that the power spectrum upper limit, post-signal loss estimation, 
increases with the number of down-weighted modes (solid red). The more modes we use in down-weighting, the stronger the coupling between the weighting and the data, and the greater the error we have in estimating the power spectrum. \citet{switzer_et_al2013} took a similar approach in determining the optimal number of modes to down-weight in GBT data, finding similar trends and noting that removing too few modes is limited by residual foregrounds and removing too many modes is limited by large error bars and signal loss.

Optimistically, we expect there to be a ``sweet spot" as we dial our regularization knob; a level of regularization where weighting 
is beneficial compared to uniform weighting (blue). In other words, we would like a weighting scheme that down-weights eigenmodes that predominantly describe foreground modes, but not EoR modes. We see in Figure \ref{fig:sigloss_modeloop} that this occurs roughly when 
only the $\sim3$ highest-valued eigenmodes are down-weighted and the rest are given equal weights (though for the case shown, weighting only slightly outperforms uniform weighting). For a similar discussion on projecting out modes (zeroing out eigenmodes, rather than just ignoring their relative weightings as we do in this study), see \citet{switzer_et_al2013}. 

We also saw in Section \ref{sec:otherweight} how adding the identity matrix to the empirical covariance can minimize signal loss. We experiment with this idea as well, shown in the bottom panel of Figure \ref{fig:sigloss_modeloop}. The dashed red and solid red lines represent power spectrum limits pre and post-signal loss estimation, respectively, as a function of the strength of $\textbf{I}$ that is added to $\widehat{\textbf{C}}$, quantified as a percentage of Tr($\widehat{\textbf{C}})\textbf{I}$ added to $\widehat{\textbf{C}}$. We parameterize this ``regularization strength" parameter as $\gamma$, namely $\widehat{\textbf{C}} \equiv \widehat{\textbf{C}} + \gamma$Tr$(\widehat{\textbf{C}})\textbf{I}$. From this plot we see that only a small percentage of Tr($\widehat{\textbf{C}})$ is needed to significantly reduce loss. We expect that as the strength of $\textbf{I}$ is increased (going to the left), both the red curves will approach the uniform-weighted case. We also notice that the post-signal loss limit hovers around the uniform-weighted limit for a large range of regularization strengths and while an overall trend from high-to-low signal loss is seen as the strength increases, there does not appear to be a clear ``minimum" that produces the least loss.

In addition to our thermal noise prediction (green) and uniform-weighted power spectrum limit (blue), one additional horizontal line is shown in Figure \ref{fig:sigloss_modeloop} 
in both panels and represents a third regularization technique. This line (black) denotes the power spectrum value, post-signal loss estimation, for inverse variance weighting (multiplying an identity 
matrix element-wise to $\widehat{\textbf{C}}$). This result is single-valued and not a function of the horizontal axis. We see that all three regularization schemes shown (solid red top panel, solid red bottom panel, black) perform similarly at 
their best (i.e., when $\sim3$ eigenmodes are down-weighted in the case of the top panel's solid red curve). However, for the remainder of this paper, we choose to use the weighting option of $\widehat{\textbf{C}} + 0.09 \,$Tr($\widehat{\textbf{C}})\textbf{I}$, or $\gamma = 0.09$, which we will denote as $\widehat{\textbf{C}}_{\rm eff}$. We choose this weighting scheme merely as a simple example of regularizing PAPER-64 covariances, noting that the power spectrum upper limit remains roughly constant for a broad range of values of $\gamma$. 

It is important to note that our signal injection methodology for assessing loss makes the assumption that we know the true signal's strength and structure. Realistically, these details about the EoR signal are unknown and our signal loss framework is limited by our simulations. Therefore, while this paper employs this methodology as an example of one way of estimating loss, Kolopanis et al. (\textit{in prep.}) use uniform weightings in order to produce more trustworthy, straightforward power spectrum limits that do not suffer from loss.

\begin{figure*}
	\centering
	\includegraphics[width=0.4\textwidth]{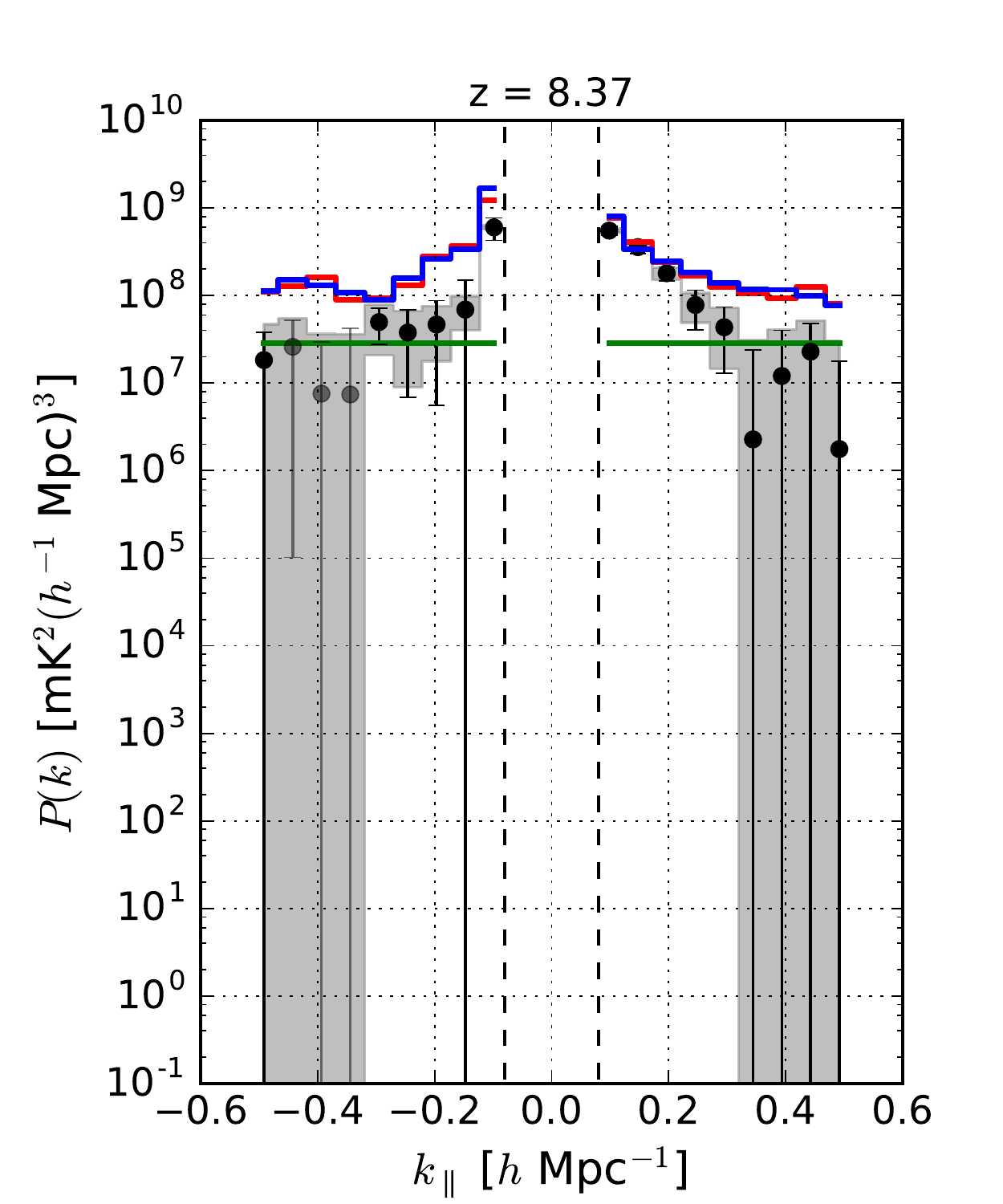}
	\includegraphics[width=0.4\textwidth]{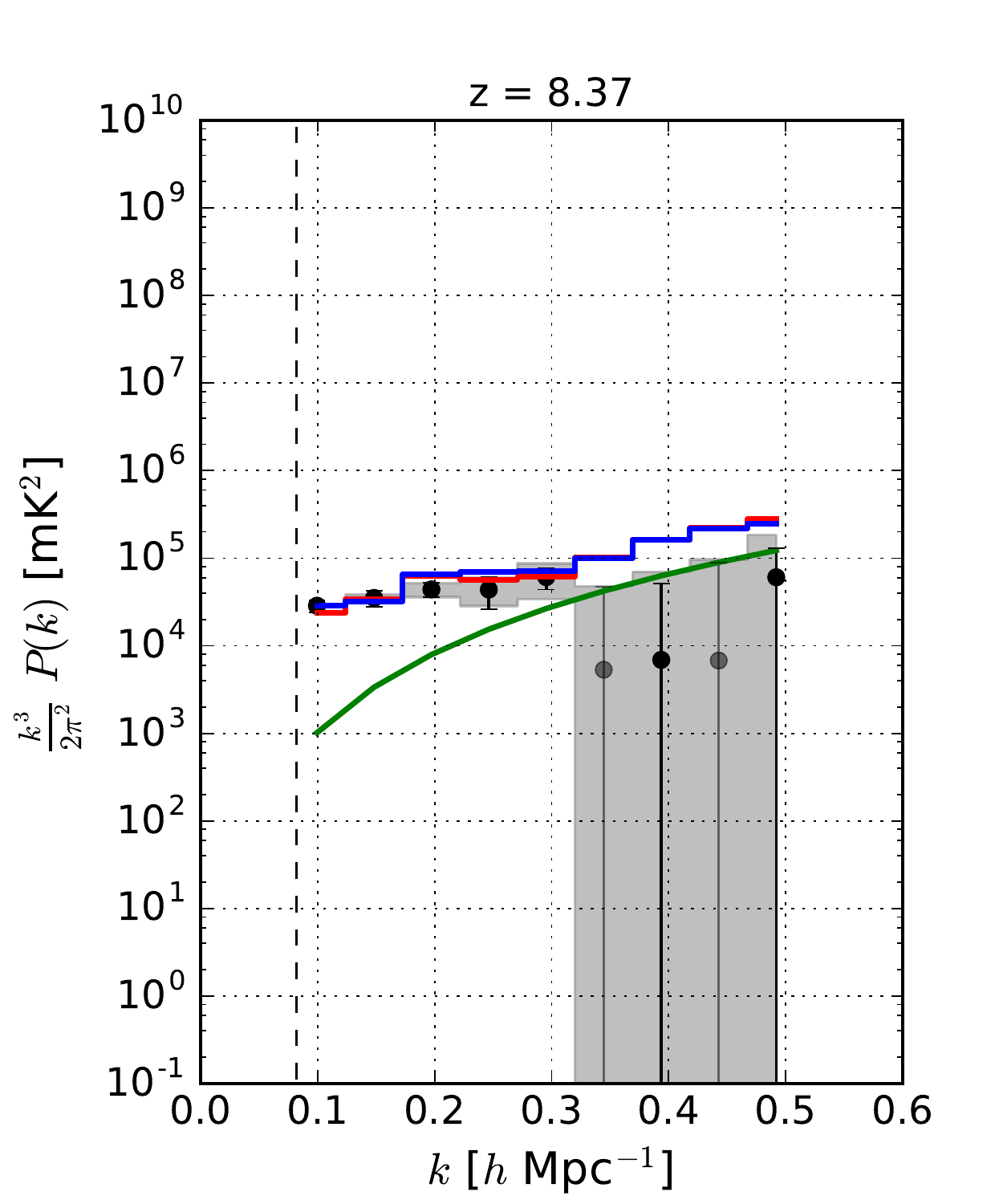}
	\caption{A power spectrum of a subset of PAPER-64 data illustrating the use of $\widehat{\textbf{C}}_{\rm eff}$ to minimize signal loss. The solid red curve is the $2\sigma$ upper limit on the EoR signal estimated from our signal injection framework. The theoretical $2\sigma$ thermal noise level prediction based on observational parameters is in green. Additionally, the power spectrum result for the uniform weighted case is shown in three different ways: power spectrum values (black and gray points as positive and negative values, respectively, with $2\sigma$ error bars from bootstrapping), the $2\sigma$ upper limit on the EoR signal using our full signal injection framework (solid blue), and the measured power spectrum values with $2\sigma$ thermal noise errors (gray shaded regions). The vertical dashed black lines signify the horizon limit for this analysis using $30$\,m baselines. This power spectrum result does not use the full dataset's sensitivity as in \citetalias{ali_et_al2015} and Kolopanis et al. (\textit{in prep.}), though we include all analysis changes which have mostly stemmed from revisions regarding signal 
loss, bootstrapping, and the theoretical error computation. We see that the regularization scheme used here produces limits similar to the unweighted limits.}
	\label{fig:ps1_data}
\end{figure*}

The power spectrum result for our subset of PAPER-64 data (using only one baseline separation type, $10$ baselines, and $\widehat{\textbf{C}}_{\rm eff}$) using the analysis presented in this paper is shown in Figure 
\ref{fig:ps1_data}. Again, the solid red curve represents our upper limit on the EoR signal using the full signal injection framework. The uniform weighted case is shown as the black and gray points, which correspond to positive and negative power spectrum values respectively (with 
$2\sigma$ errors bars from bootstrapping). It is also shown as an upper limit using the signal injection framework (solid blue), which is interestingly larger than the errors computed from bootstrapping, likely because the full injection framework takes into account additional sample variance whereas the bootstrapped errors do not. Finally, the gray shaded regions combine the measured uniform weighted power spectrum values with thermal noise errors. We show this power spectrum result as one example of how a simple regularization of an empirical covariance matrix can minimize signal loss, though we also note that this weighting does not produce more stringent limits than the uniform weighted case, thus further motivating uniform-weighting for Kolopanis et al. (\textit{in prep.}). 

In this section we have shown three simple ways of regularizing $\widehat{\textbf{C}}$ to minimize signal loss using PAPER-64 
data. There are many other weighting schemes that we leave for consideration in future work. For example, one could estimate 
$\widehat{\textbf{C}}$ using information from different subsets of baselines. For redundant arrays this might mean calculating $
\widehat{\textbf{C}}$ from a different but similar baseline type, such as the $\sim30$\,m diagonal PAPER baselines (instead of the 
horizontal E/W ones). Alternatively, covariances could be estimated from all baselines except the two being cross-multiplied 
when forming a power spectrum estimate. This method was used in \citet{parsons_et_al2014} (a similar method was also used in \citet{dillon_et_al2015}) in order to avoid suppressing the 
21\,cm signal, and it is worth noting that the PAPER-32 results are likely less impacted from the issue of signal loss underestimation 
because of this very reason (however, they are affected by the error estimation issues described in Section \ref{sec:Error}, so 
we also regard those results as suspect and superseded by those of Kolopanis et al. (\textit{in prep.})).

Another possible way to regularize $\widehat{\textbf{C}}$ is to use information from different ranges of LST. For example, one could 
calculate $\widehat{\textbf{C}}$ with data from LSTs where foregrounds are stronger (earlier or later LSTs than the ``foreground-quiet" range typically used in forming power spectra) --- doing so may yield a better description of the foregrounds that we desire to 
down-weight, especially if residual foreground chromaticity is instrumental in origin and stable in time. Fundamentally, each of these examples are similar in that they rely on a computation of $\widehat{\textbf{C}}$ from 
data that is similar but not exactly the same as the data that is being down-weighted. Ideally this would be effective in down-weighting shared contaminants yet avoid signal loss from overfitting EoR modes in the power spectrum dataset itself. 

In Section \ref{sec:CaseStudy}, we have detailed several aspects of signal loss in PAPER-64: how the loss arises, how it can be estimated from an injection framework, and ways it can be minimized. We again emphasize that these lessons learned about signal loss are largely responsible for shaping our revised analysis of PAPER data. In the remainder of this paper, we will transition to other aspects of our analysis that have been revised since \citetalias{ali_et_al2015}.


\section{Additional PAPER-64 Revisions}
\label{sec:OtherErrors}

Underestimated signal loss is the main reason for the revision of the power spectrum limits from \citetalias{ali_et_al2015}. It is interesting to note that --- had all the other aspects of the original analysis been correct --- the underestimated limits may have been more easily caught. Unfortunately, two related power spectrum components, namely the error bars on the power spectrum data points and the theoretical noise prediction, were also calculated incorrectly.

In this section, we summarize multiple inconsistencies and errors that have been found since the previous analysis in terms of error estimation. We first describe updated methods regarding bootstrapping, which determines the error bars on our limits. We then highlight an updated calculation for the theoretical noise sensitivity of PAPER-64 and illustrate how our revised calculation has been verified through simulations. 

\subsection{Bootstrapping}
\label{sec:Boot}

Broadly speaking, we desire robust methods for determining accurate 
confidence intervals for our measurements. For PAPER's analysis, we choose a data-driven method of error estimation, computing error bars that have been derived from the inherent 
variance of our measurements. A common technique used to do this is bootstrapping, which we first define below and then discuss its application to PAPER.

Bootstrapping uses sampling with replacement to estimate a posterior distribution. For example, bootstrap measurements (of power spectra, for example) can be made from different random samples of data. Each of these bootstraps is a different realization drawn from some underlying distribution, and realizations are correlated with each other to a degree set by the fraction of sampled points that are held in common between them. Through the process of re-sampling and averaging along different axes of a dataset, such as along baselines or times, we can estimate error bars for our results which represent the underlying distribution of values that are allowed by our measurements (\citealt{efron_tibshirani1994}; \citealt{andrae2010}).

One major caveat of bootstrapping arises when working with correlated data. If, for example, a dataset has many repeated 
values inside it, this would be reflected in each bootstrap. The same value would be present multiple times within a bootstrap 
and also be present between bootstraps, purely because it has a more likely chance of being drawn if there are repeats of 
itself. Therefore, bootstrapping correlated data results in a smaller variation between bootstraps, and hence, underestimates 
errors.

This is the precisely how errors were underestimated in PAPER-64. Because of fringe-rate filtering, which averages data in time to increase sensitivity, PAPER-64 data is correlated along the time axis. Hence, there are fewer independent samples after filtering, thus decreasing the variance of the bootstraps.

More specifically, the PAPER-64 pipeline outputs $20$ bootstraps (over baselines), each a $2$-dimensional power 
spectrum that is a function of $k$ and time. In \citetalias{ali_et_al2015}, a second round of bootstrapping occurred over the time axis, and a total of $400$ bootstraps were created in this step, each comprised of randomly selected values sampled with replacement (i.e., each of these bootstraps contained the same number of values as the number of time integrations, which, at $\sim$
$700$, greatly exceeds the approximate number of independent samples after fringe-rate filtering).
Means were then taken of the values in each bootstrap. Finally, power 
spectrum limits were computed by taking the mean and standard deviation over all the bootstraps. We emphasize again that in 
this previous analysis, the number of elements sampled per bootstrap greatly 
exceeded the number of independent LST samples, underestimating errors. A random draw of $700$ 
measurements from this dataset has many repeated values, and the variance between hundreds of these random 
samples is smaller than the true underlying variance of the data. 

\begin{figure}
	\centering
	\includegraphics[trim={0.3cm 0cm 0.3cm 0.3cm},width=\columnwidth]{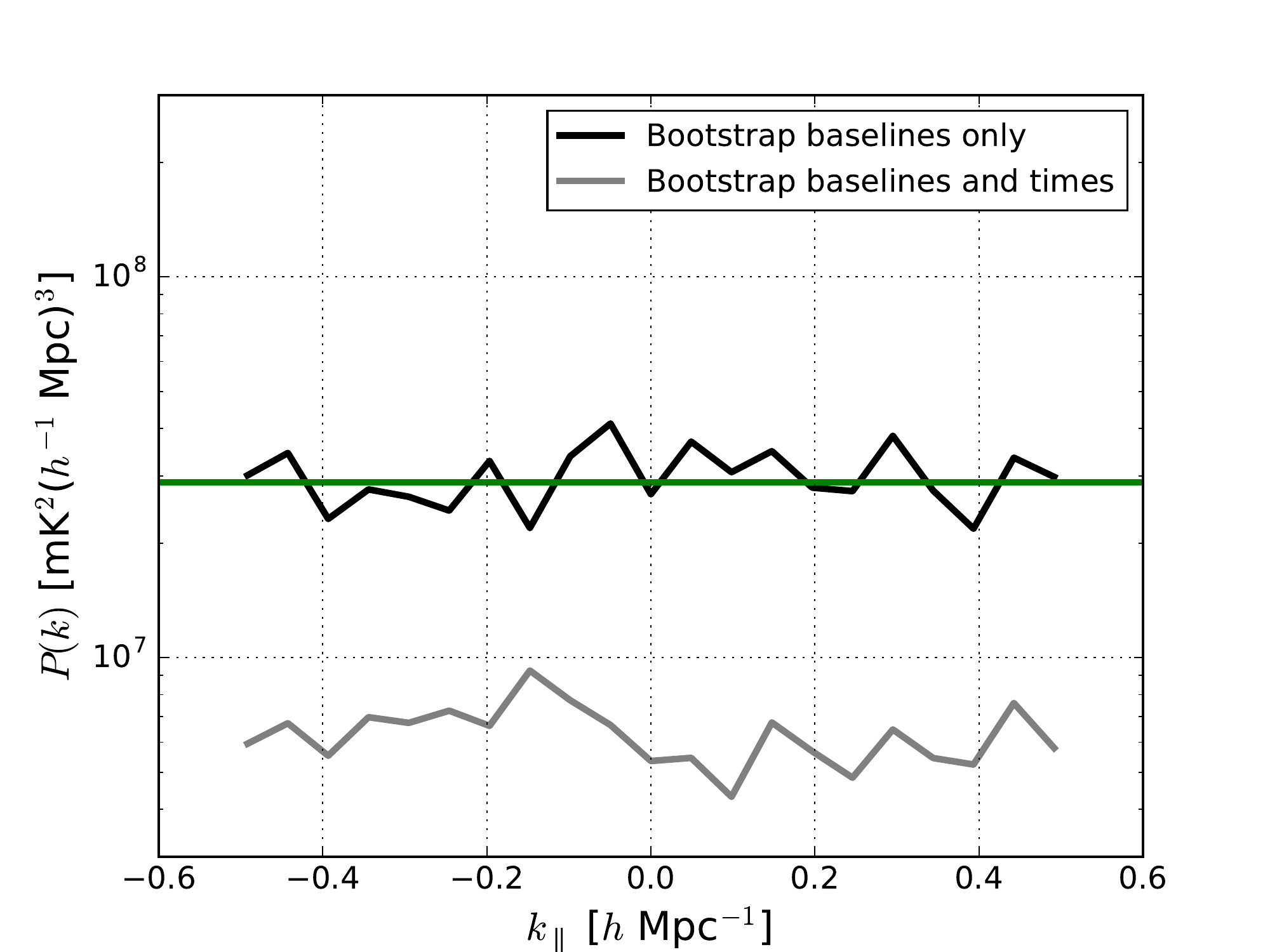}
	\caption{$2\sigma$ power spectrum errors (from bootstrap variances) for a noise simulation (computed via Equation \eqref{eq:noise} using PAPER-64 observing parameters) using two different bootstrapping 
methods. The noise is fringe-rate filtered and a weighting matrix of $\textbf{I}$ (uniform-weighted) is used in order to disentangle the 
effects of bootstrapping from signal loss. The bootstrapping method used in \citetalias{ali_et_al2015} is shown in gray, where bootstrapping occurs along both the baseline and time axes. This underestimates errors by sampling more values than independent ones in the dataset (fringe-rate filtering reduces the number of independent samples along time). We use the method illustrated by the black curve in our updated analysis, where bootstrapping only occurs along the baseline axis. We find that these revised limits agree with the $2\sigma$ analytic prediction for noise (green).}
	\label{fig:data_errors}
\end{figure}

Given our new understanding of the sensitivity of bootstraps to the number of elements sampled, we have removed the second 
bootstrapping step along time entirely and now simply bootstrap over the baseline axis. Power spectrum $2\sigma$ errors (computed from bootstrap variances) with and without this bootstrapping change for a fringe-rate filtered noise simulation are shown in Figure 
\ref{fig:data_errors} in black and gray, respectively. The estimates are uniformly weighted in order to disentangle the effects of bootstrapping from signal loss. As 
shown in the figure, when more elements are drawn for each bootstrap than the number of 
independent samples (by over-sampling elements along the time axis), repeated values begin to crop up and the apparent variation between bootstraps drops, resulting in limits (gray) below the predicted noise level (green). Using the revised bootstrapping method, where bootstrapping only occurs over the baseline axis, the limits (black) are shown to agree with the analytic prediction for noise. While Figure \ref{fig:data_errors} implies that errors, computed prior to our bootstrapping change (gray), are underestimated by a factor of $\sim$ $5$ in mK$^{2}$ for the noise simulation (whose creation details are outlined in the next section), in practice this factor is lower for the case of real data (a factor of $\sim$ $3$ in mK$^{2}$ instead), possibly due to the data being less correlated in time than the fringe-rate filtered noise in the simulation. 

In addition to learning how sample independence affects bootstrapped errors, we have made three additional changes to our bootstrapping procedure since \citetalias{ali_et_al2015}, summarized here:

\begin{itemize}

\item{A second change to our bootstrapping procedure is that we now bootstrap over baseline cross-products, instead of the baselines themselves. In the previous analysis, baselines were bootstrapped prior to forming cross power spectra, and using this particular ordering of operations (bootstrapping, then cross-multiplication) yields variances that have been found to disagree with predicted errors from bootstrapping using simulations. On the contrary, bootstrapping over cross power spectra ensures that we are estimating the variance of our quantity of interest (i.e., the power spectrum). This change, while fundamental in retaining the integrity of the bootstrapping method in general, alters the resulting power spectrum errors by factors of $<2$ in practice.}

\item{In \citetalias{ali_et_al2015}, individual baselines were divided into five independent groups, where no baselines were repeated in each group. Then, baselines within each group were averaged together, and the groups were cross-multiplied to form power spectra. This grouping method was used to reduce computational time, however upon closer examination it has been found that the initial grouping introduces an element of randomness into the final measurements --- more specifically, the power spectrum value fluctuates depending on how baselines are assigned into their initial groups. Our new approach removes this element of randomness at the cost of computational expense, as we now perform all baseline cross-products.}

\item{Finally, the last change from the \citetalias{ali_et_al2015} method is that our power spectrum points (previously computed as the mean of all bootstraps), are now computed as the power spectrum estimate resulting from not bootstrapping at all. More specifically, we compute one estimate without sampling, and this estimate is propagated through our signal loss computation (this estimate is $\widehat{\textbf{P}}_{x}$). The difference between taking the mean of the bootstrapped values and using the estimate from the no-bootstrapping case is small, but doing the latter ensures that we are forming results that reflect the estimate preferred by all our data.}

\end{itemize}

In summary, we have learned several lessons regarding bootstrapping and have revised our analysis procedure in order to determine error bars that correctly reflect the variance in our power spectrum estimates. Bootstrapping can be an effective and straightforward way to estimate errors of a dataset, however, bootstrapping as a means of estimating power spectrum errors from real fringe-rate filtered data requires knowledge of the number of independent samples, which is not always a trivial task. We have thus avoided this issue by removing one of our bootstrap axes, as well as updated several other details of our procedure to ensure accurate re-sampling and error estimation.

\subsection{Theoretical Error Estimation}
\label{sec:Error}

One useful way of cross-checking measured power spectrum values and errors is to compute a theoretical estimation of thermal noise based on observational parameters. Although a theoretical model often differs from true errors, it is helpful to understand the ideal case and the factors that affect its sensitivity. Upon re-analysis of PAPER-64, we have discovered that this estimate was also underestimated in previous analyses. 

To compute our theoretical noise estimate, we use an analytic sensitivity calculation. Through detailed studies using several independently generated noise simulations, what we found was that our simulations all agreed but were discrepant with the previous calculations. The analytic 
calculation is only an approximation and attempts to combine a large number of pieces of information in an approximate way; however, when re-considering some of the approximations, the differences were large enough (factors of $10$ in some cases) to warrant a 
careful investigation. What follows here is an 
accounting of the differences which have been discovered. We note that our theoretical error estimate, which is plotted as the solid green curve in many of the previous power spectrum plots in this paper, is computed with these changes accounted for.

The noise prediction $n(k)$ (\citealt{parsons_et_al2012a}; \citealt{pober_et_al2013}) for a power spectral analysis of 
interferometric 21\,cm data, in temperature-units, is:

\begin{equation}
\label{eq:sense}
N(k) = \frac{X^{2}Y \Omega_{\rm eff} T_{\rm sys}^{2}}{\sqrt{2N_{\rm lst}N_{\rm seps}}\,t_{\rm int}N_{\rm days}N_{\rm bls}N_{\rm pols}}.
\end{equation}
We will now explain each factor in Equation \eqref{eq:sense} and highlight key differences from the numbers used in \citetalias{ali_et_al2015}.

\begin{figure}
	\centering
	\includegraphics[width=\columnwidth]{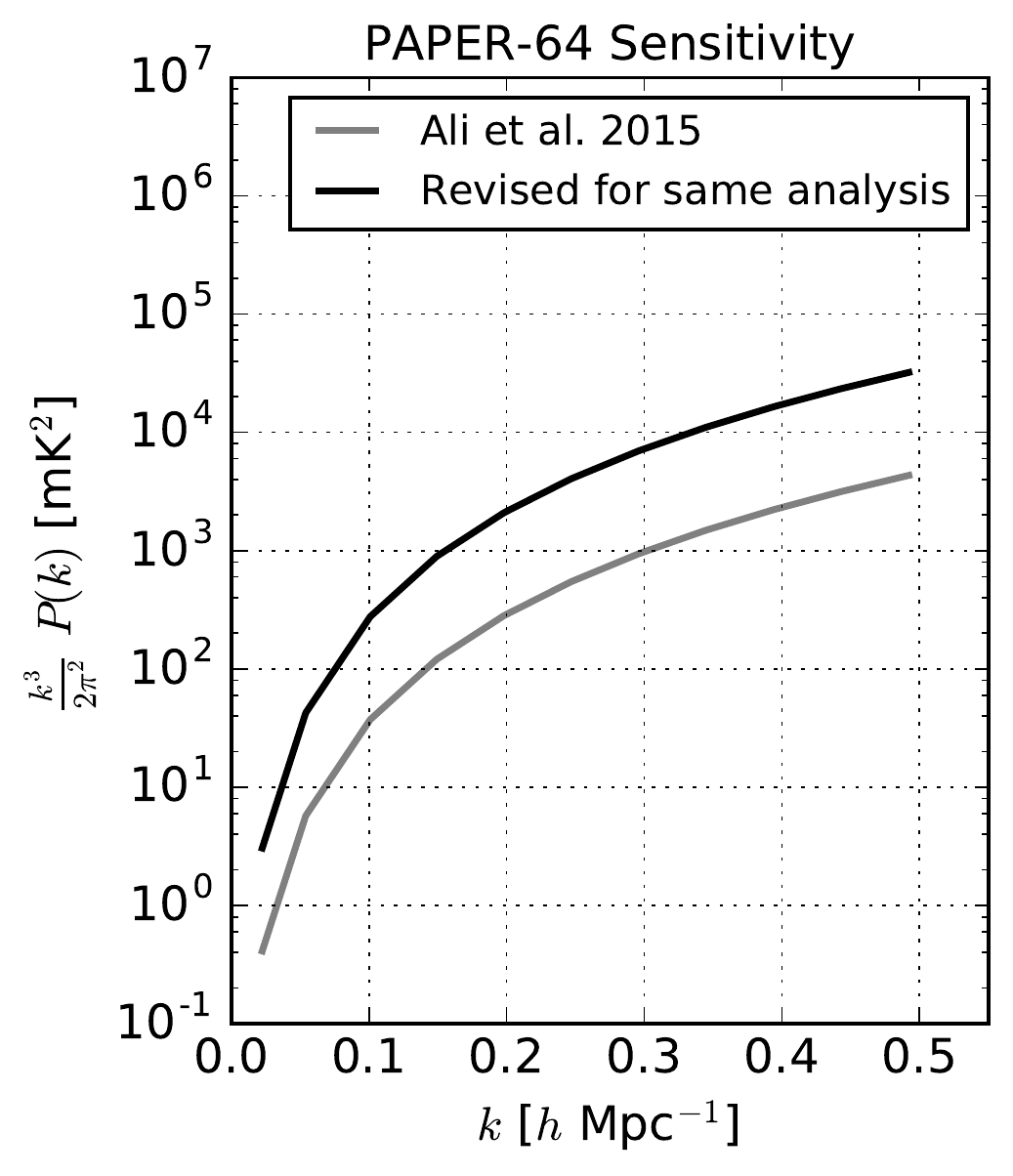}
	\caption{An updated prediction for the thermal noise level of PAPER-64 data (black) is shown in comparison to previously 
published sensitivity limits (gray), both computed for the parameters and methods used in \citetalias{ali_et_al2015}. Major factors that contribute to the discrepancy are $
\Omega_{\rm eff}$, $N_{\rm days}$ and $N_{\rm bls}$, as in Equation \eqref{eq:sense} and described in Section \ref{sec:Error}, which when combined decreases our 
sensitivity (higher noise floor) by a factor of $\sim7$ in mK$^{2}$.}
	\label{fig:sense_check}
\end{figure}

\begin{itemize}
\item $X^{2}Y$: Conversion factors from observing coordinates (angles on the sky and frequency) to cosmological coordinates (co-moving 
distances). For $z=8.4$, $X^{2}Y = 5 \times 10^{11} \, h^{-3}$ Mpc$^{3}$ str$^{-1}$ GHz$^{-1}$.
\item $\Omega_{\rm eff}$: The effective primary beam area in steradians (\citealt{parsons_et_al2010}; \citealt{pober_et_al2012}). 
The effective beam area changes with the application of a fringe-rate filter, since different parts of the beam are up-weighted and down-weighted. Using numbers from Table 1 in \citet{parsons_et_al2016}, $\Omega_{\rm eff} = 0.74^{2}/0.24$ for an optimal fringe-rate 
filter and the PAPER primary beam. 
\item $T_{\rm sys}$: The system temperature is set by:

\begin{equation}
\label{eq:sys}
T_{\rm sys} = 180\Big(\frac{\nu}{0.18}\Big)^{-2.55} + T_{\rm rcvr},
\end{equation}

where $\nu$ are frequencies in GHz (\citealt{thompson_et_al2001}). We use a receiver temperature of $144$\,K, yielding $T_{\rm sys} = 431$\,K at $150$\,MHz. 
This is lower than the $T_{\rm sys}$ of $500$\,K used in \citetalias{ali_et_al2015} because of several small miscalculation errors that were 
identified\footnote{For example, there was a missing a square root in going from a variance to a standard deviation.}.
\item $\sqrt{2}$: This factor in the denominator of the sensitivity equation comes from taking the real part of the power spectrum 
estimates after cross-multiplying two independent visibility measurements. In \citetalias{ali_et_al2015}, a factor of $2$ was mistakenly used.
\item $N_{\rm lst}$: The number of independent LST bins that go into a power spectrum estimation. The sensitivity scales as the square root 
because we integrate incoherently over time. For PAPER-64, $N_{\rm lst} = 8$.
\item $N_{\rm seps}$: The number of baseline separation types (where baselines of a unique separation type have the same orientation and length) averaged incoherently in a final power spectrum estimate. For the 
analysis in this paper, we only use one type of baseline (PAPER's 30\,m East/West baselines). However, both the updated limits in Kolopanis et al. (\textit{in prep.}) and the sensitivity prediction in Figure \ref{fig:sense_check} use three separation types ($N_{\rm seps}=3$) to match \citetalias{ali_et_al2015}.
\item $t_{\rm int}$: Length of an independent integration of the data. It is crucial to adapt this number if filtering is applied along the time axis (i.e., a 
fringe-rate filter). We compute the effective integration time of our fringe-rate filtered data by scaling the original integration time $t_{i}$
using the following:
\begin{equation}
t_{\rm int} = t_{i} \frac{\int1 \, df}{\int w^{2}(f) \,df},
\end{equation}
where $t_{i}=43$ seconds, $t_{\rm int}$ is the fringe-rate filtered integration time, $w$ is the fringe-rate profile, and the integral is 
taken over all fringe-rates. For PAPER-64, this number is $t_{\rm int} = 3857$\,s. 
\item $N_{\rm days}$: The total number of days of data analyzed. In \citetalias{ali_et_al2015}, this number was set to $135$. However, because we 
divide our data in half (to form ``even" and ``odd" datasets, or $N_{\rm datasets} = 2$), this number should reflect the number of days in each individual dataset instead of the total. Additionally, this number should be adjusted to reflect the actual number of cross-multiplications that occur between datasets (``even" with ``odd" and ``odd" with ``even", but not ``odd" with ``odd" or ``even" with ``even" in order to avoid noise biases). Finally, because our LST coverage is not $100\%$ complete (it doesn't overlap for every single day), we incorporate a root-mean-square statistic in computing a realistic value of 
$N_{\rm days}$. Our expression therefore becomes:
\begin{equation}
N_{\rm days} = \sqrt{\langle N_{i}^{2}\rangle} \sqrt{(N_{\rm datasets}^{2}-N_{\rm datasets})}
 \end{equation}
\noindent where $i$ indexes LST and frequency channel over all datasets (\citealt{jacobs_et_al2015}). For PAPER-64, our revised estimate of $N_{\rm days}$ is $\sim47$ 
days.
\item $N_{\rm bls}$: The number of baselines contributing to the sensitivity of a power spectrum estimate. In \citetalias{ali_et_al2015}, this number was 
the total number of $30$\,m East/West baselines used in the analysis. However, using the total number of baselines ($N_{\rm bls\_total} = 51$) neglects 
the fact that the \citetalias{ali_et_al2015} analysis averages baselines into groups for computational speed-up when cross-multiplying data. Our revised estimate for the parameter is:
\begin{equation}
N_{\rm bls} = \frac{N_{\rm bls\_total}}{N_{\rm gps}}\sqrt{\frac{N_{\rm gps}^{2}-N_{\rm gps}}{2}},
\end{equation}
\noindent where, in the \citetalias{ali_et_al2015} analysis, $N_{\rm gps} = 5$. Each baseline group averages down linearly as the number of baselines 
entering the group ($N_{\rm bls\_total}/N_{\rm gps}$) and then as the square root of the number of cross-multiplied pairs \Big($\sqrt{\frac{N_{\rm gps}^{2} - 
N_{\rm gps}}{2}}$\Big). A revised \citetalias{ali_et_al2015} analysis should therefore use $N_{\rm bls} \sim 32$ instead of $51$, and this change is taken into account in Figure \ref{fig:sense_check}. However, the analysis in this paper and in Kolopanis et al. (\textit{in prep.}) no longer averages baselines into groups ($N_{\rm gps} = 1$). For the subset of data presented in this paper, $N_{\rm bls} = 10$.
\item $N_{\rm pols}$: The number of polarizations averaged together. For the case of Stokes I, $N_{\rm pols}=2$.
\end{itemize}

An additional factor of $\sqrt{2}$ is gained in sensitivity when folding together positive and negative $k$'s to form $\Delta^{2}(k)$.

Our revised sensitivity estimate for the \citetalias{ali_et_al2015} analysis of PAPER-64 is shown in Figure \ref{fig:sense_check}. 
Together, the revised parameters yield a decrease in sensitivity (higher noise floor) by a factor of $\sim7$ in mK$^{2}$. 

To verify our thermal noise prediction, we form power spectra estimates using a pure noise simulation. We create Gaussian 
random noise assuming a constant $T_{\rm rcvr}$ (translated into $T_{\rm sys}$ via Equation \eqref{eq:sys}) but accounting for the true $N_{\rm days}$ as determined 
by LST sampling counts for each time and frequency in the LST-binned data. We convert $T_{\rm sys}$ into a root-mean-square variance statistic 
using:

\begin{equation}
\label{eq:noise}
T_{\rm rms} = \frac{T_{\rm sys}}{\sqrt{\Delta\nu \Delta t N_{\rm days} N_{\rm pols}}},
\end{equation}

\noindent where $\Delta\nu$ is the channel spacing, $\Delta t$ is the integration time, $N_{\rm days}$ is the number of daily counts for a 
particular time and frequency that went into our LST-binned set, and $N_{\rm pols}$ is the number of polarizations ($2$ for Stokes 
I). This temperature sets the variance of the Gaussian random noise.

\begin{figure*}
	\centering
	\includegraphics[width=0.4\textwidth]{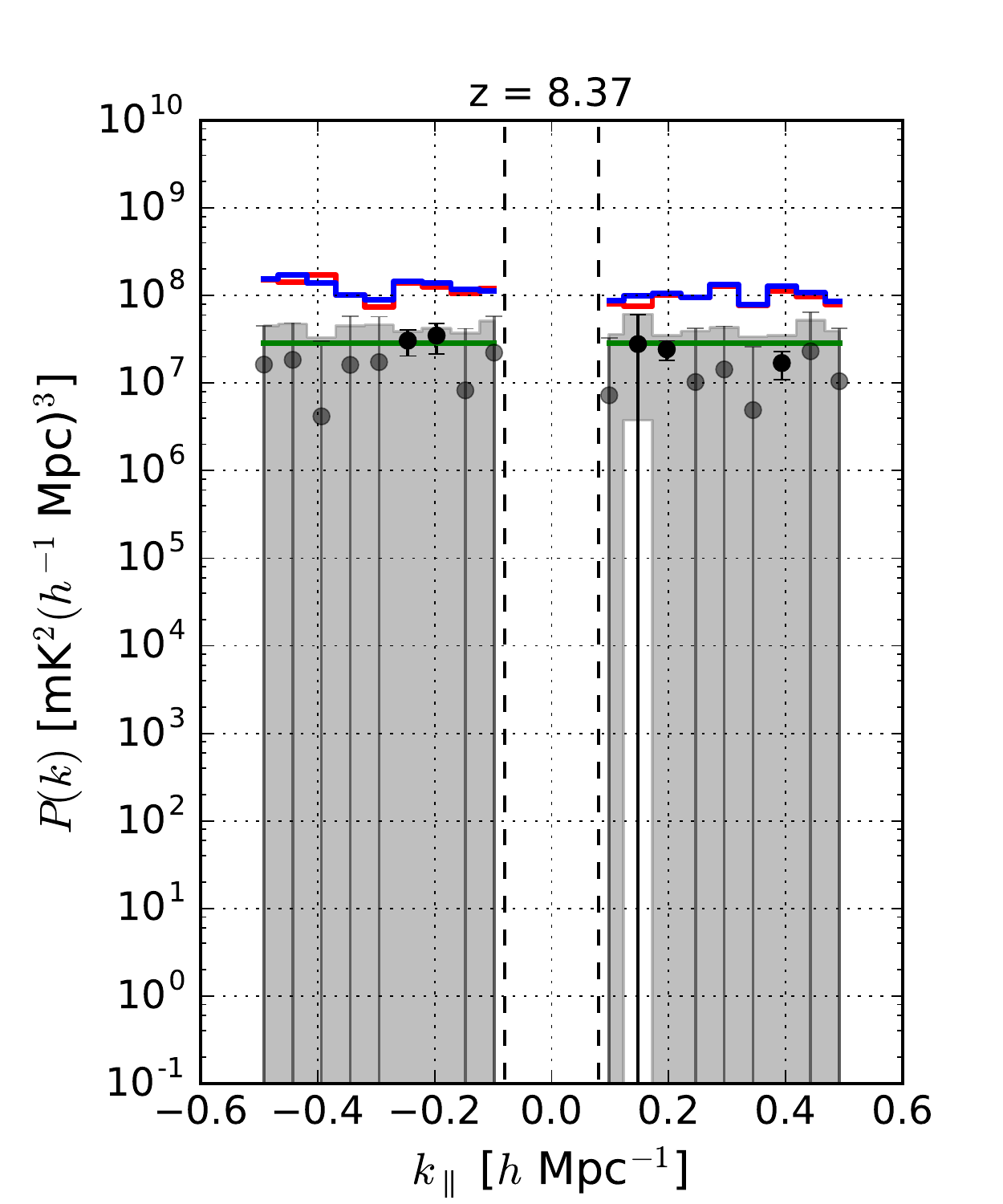}
	\includegraphics[width=0.4\textwidth]{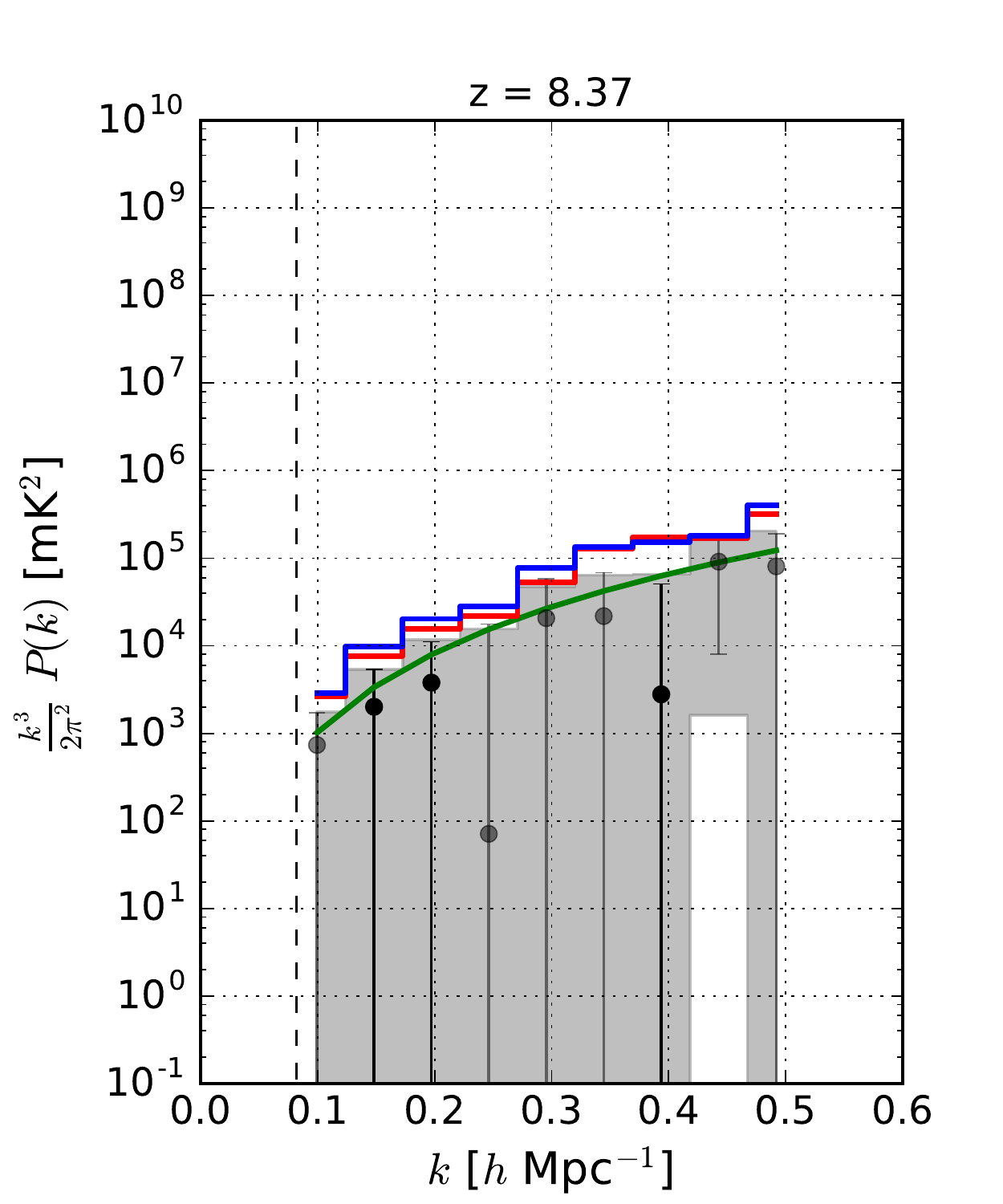}
	\caption{The power spectrum for a noise simulation that mimics the noise level of a subset of PAPER-64 data, where the solid red curve is the $2\sigma$ upper limit on the EoR signal estimated from our signal injection framework using $\widehat{\textbf{C}}_{\rm eff}$. The theoretical $2\sigma$ thermal noise level prediction based on observational parameters (calculated by Equation \eqref{eq:sense}) is in green. Additionally, the power spectrum result for the uniform weighted case is shown in three different ways: power spectrum values (black and gray points as positive and negative values, respectively, with $2\sigma$ error bars from bootstrapping), the $2\sigma$ upper limit on the EoR signal using our full signal injection framework (solid blue), and the measured power spectrum values with $2\sigma$ thermal noise errors (gray shaded regions). The vertical dashed black lines signify the horizon limit for this analysis using $30$\,m baselines. We highlight that the bootstrapped data points and thermal noise prediction show good agreement, while the limits from the full injection framework (red and blue) are inflated due to the additional inclusion of sample variance that comes from the injection simulations.}
	\label{fig:ps_noise}
\end{figure*}

Power spectrum results for the noise simulation, which uses our full power spectrum pipeline, are shown in Figure 
\ref{fig:ps_noise}. We highlight that the bootstrapped data (black and gray points, with $2\sigma$ error bars) and thermal noise prediction (solid green) show good agreement, as bootstrapping provides an accurate estimate of the noise variance. However, the limits from the full signal loss framework (weighted and unweighted in red and blue, respectively) are inflated, likely due to the additional inclusion of sample variance that comes from the EoR simulations. While the noise simulation provides an important indicator about the accuracy of our theoretical noise calculation, we note that the calculation did not take into account additional sources of error associated with earlier analysis steps (for example, \citet{trott_wayth_2017} show how calibration specifically can add errors to visibilities). Additionally, we recommend that future work investigate possible error correlations between baseline pairs and any interaction effects between signal and noise that may effect error calculations. Because of these reasons, we therefore interpret our noise prediction as the sensitivity floor for our measurements.

\section{Conclusion}
\label{sec:Con}

Although current 21\,cm published power spectrum upper limits lie several orders of magnitude above predicted EoR levels, 
ongoing analyses of deeper sensitivity datasets from PAPER, MWA, and LOFAR, as well as next generation instruments like 
HERA, are expected to continue to push towards EoR sensitivities. As the field progresses towards a detection, we have shown 
that it is crucial for future analyses to have a rigorous understanding of signal loss in an analysis pipeline and be able to accurately 
and robustly calculate both power spectrum and theoretical errors.

In particular, in this paper we have investigated the subtleties and tradeoffs of common 21\,cm power spectrum techniques on 
signal loss and error estimation, which can be summarized as follows:

\begin{itemize}
\item Substantial signal loss can result when weighting data using empirically estimated covariances due to couplings with the data realizations (Section 
\ref{sec:SiglossOverview}). Loss of the 21\,cm signal is especially significant the fewer number of independent modes that
exist in the data. Hence, there exists a trade-off between sensitivity driven 
time-averaging techniques such as fringe-rate filtering and signal loss when using empirically estimated covariances. 
\item Signal injection and recovery simulations can be used to quantify signal loss (Section \ref{sec:siglossmethod}). However, a 
signal-only simulation (i.e., comparing a uniformly weighted vs. weighted power spectrum of EoR only) can underestimate loss by 
failing to account for correlations between the data and signal which can be large and negative.
\item Errors that are estimated via bootstrapping can be underestimated if samples in the dataset are significantly correlated 
(Section \ref{sec:Boot}). However, if the number of independent samples in a dataset is well-determined, bootstrapping is a 
simple and accurate way of estimating errors.
\end{itemize}

As a consequence of our investigations, we have also used a subset of PAPER-64 data to make a new power spectrum analysis. This serves as an illustrative example of using a signal injection framework, correctly computing errors via bootstrapping, and accurately estimating thermal noise. Our revised PAPER-64 limits are presented in Kolopanis et al. (\textit{in prep.}), which supersede all previously published PAPER limits. Because of the many challenges associated with signal loss and its estimation as described in this paper, Kolopanis et al. (\textit{in prep.}) use a straightforward power spectrum estimation approach that is not lossy. However, the main reasons for a previously underestimated limit (\citealt{ali_et_al2018}) and 
ways in which our new analysis differs can still be summarized by the following:

\begin{itemize}
\item Signal loss, previously found to be $<2\%$ in \citetalias{ali_et_al2015}, was underestimated by a factor of $>$$1000$ for the case of empirically estimated inverse 
covariance weighting. Using a regularized covariance weighting method can minimize loss (Section 
\ref{sec:Weight}), however, because a regularized weighting method is not as aggressive as the former, it produces limits that are still higher than the lossy empirical inverse covariance limits. Underestimated signal loss therefore represents the bulk of our revision. 
\item Power spectrum errors, originally computed by bootstrapping, were underestimated for the data by a factor of $\sim2$ in mK due to oversampling data whose effective number of independent samples was reduced from fringe-rate filtering (Section \ref{sec:Boot}). Several other errors were also found regarding error estimation, though with smaller effects.
\item Several factors used in an analytic expression to predict the noise-level in PAPER-64 data were revised, yielding a 
decrease in predicted sensitivity level by a factor of $\sim3$ in mK (Section \ref{sec:Error}). We note that our sensitivity prediction is revised by a factor less than our overall
power spectrum result, implying that if taken at face value, the theoretical prediction for noise in \citetalias{ali_et_al2015} was too high for its data 
points.
\end{itemize}

The future of 21\,cm cosmology is exciting, as new experiments have sensitivities that expect to reach and surpass EoR levels, improved 
foreground mitigation and removal strategies are being developed, and simulations are being designed to better understand 
instruments. On the power spectrum analysis side, robust signal loss simulations and precise error calculations will play critical roles in accurate 21\,cm results. With strong foundations being established now, it is safe to say that we can expect to learn much about reionization and our early Universe in the coming years.


\section{Acknowledgements}
CC would like to acknowledge the UC Berkeley Chancellor's Fellowship and National Science Foundation Graduate Research 
Fellowship (Division of Graduate Education award 1106400). She would also like to thank Phil Bull, Bryna Hazelton, Miguel Morales, and Eric Switzer for helpful discussions. PAPER and HERA 
are supported by grants from the National Science Foundation (awards 1440343, and 1636646). ARP, DCJ, and JEA would 
also like to acknowledge NSF support (awards 1352519, 1401708, and 1455151, respectively). AL acknowledges support for this work by NASA through Hubble Fellowship grant \#HST-HF2-51363.001-A awarded by the Space Telescope Science Institute, which is operated by the Association of Universities for Research in Astronomy, Inc., for NASA, under contract NAS5-26555. SAK is supported by the University of Pennsylvania School of Arts and Sciences Dissertation Completion Fellowship. JSD acknowledges NSF AAPF
award 1701536. GB acknowledges support from the Royal Society and the Newton Fund under grant NA150184. This work is based on research supported in part by the National Research Foundation of South Africa (award 103424). We graciously thank SKA-SA for site infrastructure and observing support.
\label{sec:Ack}

\bibliographystyle{apj}
\bibliography{refs}

\begin{thebibliography}{76}
\expandafter\ifx\csname natexlab\endcsname\relax\def\natexlab#1{#1}\fi

\bibitem[{{Ali} {et~al.}(2008){Ali}, {Bharadwaj}, \&
  {Chengalur}}]{ali_et_al2008}
{Ali}, S.~S., {Bharadwaj}, S., \& {Chengalur}, J.~N. 2008, \mnras, 385, 2166

\bibitem[{{Ali} {et~al.}(2015){Ali}, {Parsons}, {Zheng}, {Pober}, {Liu},
  {Aguirre}, {Bradley}, {Bernardi}, {Carilli}, {Cheng}, {DeBoer}, {Dexter},
  {Grobbelaar}, {Horrell}, {Jacobs}, {Klima}, {MacMahon}, {Maree}, {Moore},
  {Razavi}, {Stefan}, {Walbrugh}, \& {Walker}}]{ali_et_al2015}
{Ali}, Z.~S., {et~al.} 2015, \apj, 809, 61

\bibitem[{Ali {et~al.}(2018)Ali, Parsons, Zheng, Pober, Liu, Aguirre, Bradley,
  Bernardi, Carilli, Cheng, DeBoer, Dexter, Grobbelaar, Horrell, Jacobs, Klima,
  MacMahon, Maree, Moore, Razavi, Stefan, Walbrugh, \& Walker}]{ali_et_al2018}
Ali, Z.~S., {et~al.} 2018, The Astrophysical Journal, 863, 201

\bibitem[{{Andrae}(2010)}]{andrae2010}
{Andrae}, R. 2010, ArXiv e-prints

\bibitem[{{Barkana} \& {Loeb}(2001)}]{barkana_and_loeb2001}
{Barkana}, R., \& {Loeb}, A. 2001, \physrep, 349, 125

\bibitem[{{Barkana} \& {Loeb}(2008)}]{barkana_and_loeb2008}
---. 2008, Monthly Notices of the Royal Astronomical Society, 384, 1069

\bibitem[{{Bernardi} {et~al.}(2009){Bernardi}, {de Bruyn}, {Brentjens},
  {Ciardi}, {Harker}, {Jeli{\'c}}, {Koopmans}, {Labropoulos}, {Offringa},
  {Pandey}, {Schaye}, {Thomas}, {Yatawatta}, \& {Zaroubi}}]{bernardi_et_al2009}
{Bernardi}, G., {et~al.} 2009, \aap, 500, 965

\bibitem[{{Bernardi} {et~al.}(2010){Bernardi}, {de Bruyn}, {Harker},
  {Brentjens}, {Ciardi}, {Jeli{\'c}}, {Koopmans}, {Labropoulos}, {Offringa},
  {Pandey}, {Schaye}, {Thomas}, {Yatawatta}, \& {Zaroubi}}]{bernardi_et_al2010}
---. 2010, \aap, 522, A67+

\bibitem[{Bernardi {et~al.}(2013)Bernardi, Greenhill, Mitchell, Ord, Hazelton,
  Gaensler, de~Oliveira-Costa, Morales, Shankar, Subrahmanyan, Wayth, Lenc,
  Williams, Arcus, Arora, Barnes, Bowman, Briggs, Bunton, Cappallo, Corey,
  Deshpande, deSouza, Emrich, Goeke, Herne, Hewitt, Johnston-Hollitt, Kaplan,
  Kasper, Kincaid, Koenig, Kratzenberg, Lonsdale, Lynch, McWhirter, Morgan,
  Oberoi, Pathikulangara, Prabu, Remillard, Rogers, Roshi, Salah, Sault,
  Srivani, Stevens, Tingay, Waterson, Webster, Whitney, Williams, \&
  Wyithe}]{bernardi_et_al2013}
Bernardi, G., {et~al.} 2013, The Astrophysical Journal, 771, 105

\bibitem[{{Bernardi} {et~al.}(2016){Bernardi}, {Zwart}, {Price}, {Greenhill},
  {Mesinger}, {Dowell}, {Eftekhari}, {Ellingson}, {Kocz}, \&
  {Schinzel}}]{bernardi_et_al2016}
{Bernardi}, G., {et~al.} 2016, \mnras, 461, 2847

\bibitem[{{Bond} {et~al.}(1998){Bond}, {Jaffe}, \& {Knox}}]{bond_et_al1998}
{Bond}, J.~R., {Jaffe}, A.~H., \& {Knox}, L. 1998, \prd, 57, 2117

\bibitem[{{Bowman} \& {Rogers}(2010)}]{bowman2010}
{Bowman}, J.~D., \& {Rogers}, A.~E.~E. 2010, \nat, 468, 796

\bibitem[{{Bowman} {et~al.}(2018){Bowman}, {Rogers}, {Monsalve}, {Mozdzen}, \&
  {Mahesh}}]{bowman_et_al2018}
{Bowman}, J.~D., {Rogers}, A.~E.~E., {Monsalve}, R.~A., {Mozdzen}, T.~J., \&
  {Mahesh}, N. 2018, \nat, 555, 67

\bibitem[{{Burns} {et~al.}(2012){Burns}, {Lazio}, {Bale}, {Bowman}, {Bradley},
  {Carilli}, {Furlanetto}, {Harker}, {Loeb}, \& {Pritchard}}]{burns2012}
{Burns}, J.~O., {et~al.} 2012, Advances in Space Research, 49, 433

\bibitem[{{Chang} {et~al.}(2010){Chang}, {Pen}, {Bandura}, \&
  {Peterson}}]{chang_et_al2010}
{Chang}, T.-C., {Pen}, U.-L., {Bandura}, K., \& {Peterson}, J.~B. 2010, \nat,
  466, 463

\bibitem[{{de Oliveira-Costa} {et~al.}(2008){de Oliveira-Costa}, {Tegmark},
  {Gaensler}, {Jonas}, {Landecker}, \& {Reich}}]{deOliveiraCosta_et_al2008}
{de Oliveira-Costa}, A., {Tegmark}, M., {Gaensler}, B.~M., {Jonas}, J.,
  {Landecker}, T.~L., \& {Reich}, P. 2008, \mnras, 388, 247

\bibitem[{DeBoer {et~al.}(2017)DeBoer, Parsons, Aguirre, Alexander, Ali,
  Beardsley, Bernardi, Bowman, Bradley, Carilli, Cheng, de~Lera~Acedo, Dillon,
  Ewall-Wice, Fadana, Fagnoni, Fritz, Furlanetto, Glendenning, Greig,
  Grobbelaar, Hazelton, Hewitt, Hickish, Jacobs, Julius, Kariseb, Kohn,
  Lekalake, Liu, Loots, MacMahon, Malan, Malgas, Maree, Martinot, Mathison,
  Matsetela, Mesinger, Morales, Neben, Patra, Pieterse, Pober, Razavi-Ghods,
  Ringuette, Robnett, Rosie, Sell, Smith, Syce, Tegmark, Thyagarajan, Williams,
  \& Zheng}]{deboer_et_al2017}
DeBoer, D.~R., {et~al.} 2017, Publications of the Astronomical Society of the
  Pacific, 129, 045001

\bibitem[{{Dillon} {et~al.}(2013){Dillon}, {Liu}, \&
  {Tegmark}}]{dillon_et_al2013a}
{Dillon}, J.~S., {Liu}, A., \& {Tegmark}, M. 2013, \prd, 87, 043005

\bibitem[{Dillon \& Parsons(2016)}]{dillon_parsons2016}
Dillon, J.~S., \& Parsons, A.~R. 2016, The Astrophysical Journal, 826, 181

\bibitem[{Dillon {et~al.}(2014)Dillon, Liu, Williams, Hewitt, Tegmark, Morgan,
  Levine, Morales, Tingay, Bernardi, Bowman, Briggs, Cappallo, Emrich,
  Mitchell, Oberoi, Prabu, Wayth, \& Webster}]{dillon_et_al2014}
Dillon, J.~S., {et~al.} 2014, Phys. Rev. D, 89, 023002

\bibitem[{Dillon {et~al.}(2015)Dillon, Neben, Hewitt, Tegmark, Barry,
  Beardsley, Bowman, Briggs, Carroll, de~Oliveira-Costa, Ewall-Wice, Feng,
  Greenhill, Hazelton, Hernquist, Hurley-Walker, Jacobs, Kim, Kittiwisit, Lenc,
  Line, Loeb, McKinley, Mitchell, Morales, Offringa, Paul, Pindor, Pober,
  Procopio, Riding, Sethi, Shankar, Subrahmanyan, Sullivan, Thyagarajan,
  Tingay, Trott, Wayth, Webster, Wyithe, Bernardi, Cappallo, Deshpande,
  Johnston-Hollitt, Kaplan, Lonsdale, McWhirter, Morgan, Oberoi, Ord, Prabu,
  Srivani, Williams, \& Williams}]{dillon_et_al2015}
---. 2015, Phys. Rev. D, 91, 123011

\bibitem[{{Dodelson} \& {Schneider}(2013)}]{dodelson_schneider2013}
{Dodelson}, S., \& {Schneider}, M.~D. 2013, \prd, 88, 063537

\bibitem[{Efron \& Tibshirani(1994)}]{efron_tibshirani1994}
Efron, B., \& Tibshirani, R. 1994, An Introduction to the Bootstrap, Chapman \&
  Hall/CRC Monographs on Statistics \& Applied Probability (Taylor \& Francis)

\bibitem[{{Furlanetto} {et~al.}(2006){Furlanetto}, {Oh}, \&
  {Briggs}}]{furlanetto_et_al2006}
{Furlanetto}, S.~R., {Oh}, S.~P., \& {Briggs}, F.~H. 2006, \physrep, 433, 181

\bibitem[{{Ghosh} {et~al.}(2011){Ghosh}, {Bharadwaj}, {Ali}, \&
  {Chengalur}}]{ghosh_et_al2011}
{Ghosh}, A., {Bharadwaj}, S., {Ali}, S.~S., \& {Chengalur}, J.~N. 2011, \mnras,
  418, 2584

\bibitem[{{Hartlap} {et~al.}(2007){Hartlap}, {Simon}, \&
  {Schneider}}]{hartlap_et_al2007}
{Hartlap}, J., {Simon}, P., \& {Schneider}, P. 2007, \aap, 464, 399

\bibitem[{Jacobs {et~al.}(2015)Jacobs, Pober, Parsons, Aguirre, Ali, Bowman,
  Bradley, Carilli, DeBoer, Dexter, Gugliucci, Klima, Liu, MacMahon, Manley,
  Moore, Stefan, \& Walbrugh}]{jacobs_et_al2015}
Jacobs, D.~C., {et~al.} 2015, \apj, 801, 51

\bibitem[{{Jacobs} {et~al.}(2016){Jacobs}, {Hazelton}, {Trott}, {Dillon},
  {Pindor}, {Sullivan}, {Pober}, {Barry}, {Beardsley}, {Bernardi}, {Bowman},
  {Briggs}, {Cappallo}, {Carroll}, {Corey}, {de Oliveira-Costa}, {Emrich},
  {Ewall-Wice}, {Feng}, {Gaensler}, {Goeke}, {Greenhill}, {Hewitt},
  {Hurley-Walker}, {Johnston-Hollitt}, {Kaplan}, {Kasper}, {Kim},
  {Kratzenberg}, {Lenc}, {Line}, {Loeb}, {Lonsdale}, {Lynch}, {McKinley},
  {McWhirter}, {Mitchell}, {Morales}, {Morgan}, {Neben}, {Thyagarajan},
  {Oberoi}, {Offringa}, {Ord}, {Paul}, {Prabu}, {Procopio}, {Riding}, {Rogers},
  {Roshi}, {Udaya Shankar}, {Sethi}, {Srivani}, {Subrahmanyan}, {Tegmark},
  {Tingay}, {Waterson}, {Wayth}, {Webster}, {Whitney}, {Williams}, {Williams},
  {Wu}, \& {Wyithe}}]{Jacobs2016}
{Jacobs}, D.~C., {et~al.} 2016, \apj, 825, 114

\bibitem[{Jaynes(1968)}]{jaynes1968}
Jaynes, E. 1968, IEEE Transactions on Systems Science and Cybernetics, 4, 227

\bibitem[{{Jeli{\'c}} {et~al.}(2008){Jeli{\'c}}, {Zaroubi}, {Labropoulos},
  {Thomas}, {Bernardi}, {Brentjens}, {de Bruyn}, {Ciardi}, {Harker},
  {Koopmans}, {Pandey}, {Schaye}, \& {Yatawatta}}]{jelic_et_al2008}
{Jeli{\'c}}, V., {et~al.} 2008, \mnras, 389, 1319

\bibitem[{{Joachimi}(2017)}]{joachimi_2017}
{Joachimi}, B. 2017, \mnras, 466, L83

\bibitem[{{Kohn} {et~al.}(2016){Kohn}, {Aguirre}, {Nunhokee}, {Bernardi},
  {Pober}, {Ali}, {Bradley}, {Carilli}, {DeBoer}, {Gugliucci}, {Jacobs},
  {Klima}, {MacMahon}, {Manley}, {Moore}, {Parsons}, {Stefan}, \&
  {Walbrugh}}]{kohn_et_al2016}
{Kohn}, S.~A., {et~al.} 2016, \apj, 823, 88

\bibitem[{{Koopmans} {et~al.}(2015){Koopmans}, {Pritchard}, {Mellema},
  {Aguirre}, {Ahn}, {Barkana}, {van Bemmel}, {Bernardi}, {Bonaldi}, {Briggs},
  {de Bruyn}, {Chang}, {Chapman}, {Chen}, {Ciardi}, {Dayal}, {Ferrara},
  {Fialkov}, {Fiore}, {Ichiki}, {Illiev}, {Inoue}, {Jelic}, {Jones}, {Lazio},
  {Maio}, {Majumdar}, {Mack}, {Mesinger}, {Morales}, {Parsons}, {Pen},
  {Santos}, {Schneider}, {Semelin}, {de Souza}, {Subrahmanyan}, {Takeuchi},
  {Vedantham}, {Wagg}, {Webster}, {Wyithe}, {Datta}, \&
  {Trott}}]{koopmans_et_al2015}
{Koopmans}, L., {et~al.} 2015, Advancing Astrophysics with the Square Kilometre
  Array (AASKA14), 1

\bibitem[{Liu \& Parsons(2016)}]{liu_parsons_2016}
Liu, A., \& Parsons, A.~R. 2016, Monthly Notices of the Royal Astronomical
  Society, 457, 1864

\bibitem[{{Liu} {et~al.}(2014{\natexlab{a}}){Liu}, {Parsons}, \&
  {Trott}}]{liu_et_al2014a}
{Liu}, A., {Parsons}, A.~R., \& {Trott}, C.~M. 2014{\natexlab{a}}, \prd, 90,
  023018

\bibitem[{{Liu} {et~al.}(2014{\natexlab{b}}){Liu}, {Parsons}, \&
  {Trott}}]{liu_et_al2014b}
---. 2014{\natexlab{b}}, \prd, 90, 023019

\bibitem[{Liu \& Tegmark(2011)}]{liu_tegmark2011}
Liu, A., \& Tegmark, M. 2011, Phys. Rev. D, 83, 103006

\bibitem[{{Loeb} \& {Furlanetto}(2013)}]{loeb_furlanetto_2013}
{Loeb}, A., \& {Furlanetto}, S. 2013, The First Galaxies in the Universe
  (Princeton University Press)

\bibitem[{{Masui} {et~al.}(2013){Masui}, {Switzer}, {Banavar}, {Bandura},
  {Blake}, {Calin}, {Chang}, {Chen}, {Li}, {Liao}, {Natarajan}, {Pen},
  {Peterson}, {Shaw}, \& {Voytek}}]{masui_et_al2013}
{Masui}, K.~W., {et~al.} 2013, \apjl, 763, L20

\bibitem[{Moore {et~al.}(2013)Moore, Aguirre, Parsons, Jacobs, \&
  Pober}]{moore_et_al2013}
Moore, D.~F., Aguirre, J.~E., Parsons, A.~R., Jacobs, D.~C., \& Pober, J.~C.
  2013, The Astrophysical Journal, 769, 154

\bibitem[{{Morales} \& {Wyithe}(2010)}]{morales_and_wyithe2010}
{Morales}, M.~F., \& {Wyithe}, J.~S.~B. 2010, \araa, 48, 127

\bibitem[{{Paciga} {et~al.}(2013{\natexlab{a}}){Paciga}, {Albert}, {Bandura},
  {Chang}, {Gupta}, {Hirata}, {Odegova}, {Pen}, {Peterson}, {Roy}, {Shaw},
  {Sigurdson}, \& {Voytek}}]{paciga_et_al2013}
{Paciga}, G., {et~al.} 2013{\natexlab{a}}, \mnras

\bibitem[{{Paciga} {et~al.}(2013{\natexlab{b}}){Paciga}, {Albert}, {Bandura},
  {Chang}, {Gupta}, {Hirata}, {Odegova}, {Pen}, {Peterson}, {Roy}, {Shaw},
  {Sigurdson}, \& {Voytek}}]{Paciga2013}
---. 2013{\natexlab{b}}, \mnras, 433, 639

\bibitem[{{Padmanabhan} {et~al.}(2016){Padmanabhan}, {White}, {Zhou}, \&
  {O'Connell}}]{padmanabhan_et_al2016}
{Padmanabhan}, N., {White}, M., {Zhou}, H.~H., \& {O'Connell}, R. 2016, \mnras,
  460, 1567

\bibitem[{{Parsons} {et~al.}(2012{\natexlab{a}}){Parsons}, {Pober}, {McQuinn},
  {Jacobs}, \& {Aguirre}}]{parsons_et_al2012a}
{Parsons}, A., {Pober}, J., {McQuinn}, M., {Jacobs}, D., \& {Aguirre}, J.
  2012{\natexlab{a}}, \apj, 753, 81

\bibitem[{{Parsons} {et~al.}(2016){Parsons}, {Liu}, {Ali}, \&
  {Cheng}}]{parsons_et_al2016}
{Parsons}, A.~R., {Liu}, A., {Ali}, Z.~S., \& {Cheng}, C. 2016, \apj, 820, 51

\bibitem[{{Parsons} {et~al.}(2012{\natexlab{b}}){Parsons}, {Pober}, {Aguirre},
  {Carilli}, {Jacobs}, \& {Moore}}]{parsons_et_al2012b}
{Parsons}, A.~R., {Pober}, J.~C., {Aguirre}, J.~E., {Carilli}, C.~L., {Jacobs},
  D.~C., \& {Moore}, D.~F. 2012{\natexlab{b}}, \apj, 756, 165

\bibitem[{{Parsons} {et~al.}(2010){Parsons}, {Backer}, {Foster}, {Wright},
  {Bradley}, {Gugliucci}, {Parashare}, {Benoit}, {Aguirre}, {Jacobs},
  {Carilli}, {Herne}, {Lynch}, {Manley}, \& {Werthimer}}]{parsons_et_al2010}
{Parsons}, A.~R., {et~al.} 2010, \aj, 139, 1468

\bibitem[{{Parsons} {et~al.}(2014){Parsons}, {Liu}, {Aguirre}, {Ali},
  {Bradley}, {Carilli}, {DeBoer}, {Dexter}, {Gugliucci}, {Jacobs},
  {et~al.}}]{parsons_et_al2014}
---. 2014, \apj, 788, 106

\bibitem[{{Patil} {et~al.}(2016){Patil}, {Yatawatta}, {Zaroubi}, {Koopmans},
  {de Bruyn}, {Jeli{\'c}}, {Ciardi}, {Iliev}, {Mevius}, {Pandey}, \&
  {Gehlot}}]{Patil2016}
{Patil}, A.~H., {et~al.} 2016, \mnras, 463, 4317

\bibitem[{{Patra} {et~al.}(2015){Patra}, {Subrahmanyan}, {Sethi}, {Udaya
  Shankar}, \& {Raghunathan}}]{patra2015}
{Patra}, N., {Subrahmanyan}, R., {Sethi}, S., {Udaya Shankar}, N., \&
  {Raghunathan}, A. 2015, \apj, 801, 138

\bibitem[{{Paz} \& {S{\'a}nchez}(2015)}]{paz_sanchez2015}
{Paz}, D.~J., \& {S{\'a}nchez}, A.~G. 2015, \mnras, 454, 4326

\bibitem[{{Pearson} \& {Samushia}(2016)}]{pearson_samushia2016}
{Pearson}, D.~W., \& {Samushia}, L. 2016, \mnras, 457, 993

\bibitem[{{Peterson}(2004)}]{peterson_et_al2004}
{Peterson}, U.-L.~P.~X.-P.~W.~J. 2004, ArXiv Astrophysics e-prints

\bibitem[{{Pober} {et~al.}(2012){Pober}, {Parsons}, {Jacobs}, {Aguirre},
  {Bradley}, {Carilli}, {Gugliucci}, {Moore}, \& {Parashare}}]{pober_et_al2012}
{Pober}, J.~C., {et~al.} 2012, \aj, 143, 53

\bibitem[{Pober {et~al.}(2013)Pober, Parsons, Aguirre, Ali, Bradley, Carilli,
  DeBoer, Dexter, Gugliucci, Jacobs, Klima, MacMahon, Manley, Moore, Stefan, \&
  Walbrugh}]{pober_et_al2013b}
Pober, J.~C., {et~al.} 2013, The Astrophysical Journal Letters, 768, L36

\bibitem[{{Pober} {et~al.}(2013){Pober}, {Parsons}, {DeBoer}, {McDonald},
  {McQuinn}, {Aguirre}, {Ali}, {Bradley}, {Chang}, \&
  {Morales}}]{pober_et_al2013}
{Pober}, J.~C., {et~al.} 2013, \aj, 145, 65

\bibitem[{{Pober} {et~al.}(2014){Pober}, {Liu}, {Dillon}, {Aguirre}, {Bowman},
  {Bradley}, {Carilli}, {DeBoer}, {Hewitt}, {Jacobs},
  {et~al.}}]{pober_et_al2014}
---. 2014, \apj, 782, 66

\bibitem[{{Pope} \& {Szapudi}(2008)}]{pope_szapudi2008}
{Pope}, A.~C., \& {Szapudi}, I. 2008, \mnras, 389, 766

\bibitem[{{Pritchard} \& {Loeb}(2010)}]{pritchard_and_loeb2010}
{Pritchard}, J.~R., \& {Loeb}, A. 2010, \prd, 82, 023006

\bibitem[{{Pritchard} \& {Loeb}(2012)}]{pritchard_loeb2012}
---. 2012, Reports on Progress in Physics, 75, 086901

\bibitem[{{Santos} {et~al.}(2005){Santos}, {Cooray}, \&
  {Knox}}]{santos_et_al2005}
{Santos}, M.~G., {Cooray}, A., \& {Knox}, L. 2005, \apj, 625, 575

\bibitem[{{Sellentin} \& {Heavens}(2016)}]{sellentin_heavens2016}
{Sellentin}, E., \& {Heavens}, A.~F. 2016, \mnras, 456, L132

\bibitem[{{Sokolowski} {et~al.}(2015){Sokolowski}, {Tremblay}, {Wayth},
  {Tingay}, {Clarke}, {Roberts}, {Waterson}, {Ekers}, {Hall}, {Lewis},
  {Mossammaparast}, {Padhi}, {Schlagenhaufer}, {Sutinjo}, \&
  {Tickner}}]{sokolowski2015}
{Sokolowski}, M., {et~al.} 2015, \pasa, 32, e004

\bibitem[{{Switzer} {et~al.}(2015){Switzer}, {Chang}, {Masui}, {Pen}, \&
  {Voytek}}]{switzer_et_al2015}
{Switzer}, E.~R., {Chang}, T.-C., {Masui}, K.~W., {Pen}, U.-L., \& {Voytek},
  T.~C. 2015, \apj, 815, 51

\bibitem[{{Switzer} {et~al.}(2013){Switzer}, {Masui}, {Bandura}, {Calin},
  {Chang}, {Chen}, {Li}, {Liao}, {Natarajan}, {Pen}, {Peterson}, {Shaw}, \&
  {Voytek}}]{switzer_et_al2013}
{Switzer}, E.~R., {et~al.} 2013, \mnras, 434, L46

\bibitem[{{Taylor} \& {Joachimi}(2014)}]{taylor_joachimi_etal2014}
{Taylor}, A., \& {Joachimi}, B. 2014, \mnras, 442, 2728

\bibitem[{{Tegmark}(1997)}]{tegmark_et_al1997a}
{Tegmark}, M. 1997, \prd, 55, 5895

\bibitem[{{Thompson} {et~al.}(2001){Thompson}, {Moran}, \&
  {Swenson}}]{thompson_et_al2001}
{Thompson}, A.~R., {Moran}, J.~M., \& {Swenson}, Jr., G.~W. 2001,
  {Interferometry and Synthesis in Radio Astronomy, 2nd Edition}

\bibitem[{{Tingay} {et~al.}(2013){Tingay}, {Goeke}, {Bowman}, {Emrich}, {Ord},
  {Mitchell}, {Morales}, {Booler}, {Crosse}, {Wayth},
  {et~al.}}]{tingay_et_al2013}
{Tingay}, S.~J., {et~al.} 2013, \pasa, 30, 7

\bibitem[{Trott \& Wayth(2017)}]{trott_wayth_2017}
Trott, C.~M., \& Wayth, R.~B. 2017, Publications of the Astronomical Society of
  Australia, 34, e061

\bibitem[{{Trott} {et~al.}(2012){Trott}, {Wayth}, \&
  {Tingay}}]{trott_et_al2012}
{Trott}, C.~M., {Wayth}, R.~B., \& {Tingay}, S.~J. 2012, \apj, 757, 101

\bibitem[{Trott {et~al.}(2016)Trott, Pindor, Procopio, Wayth, Mitchell,
  McKinley, Tingay, Barry, Beardsley, Bernardi, Bowman, Briggs, Cappallo,
  Carroll, de~Oliveira-Costa, Dillon, Ewall-Wice, Feng, Greenhill, Hazelton,
  Hewitt, Hurley-Walker, Johnston-Hollitt, Jacobs, Kaplan, Kim, Lenc, Line,
  Loeb, Lonsdale, Morales, Morgan, Neben, Thyagarajan, Oberoi, Offringa, Ord,
  Paul, Pober, Prabu, Riding, Shankar, Sethi, Srivani, Subrahmanyan, Sullivan,
  Tegmark, Webster, Williams, Williams, Wu, \& Wyithe}]{trott_et_al2016}
Trott, C.~M., {et~al.} 2016, The Astrophysical Journal, 818, 139

\bibitem[{{van Haarlem} {et~al.}(2013){van Haarlem}, {Wise}, {Gunst}, {Heald},
  {McKean}, {Hessels}, {de Bruyn}, {Nijboer}, {Swinbank}, {Fallows},
  {et~al.}}]{van_haarlem_et_al2013}
{van Haarlem}, M.~P., {et~al.} 2013, \aap, 556, A2

\bibitem[{{Voytek} {et~al.}(2014){Voytek}, {Natarajan}, {J{\'a}uregui
  Garc{\'{\i}}a}, {Peterson}, \& {L{\'o}pez-Cruz}}]{voytek2014}
{Voytek}, T.~C., {Natarajan}, A., {J{\'a}uregui Garc{\'{\i}}a}, J.~M.,
  {Peterson}, J.~B., \& {L{\'o}pez-Cruz}, O. 2014, \apjl, 782, L9

\bibitem[{{Wu}(2009)}]{wu2009}
{Wu}, X. 2009, in Bulletin of the American Astronomical Society, Vol.~41,
  American Astronomical Society Meeting Abstracts \#213, 474

\end{thebibliography}

\end{document}